%% file: susyRev_ijmpa.tex
\documentclass[10pt,a4paper]{article}
\usepackage{amsmath}
\usepackage{amssymb}
\usepackage[compress]{cite}
\usepackage{susyRev_ijmpa-defs}
\usepackage{xspace}
\usepackage{xcolor,colortbl}
\usepackage{hyperref}
\usepackage{subfigure}
\usepackage{hepnames}
\usepackage{tikz}
\usepackage{authblk}
\usepackage{doi}

\begin{document}

\title{Status of searches for electroweak-scale supersymmetry  after LHC~Run~2}


\author[1]{Wolfgang Adam\thanks{wolfgang.adam@oeaw.ac.at}}
\author[2]{Iacopo Vivarelli\thanks{Corresponding author, i.vivarelli@sussex.ac.uk}}

\affil[1]{\small{Institute of High Energy Physics, Austrian Academy of Sciences,
  Nikolsdorferg. 18, A-1050 Wien, Austria}}
\affil[2]{\small{Physics and Astronomy Department, 4A7 Pevensey 2,
University of Sussex, BN1 9QH Falmer, United Kingdom}}

\date{}

\maketitle

\begin{abstract}
The second period of datataking at the Large Hadron Collider (LHC) has provided a large dataset of proton-proton collisions that is unprecedented in terms of its centre-of-mass energy of 13\unit{TeV} and integrated luminosity of almost 140\,\ifb.
These data constitute a formidable laboratory for the search for new particles predicted by models of supersymmetry.
The analysis activity is still ongoing, but a host of results on supersymmetry has already been released by the general purpose LHC experiments ATLAS and CMS.
In this paper, we provide a map into this remarkable body of research, which spans a multitude of experimental signatures and phenomenological scenarios.
In the absence of conclusive evidence for the production of supersymmetric particles we discuss the constraints obtained in the context of various models.
We finish with a short outlook on the new opportunities for the next runs that will be provided by the upgrade of detectors and accelerator.

\end{abstract}

\clearpage

\tableofcontents

\section{Introduction}
\input{tex/introduction}

\section{Models of signal}
\label{sec:models}
\input{tex/models}

\section{Experimental techniques in the search for supersymmetry}
\label{sec:exp_techniques}
\input{tex/exp_techniques}

\section{Simplified model results}
\input{tex/simplified_model_results}

\section{Evolution of results and interpretations in complete models}
\input{tex/model_interpr}

\section{Summary and outlook}
\input{tex/conclusions}

\section*{Acknowledgments}

We are very grateful to the conveners of ATLAS and CMS Supersymmetry physics groups for helpful feedback.

\clearpage
\appendix

\input{tex/analysisSummary_pdgFormat}

\clearpage
\bibliography{susyRev_ijmpa}{}
\bibliographystyle{susyRev_ijmpa}

\end{document}

%% file: tex/introduction.tex
Among all theoretical paradigms developed over the years by the theory community, electroweak-scale Supersymmetry~\cite{Golfand:1971iw,Volkov:1973ix,Wess:1974tw,Wess:1974jb,Ferrara:1974pu,Salam:1974ig,Nilles:1983ge,Barbieri:1982eh} (SUSY) is certainly the one that has received most attention within high-energy physics. The ability of simple (minimal) supersymmetric extensions of the Standard Model (SM) to stabilise the Higgs boson mass under quantum corrections, to hint at a potential GUT scale for the unification of the gauge couplings, and to provide a potential dark matter candidate is certainly inspiring. A testament to this interest is the remarkable experimental effort that has been devoted over four decades to prove the existence of supersymmetric particles\cite{Dawson:1983fw,Haber:1984rc}: a significant fraction of the programme for direct search of production of new particles at LEP, Tevatron, and now the Large Hadron Collider specifically focuses on signatures and final states predicted by supersymmetric models.  \\

However, such monumental effort has not been awarded with a long-sought discovery. There is no conclusive evidence of direct production of supersymmetric particles at a collider, nor is there any unambiguous significant deviation from the predictions of the SM that can be directly attributed to supersymmetric modifications of the SM. The consequent exclusion limits from the LHC experiments in particular have caused a paradigm-shifting impact on the view of Supersymmetry in the community. Models of Supersymmetry that were considered mainstream before the LHC startup have been completely excluded by the experimental evidence. Some of the key predictions of electroweak-scale Supersymmetry\cite{Barbieri:1987fn,deCarlos:1993yy,Witten:1981nf,Dine:1981za,Sakai:1981gr,Kaul:1981hi} (existence of TeV-scale partners of the gluon, top quark, Higgs bosons) have been challenged and progressively constrained by the experimental results. Truly the LHC is fulfilling its mission of full exploration of the TeV-scale, and the answer that so far this remarkable machine is providing is one that is forcing the theory community to fully re-think the approach to model building.

The two LHC collaborations operating the general-purpose detectors (ATLAS and CMS) have recently published a wealth of results using the dataset of proton-proton collisions produced at a centre-of-mass energy of $\sqrt{s} = 13$ TeV. Such collisions have been collected by the two experiments in the period 2015-2018, during the so-called Run 2, which followed Run 1 that took place between 2009 and 2012 at $\sqrt{s} = 7$ and 8 TeV. The Run 2 dataset corresponds to an integrated luminosity of about 140 \ifb of proton-proton collisions per experiment. The time the LHC will take to significantly increase the dataset size is now substantially longer than in the past: it will take a large fraction of Run 3, foreseen to start in 2022, to double the integrated luminosity. At the same time, no big increase in the centre-of-mass energy of the collisions is expected. While the ability of the skilled experimental particle physicists working in the two collaborations will certainly improve the current constrains on the existence of supersymmetric particles, it seems clear that many of the current results will have a long lifetime. 

This paper wants to be a map into a remarkable body of experimental research that ATLAS and CMS have produced in the field of SUSY. We will mostly limit ourselves to published results, although in some case we refer to preliminary collaboration results: the latter are not yet peer-reviewed, and may be subject to (usually small) changes during the review process. We will start from the collaborations' summary plots, which attempt to deliver a highly summarised and often schematic message to the community, and we will highlight the assumptions which are made, connect them to other searches produced by the collaborations, and eventually try to present a global view of what the status of the search for SUSY particles after Run 2 is. We hope that the paper will help colleagues to orient themselves and appreciate the coherence of an experimental effort that can often be perceived as hard to follow by those not directly involved in it. 

Before diving into the review, we remind the reader about the main features of the simplest supersymmetric extensions to the SM. A complete theoretical overview of Supersymmetry goes well beyond the scope of this review, but there are excellent textbooks and long papers (a few examples are listed in the references\cite{baer_tata_2006, Martin:1997ns}), which can satisfy even the most eager reader. 

Supersymmetric extensions to the SM introduce a bosonic (fermionic) degree of freedom for every fermionc (bosonic) degree of freedom of the SM. For every SM fermion, two scalar fields are introduced that mix to yield two mass eigenstates, numbered in order of increasing mass. So, for example, two scalar fields (indicated with $\tilde{f}_{\mathrm{L}}$ and $\tilde{f}_{\mathrm{R}}$) correspond to the two chiral states of the generic fermionic field field, $f_{\mathrm{L}}$ and $f_{\mathrm{R}}$, and mix together to yield the mass eigenstates $\tilde{f}_1$ and $\tilde{f}_2$\footnote{In the following we will refer to the mass eigenstates as "SUSY particles" or "sparticles". Unless differently stated, the symbol $\tilde{f}$ will indicate the lightest mass eigenstate of the supersymemtric partner of the SM fermion $f$.}. In the following, the notation $\tilde{f}_{\mathrm{L,R}}$ indicates that the two chirality states\footnote{For the sake of brevity, we will slightly abuse the term ``chirality'' to indicate the scalar superpartners of the SM fermionic chirality states.} are assumed to be mass-degenerate. In supersymmetric models, a single Higgs doublet is not sufficient to give mass to the up-type and down-type fermions without breaking Supersymmetry, therefore a second Higgs doublet is introduced, leading to the prediction of five Higgs boson states. The ratio between the vacuum expectation values of the two Higgs doublets is indicated with $\tan \beta$. In the following, unless differently stated, it is assumed that the lightest of these states, indicated with $h$, is an SM-like Higgs boson. The supersymmetric partners of these Higgs boson states (the higgsinos, $\tilde{\mathbf{H}}$) mix with the supersymmetric partners of the $B$ (the bino, $\tilde{B}$)  and $\mathbf{W}$ (the wino triplet, $\tilde{\mathbf{W}}$) SM fields\footnote{The  $B$ and $\mathbf{W}$ fields of the SM mix to yield the mass eigenstates $\gamma$, $Z$, $W^{\pm}$.} to yield eight ``electroweakino'' states, four neutralinos $\ninoone....\ninofour$ and two pairs of charginos $\chinoonepm, \chinotwopm$. Experimental evidence tells us that SUSY is not an exact symmetry of nature: the supersymmetric particles' mass scales are unknown. However, supersymmetric extensions of the Standard Model are often invoked to solve the hierarchy problem\cite{PhysRevD.13.974,PhysRevD.14.1667,PhysRevD.19.1277,PhysRevD.20.2619} and stabilise the $h$ boson mass at the electroweak scale. In order to do so, a subset of the new SUSY particles (most notably the higgsinos, stops, gluinos) are required to be at the TeV scale\cite{Barbieri:1987fn,deCarlos:1993yy,Ellis:1986yg,Papucci:2011wy,Brust:2011tb}. Although there is still a debate in the community about how strict these bounds actually are on stops and gluinos\cite{BAER2017451}, the prediction that higgsinos should not be heavier than a few hundred GeV seems well established~\cite{Barbieri:2009ev,Baer:2011ec,Papucci:2011wy,Baer:2012up,Han:2013usa}.

To explicitly forbid a too fast proton decay, the conservation of a multiplicative quantum number, called \Rparity, can be imposed~\cite{Farrar:1978xj}: the \Rparity is $+1$ for SM particles and $-1$   for supersymmetric particles. \Rparity conservation (RPC) imposes that SUSY particles are always present in even numbers at an interaction vertex. Therefore, SUSY particles are produced in pairs, and the lightest supersymmetric particle (LSP) is stable. If the LSP is also weakly interacting, it provides the typical collider signature of missing transverse momentum in the final state. Explicit \Rparity violation can be included by adding the following terms to the lagrangian\cite{Mohapatra:2015fua}: 

\begin{eqnarray}
W_{\mathrm{RPV}} = \lambda_{ijk} L_iL_je_k + \lambda'_{ijk} Q_iL_j d_k + \lambda''_{ijk} u_id_jd_k + m_i L_iH_u, 
\end{eqnarray}

\hfill \break
\noindent where $L,Q,e,u,d$ are matter multiplets, and $H_u$ is the up-type Higgs field. The RPV couplings $(\lambda, \lambda', \lambda'')$, also referred to as LLE, LQD and UDD couplings, determine the phenomenology of the RPV sparticle decays. Typically one has to assume small values of the RPV couplings, in order to preserve consistency with existing constraints with lepton and baryon number conservation: this implies that the RPV coupling becomes phenomenologically relevant only in absence of competing RPC decays. The most important consequence of the introduction of RPV couplings is that the lightest supersymmetric particle is not stable.

This paper focuses on the search for pair production of supersymmetric particles. However, while these searches have a very clear theoretical drive, the experimental signatures investigated have an applicability that is significantly wider than the initial purpose. For example, signatures of production of invisible particles in association with jets, top quarks, leptons, etc., are ubiquitous in more generic models of dark matter\cite{Buckley:2014fba,Abercrombie:2015wmb,LHCDarkMatterWorkingGroup:2018ufk}: the main differences arise in terms of cross sections (well-predicted in SUSY, depending on the choice of the essentially arbitrary couplings in more generic dark matter models) and sometimes in terms of kinematics of the final state (typically the assumed intermediate states are different, leading to a different share of energy-momentum between the final state particles). Another example is the production of leptoquarks\cite{dimop_suss,techni2,techni3}. Pair-produced leptoquarks can decay into a quark and a neutrino, leading to two quarks and missing transverse momentum final states, which are studied as part of supersymmetric searches.

 At the same time, it is a fact that many signatures studied as part of other search programmes or precision measurements are relevant to set constraints on SUSY models. This is certainly the case for a large fraction of the programme for the search of exotic long-lived signatures or resonances, which provide strong constraints on key SUSY scenarios (in particular models with non-prompt decays of SUSY particles, or with RPV couplings leading to resonant SUSY particle decays). It is also the case for a large set of precision measurements that provides constraints on supersymmetric scenarios: a classical example in this sense is the measurement of spin correlations in \ttbar final states constraining stop pair production\cite{CMS:2019nrx,ATLAS:2019zrq} in scenarios with $\mass{\stopone} \approx \mass{t}$ and $\mass{\ninoone} \approx 0$.     

It is therefore not simple, and to some extent arbitrary, to decide what should be part of a SUSY review. The choice we have made is to limit ourselves to the search for the direct production of supersymmetric particles, leaving out everything else. According to this choice, all searches for extended Higgs sectors in particular via 2HDM models\cite{PhysRevD.8.1226}, which are certainly very relevant as constraints in particular for the MSSM, are not discussed. Recent measurements of discrepancies in the muon gyromagnetic factor with respect to the commonly accepted theoretical value\cite{PhysRevLett.126.141801}, confirming previously observed discrepancies\cite{PhysRevD.73.072003}, are certainly intriguing, and, if confirmed, they would be easily accommodated in a SUSY framework. However, potential SUSY interpretations of such discrepancies have a very rich  literature supporting them\cite{Chakraborti:2020vjp}: we do not discuss these aspects explicitly in this review. 

This paper is organised as follows: Sec. 2 provides a review of the main paradigms used for the definition of the signals of interest at the LHC, highlighting simplified models as the building block over which the ATLAS and CMS search strategy has been built. Section 3 provides a review of the main experimental aspects connected with the reconstruction of the final state objects and with the background estimation. Section 4 provides a review of the main searches developed by the two collaborations for the detection of supersymmetric particles produced via strong or electroweak interactions. We start from the summary plots provided by the collaborations, and discuss the hypotheses made in defining the simplified models used to derive those results. We then discuss briefly those searches that target complementary simplified models, based on different, well-motivated hypotheses.  Results relating to RPV SUSY scenarios, or scenarios involving long-lived SUSY particles, are also discussed extensively. Section 5 is instead focusing on the evolution of the limits with time, also touching upon interpretations of the search results in more complete SUSY models. Finally, Sec. 6 draws some conclusions and provides an outlook for what can be expected in the near future.

%% file: tex/models.tex
Physicists not working directly on searches for SUSY often find hard to stay up to date with the constraints that ATLAS and CMS set on the supersymmetric parameter space: indeed, the number of different SUSY results released by the two collaborations is large: about 50 different analyses have been released by each of the two collaborations during Run 2, often with multiple updates. The proliferation of constraints stems from the need of addressing a very high parameter dimensionality, which is a common trait of most supersymmetric models. 

The well-known Minimal Supersymmetric Extension of the Standard Model (MSSM)\cite{Fayet:1976et,Fayet:1977yc} is built by first extending the SM Higgs sector with a second scalar doublet (in order to be able to give mass to both the down-type and up-type fermions), then by introducing supersymmetric partner fields of the scalars, fermions and vector bosons of the SM in order to obtain a Lagrangian which is invariant under the Supersymmetry operator. The final step is the introduction of the so-called soft supersymmetry-breaking terms\cite{Martin:1997ns}, necessary to yield the EW-scale particle spectrum which has been observed in several decades of particle physics experiments, and which shows no evidence of supersymmetric partners so far.  

The soft supersymmetry-breaking terms introduce many parameters: 105 in the MSSM, including supersymmetric particle masses and field phases that cannot be absorbed by a redefinition of the fields. Clearly, some sort of guiding principle is needed, to be able to structure a suitable research strategy for the investigation of such a vast parameter space. 

Even after the application of constraints which are well motivated by experimental observations\footnote{These are the absence of new sources of CP violations as required by, e.g., results on electron and neutron dipole moments; the absence of flavour-changing neutral currents; the universality of first and second generation sfermions, as required by, e.g., the neutral kaon system.}, the number of remaining parameters is still relatively large: this is known as the phenomenological MSSM\cite{Djouadi:2002ze,Berger:2008cq,MSSMWorkingGroup:1998fiq}, or pMSSM, and contains 19 parameters if the additional hypothesis that the lightest supersymmetric particle is the lightest neutralino is made. Non-minimal extensions of the Standard Model typically add additional fields and corresponding supersymmetric counterparts, further enlarging the number of parameters. 

From an experimental point of view, the question of how to cover such a vast parameter space with a manageable number of searches is a non-trivial one. A possible answer is to rely on some sort of theoretical paradigm to define the signatures of interest. One famous example is  the five-parameter constrained MSSM\cite{Kane:1993td} (cMSSM), which dominated the SUSY search landscape for many years, but was already heavily constrained by the early LHC Run 1 data. Others have supported the SUSY search programmes of the collaborations even in more recent times. To recall the main ones and their most common event topologies: 

\begin{itemize}
\item Some final state event topologies are recurrent in classes of models with well-defined SUSY symmetry breaking patterns. For example, Gauge Mediated Supersymmetry Breaking\cite{Giudice:1998bp} (GMSB) models  are defined within the General Gauge Mediation\cite{Meade:2008wd} (GGM) framework, and  feature an interesting SUSY particle spectrum, where the LSP is almost always the gravitino ($\tilde{G}$). The specific phenomenology of a given GMSB/GGM model is often determined by the nature of the next-to-lightest supersymmetric particle (NLSP). Large classes of models feature a bino LSP, yielding final states including photons or $Z$ bosons resulting from the $\tilde{B}\rightarrow \tilde{G}$ transition. Other models feature a stau $\stauone$ as NLSP, yielding final states rich of $\tau$ leptons. As an additional example, Anomaly Mediated Symmetry Breaking (AMSB) models \cite{Giudice:1998xp,Randall:1998uk} tend to feature a pure wino LSP, often yielding long-lived charginos.
\item In Split SUSY models~\cite{Giudice:2004tc,ArkaniHamed:2004fb}, the fermionic supersymmetric partners are typically decoupled in mass from the scalar partners. This implies that the only particles which are energetically accessible at the LHC may be the electroweakinos, rather than the strongly produced states. 
\item  In Stealth SUSY models~\cite{Fan_2011}, SUSY particles decay through a compressed hidden sector which is absorbing all the invisible particles' momentum. Therefore, the available missing transverse momentum in the final state is suppressed even in presence of \Rparity conservation. 
\end{itemize}

Already in the second part of Run 1, the two collaborations started to make extensive use of simplified model spectra\cite{Alwall:2008ve,Alwall:2008ag,Alves:2011wf,Arkani-Hamed:2007gys}. A simplified model is one where only a few supersymmetric particle production and decay modes are considered, often only one. The estimate of the production cross sections is performed under the assumption that any  supersymmetric particle other than those considered in the model contributes in a negligible way, even as a quantum correction. The reader should be aware that for this approximation to be realised, often a rather extreme mass decoupling of the SUSY particles is required. For example, gluinos need to have masses of several tens of TeV, for their contribution to squark pair production cross section to be negligible for squark masses of 2 TeV. 

\begin{figure}[tb]
\begin{center}
\includegraphics[width=0.7\textwidth]{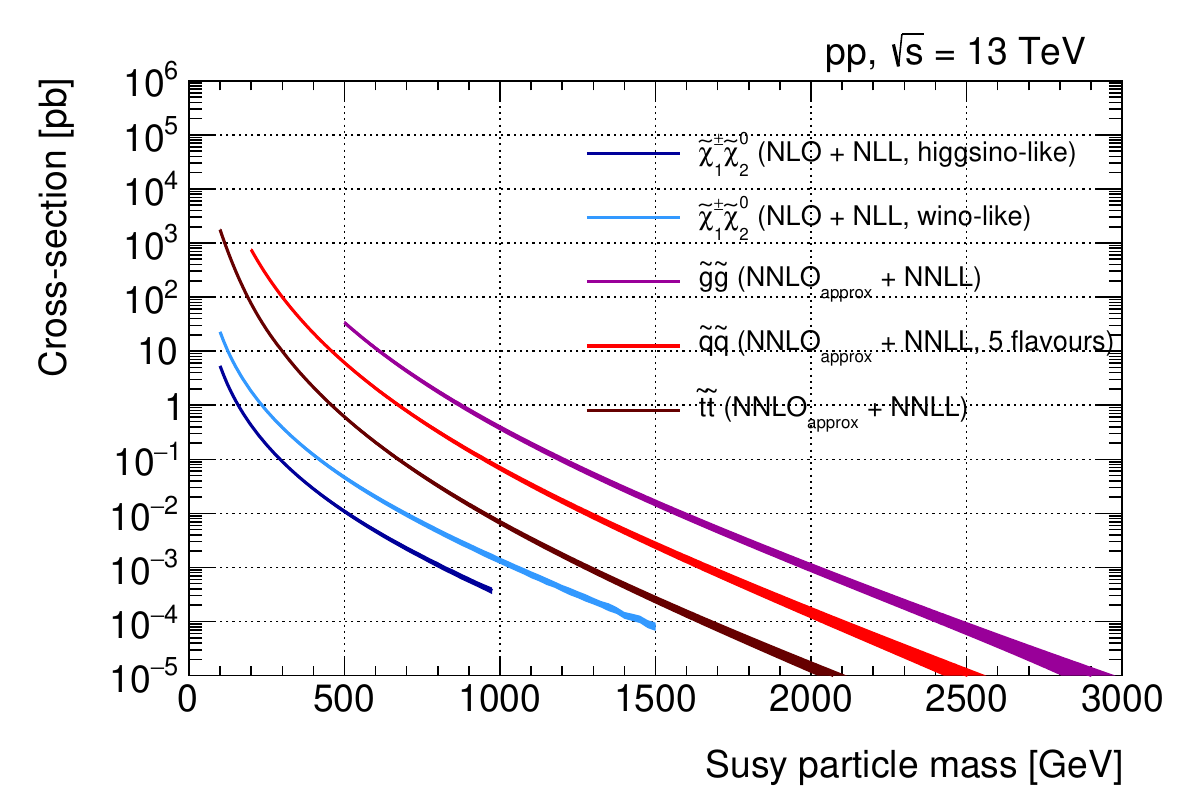}
\end{center}
\caption{Direct pair-production cross sections for a few processes mediated by the strong or electroweak processes. The width of the line represents the theoretical uncertainty. The values of the cross sections and uncertainties are those agreed upon by the LHC SUSY Cross Section Working Group~\cite{susyXsec}.}
\label{fig:xsec}
\end{figure}

Figure~\ref{fig:xsec} shows the state-of-the-art cross sections for the pair production of different SUSY particles. The strong production cross sections are available with a NNLO$_{\mathrm{approx}}$ + NNLL perturbative precision~\cite{Beenakker:2016lwe,Beenakker:2014sma,Beenakker:2013mva,Beenakker:2011sf,Beenakker:2009ha,Kulesza:2009kq,Kulesza:2008jb,Beenakker:1996ch}, while the electroweak ones are known at the NLO + NLL order~\cite{Beenakker:1999xh,Debove:2010kf,Fuks:2012qx,Fuks:2013vua,Fuks:2017rio,Fiaschi:2018hgm} in $\alpha_{\mathrm{s}}$. 

Simplified models are a convenient interface between experiment and phenomenology: they give a good representation of the experimental sensitivity whenever the assumption that it is dominated by masses and decay modes of the lowest states is realised, they give precise phenomenological predictions, and they offer the possibility for a combination of results. Classes of simplified models have been defined that address production driven by either the strong or the electroweak interactions. The particle content is typically inspired by specific physical principles (e.g. naturalness requirements, or consistency with the observed dark matter relic density) or by the need of representing specific regions of the parameter space. ATLAS and CMS have based most of their search strategy on these simplified models: searches have therefore been designed and optimised having specific signatures in mind. The results of those searches are in a sense general: they do not depend on a specific completion of the model - the extracted limits apply whenever the predicted SUSY scenario at the LHC energies approaches that of the simplified model.  The drawback of the approach is clearly that the obtained mass limits do not apply if more complex SUSY realisations, with many competing production and decay modes, are instead realised. To address this limitation, the simplified-model-inspired searches have then been combined and reinterpreted in more complete models either by the collaborations themselves \cite{ATLAS:2015wrn,CMS:2016lcl} or by the phenomenology community.

Some examples of simplified models are discussed below, together with their motivations.

\subsection{Strong production}
\label{sec:models-strong}

As clearly shown in Fig.~\ref{fig:xsec}, degenerate gluino production is the process with the highest cross section at a given value of the SUSY particle masses. Gluino pair production simplified models have been one of the most used signal categories during Run 1 and Run 2. Several different decays of the gluinos have been considered: 

\begin{itemize} 
\item If the only SUSY particles playing a role are the gluino itself and the LSP (assumed to be $\ninoone$), then the only process taking place is gluino pair production followed by the gluino decay $\gluino\rightarrow q\bar{q} \ninoone$ (Fig.~\ref{fig:feyn_diag_a}), where $q$ can be $\left(u,d,c,s\right)$ (the case where $q = \left(b,t\right)$ will be discussed separately). If more intermediate electroweak decays take part in the process (Fig.~\ref{fig:feyn_diag_b}), the possibility of longer gluino decay chains opens up, for example $\gluino\rightarrow q \bar{q}' \chinoonepm \rightarrow q \bar{q}' \ell \nu \ninoone$. Typical final states in these cases involve large jet multiplicities, missing transverse momentum from the LSP and possibly leptons. In the MSSM, gluinos are Majorana fermions, which implies a higher probability than the typical Standard Model backgrounds to produce pairs of leptons with the same charge in two decay chains. 

\item Similarly, if squarks and the LSP are the only SUSY particles taking part in the process, direct squark production leads to final states with two jets and missing transverse momentum via the decay $\squark \rightarrow q \ninoone$ (Fig.~\ref{fig:feyn_diag_c}). Analogously to the gluino case, the presence of more intermediate electroweak states (Fig.~\ref{fig:feyn_diag_d}) opens up the possibility of final states with higher jet multiplicities or leptons. Typically squark initial states lead to lower multiplicity final states than gluino initial states. The total squark pair-production cross section scales linearly with the number of squark flavours and L/R states which are assumed to be degenerate. Many analyses interpret their results assuming  a scenario with a  four-fold flavour, two-fold chirality degeneracy $\left(u,d,c,s\right)$, or under the more pessimistic assumption of a single accessible squark flavour and chirality.  

\item Naturalness arguments generically favour the existence of light stops \cite{Inoue:1982pi,Ellis:1983ed}, possibly accompanied by light sbottoms\footnote{The left-handed stop and sbottom chirality states are part of the same electroweak multiplet and share the same mass parameters.}. Moreover, the decays of third-generation squarks leads to final states with $b$-jets, providing a clear experimental handle to improve the signal separation from the background. Finally, the potential presence of top quarks in the final state offers additional kinematic handles to further suppress the background. These are the reasons that lead the collaborations to define specific simplified models of either gluino production followed by the decay via $\gluino \rightarrow b\bar{b} \ninoone$,  $\gluino \rightarrow t\bar{t} \ninoone$ for example, or stop/sbottom pair production followed by $\stopone \rightarrow t^{(*)} \ninoone$ (Fig.~\ref{fig:feyn_diag_e}), $\sbottomone \rightarrow b \ninoone$. Longer decay chains with intermediate chargino or neutralino states have also been considered.

\begin{figure}[h!]
\begin{center}
\subfigure[]{
\includegraphics[width=0.28\textwidth]{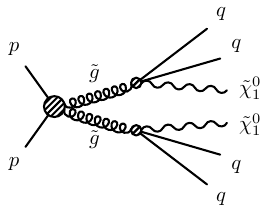}
\label{fig:feyn_diag_a}}
\subfigure[]{
\includegraphics[width=0.28\textwidth]{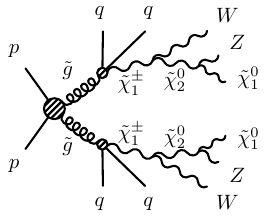}
\label{fig:feyn_diag_b}}
\subfigure[]{
\includegraphics[width=0.28\textwidth]{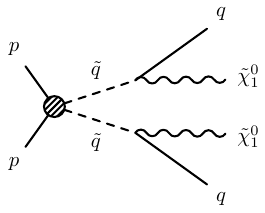}\label{fig:feyn_diag_c}
}
\subfigure[]{
\includegraphics[width=0.28\textwidth]{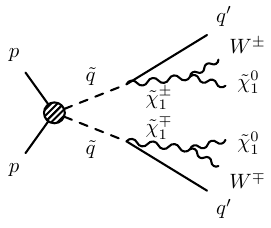}\label{fig:feyn_diag_d}
}
\subfigure[]{
\includegraphics[width=0.28\textwidth]{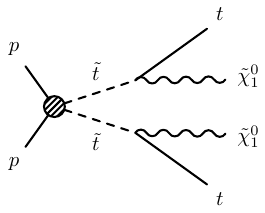}\label{fig:feyn_diag_e}
}
\subfigure[]{
\includegraphics[width=0.28\textwidth]{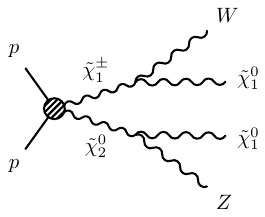}\label{fig:feyn_diag_f}
}
\subfigure[]{
\includegraphics[width=0.28\textwidth]{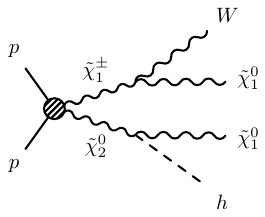}\label{fig:feyn_diag_g}
}
\subfigure[]{
\includegraphics[width=0.28\textwidth]{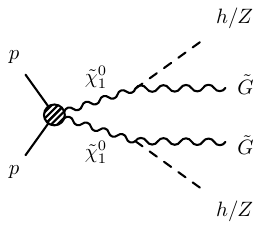}\label{fig:feyn_diag_h}
}
\subfigure[]{
\includegraphics[width=0.28\textwidth]{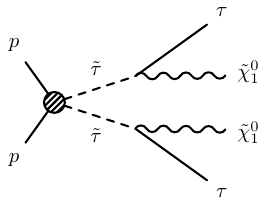}\label{fig:feyn_diag_i}
}

\end{center}
\caption{Simplified models of: gluino pair production followed by (a) $\gluino\rightarrow  q\bar{q}\ninoone$ or (b) a longer decay chain with further intermediate chargino and neutralino states; squark pair production followed by (c) $\squark \rightarrow q \ninoone$ or (d) a longer decay chain; (e) stop pair production followed by $\stop \rightarrow t^{(*)} \ninoone$; $\chinoonepm\ninotwo$ production followed by (f) $\chinoonepm\rightarrow \Wpm\ninoone, \ninotwo\rightarrow Z\ninoone$ or (g)  $\chinoonepm\rightarrow \Wpm\ninoone, \ninotwo\rightarrow h\ninoone$; (h)$\ninoone\ninoone$ production followed by $\ninoone\rightarrow h\tilde{G}$; (i) direct stau pair production followed by $\stau \rightarrow \tau \ninoone$.}
\label{fig:feyn_diag}
\end{figure}

\item Final states containing photons or $\tau$ leptons are obtained as a result of squark or gluino pair production, followed by their decay to the neutralino NLSP, then to the gravitino LSP, in GGM-inspired models. 
\item Dedicated simplified models have been defined to address the possibility of \Rparity violating couplings of either the pair-produced squarks or gluinos, or of the neutralino LSP in the decay chain. The opening of the possibility of \Rparity violation increases dramatically the possible combination of produced final state particles, and has as a consequence the absence of missing transverse momentum in the final state. 
\item If for some reason their decay partial width to the LSP is suppressed, squarks and gluinos can give rise to neutral and charged stable or quasi-stable supersymmetric hadrons, called $R$-hadrons. One example is a scenario where the gluino and the neutralino LSP are the only energetically accessible SUSY particles, and the squarks are extremely massive, so that the decay $\gluino\rightarrow q\bar{q}\ninoone$,which proceeds through a virtual squark state, is extremely suppressed. Searches of long-lived $R$-hadrons require dedicated experimental techniques, depending on their lifetime and therefore on the relevant sub-detectors.

\end{itemize}

\subsection{Electroweak production}
\label{sec:models-eweak}

The strong production of SUSY particles was already heavily constrained by the LHC Run 1 results (and, as we will see, Run 2 has strongly enhanced those constrains). However, because of the smaller involved production cross sections, it is fair to say that the electroweak production parameter space has been mostly constrained during Run 2, thanks to the higher integrated luminosity and centre of mass energy of the LHC $pp$ collisions. 

Looking again at the curves for $\chinoonepm\ninotwo$ production in Fig.~\ref{fig:xsec}, there are a few observations to be made. The first one is that, as expected from the different coupling strength, at a fixed mass the cross section is significantly lower than those for strong production. The second observation is that unlike the cross section for squark or gluino pair production, that for electroweakino production does depend on the nature of the involved electroweakinos: given the different couplings of the $B, \mathbf{W}$ and of the extended Higgs sector, the electroweakino production cross section depends on the bino, wino, higgsino admixture of their mass eigenstates. In the MSSM, for example, the electroweakino mass matrix and couplings (therefore the decay branching ratios) are completely determined by relatively few parameters: the bino mass parameter $M_1$, the wino mass parameter $M_2$, the higgsino mass parameter $\mu$, and $\tan \beta$.

A neutralino LSP, with mass of the order of $100-10000$ GeV, is one of the best candidates to explain the cold dark matter relic density observed cosmologically. Compatibility of a given set of electroweak SUSY sector parameters with the dark matter relic density constraints, and consequent particle spectra, are also sometimes taken broadly into account when designing suitable electroweak production simplified models. A detailed review of such criteria goes beyond the scope of this work, but there are comprehensive reviews available~\cite{Feng:2010gw}.

\begin{figure}[tb]
\begin{center}
\input{figures/models/ew_spectrum}
\end{center}
\caption{Sketch of electroweakino mass hierarchies used by ATLAS and CMS during Run 2.}
\label{fig:ewikino_spectrum}
\end{figure}
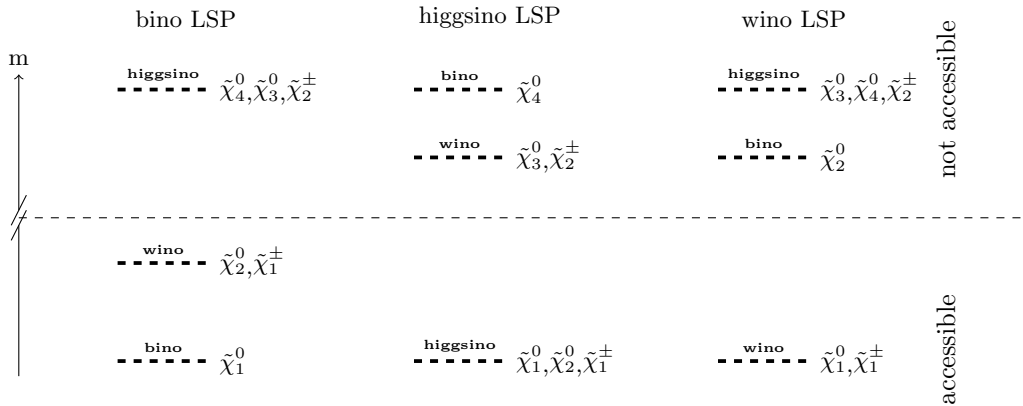

Slepton pairs would also be produced only through electroweak production. Similarly to the squark sector, the third generation (the staus) may be singled out, both because of the specific experimental challenges connected with the tau lepton decay into hadrons (in the following indicated with $\tauhad$) identification and tagging, and because of the role they may play in establishing the correct dark matter relic density through co-annihilation with neutralinos~\cite{Ellis:1998kh}. 

These considerations fully justify the set of simplified models considered by ATLAS and CMS as a benchmark for electroweak production: 

\begin{itemize}
\item Possibly the most explored of all electroweak production models is one where the $\chinoonepm$ and $\ninotwo$ are assumed to be degenerate in mass and produced with wino-like cross sections, while the $\ninoone$ is assumed to be the LSP. This model is inspired by the MSSM for scenarios where $\mu \gg M_2 > M_1$ (Fig.~\ref{fig:ewikino_spectrum} left), such that the higgsino components of the electroweak sector are decoupled and can be neglected in production. The relevant production channels are $\chinoonep\chinoonem$, $\ninotwo\chinoonepm$, often considered separately. If it is assumed that the sleptons do not take part in the decay, then the only possible decay channel for the chargino is $\chinopm \rightarrow W^{\pm(*)} \ninoone$, while those for the $\ninotwo$ are $\ninotwo\rightarrow Z^{(*)}\ninoone$ and $\ninotwo \rightarrow h \ninoone$. Therefore, in \Rparity conserving models, $\ninotwo\chinoonepm$ production gives rise to final states with missing transverse momentum associated to the $\ninoone$s, and either $WZ$ or $Wh$, while chargino pair production yields final states containing missing transverse momentum and $WW$. Models that assume chargino decays with intermediate slepton states have also been considered. They lead to final states enriched in leptons, given a branching ratio of 100\% for $\tilde{\ell}\rightarrow \ell \ninoone$ if lepton number and flavour have to be conserved in the slepton decay.  
\item Naturalness criteria require that the higgsino mass parameter $\mu$ should be a few hundred GeV maximum. This suggests the possibility of scenarios where the entire electroweakino spectrum is determined by a relatively low mass set of higgsinos, and much heavier winos and binos (Fig.~\ref{fig:ewikino_spectrum} middle). Taking again the MSSM as a reference, such scenario corresponds to a spectrum where one has a nearly degenerate triplet of electroweak states ($\chinopm, \ninotwo, \ninoone$). The exact mass splitting between these states can vary between few tens of GeV and few hundred MeV, depending mainly on the value of $M_1$ and $M_2$. This spectrum of {\it compressed} electroweak states, associated with the low cross sections, gives rise to a number of experimental challenges (low-momentum object reconstruction and corresponding background rejection). 
\item If the only relatively light mass parameter is $M_2$, then a highly compressed  multiplet of $\chinoonepm$ and $\ninoone$ is the only light state available, with mass splittings of the order of hundreds of MeV. In such wino-LSP case, the chargino is often long-lived, and decays into a low momentum, unreconstructed pion or lepton, and the invisible neutralino within the tracking volume, giving rise to a striking disappearing track signature. 
\item If \Rparity conservation with a neutralino LSP is assumed, models of selectron and smuon pair production followed by $\tilde{\ell}\rightarrow \ell \ninoone$ lead to final states containing two same-flavour, opposite-sign leptons and missing transverse momentum. Similarly, direct stau pair production leads to $\tilde{\tau}\rightarrow \tau \ninoone$, therefore to a final state containing two $\tau$ leptons and missing transverse momentum. The better performance of electron and muon reconstruction and identification compared to those for $\tau$ reconstruction and identification make the search for direct $\stau$ pair production significantly more challenging than that for $\sel$ or $\smu$. 
\item Similarly to the case of strong production, the removal of \Rparity conservation, or the possibility that the electroweakinos or sleptons are long-lived opens up the possibility of a large variety of experimental signatures. Three-body, \Rparity violating electroweakino decays give rise to final states with high jet or lepton multiplicities (depending on which RPV couplings are different from zero). Long-lived charginos, neutralinos and sleptons open up the possibiity of the production of leptons with large impact parameters, displaced vertices with leptons, creating very interesting experimental signatures to investigate.
\end{itemize}

%% file: figures/models/ew_spectrum.tex
\begin{tikzpicture}
\draw [->] (-5.2,2.2) -- ++ (0,+1.8) node [above] {\small m};	
\draw (-5.2,0) -- ++ (0,+2);	
\draw (-5.3,1.8) -- ++ (0.2,0.4);
\draw (-5.3,2) -- ++ (0.2,0.4);
\node [above ] at (-3,4.5) {\small bino LSP};
\node [above ] at (1,4.5) {\small higgsino LSP};
\node [above ] at (5,4.5) {\small wino LSP};
\draw [dashed] (-5.2,2.1) -- (8,2.1);
\node [right, rotate=90] at (7,2.5) {not accessible};
\node [right, rotate=90] at (7,-0.5) {accessible};

\node [above] at (-3.3,3.8) {\tiny \textbf{higgsino}} ;
\draw [dashed,line width=0.5mm] (-3.9,3.8) -- ++ (1.2,0) node [right] {\small \ninofour,\ninothree,\chinotwopm};
\node [above] at (-3.3,1.5) {\tiny \textbf{wino}} ;
\draw [dashed,line width=0.5mm] (-3.9,1.5) -- ++ (1.2,0) node [right] {\small \ninotwo,\chinoonepm};
\node [above] at (-3.3,0.2) {\tiny \textbf{bino}} ;
\draw [dashed,line width=0.5mm] (-3.9,0.2) -- ++ (1.2,0) node [right] {\small \ninoone};

\node [above] at (0.6,3.8) {\tiny \textbf{bino}} ;
\draw [dashed,line width=0.5mm] (0,3.8) -- ++ (1.2,0) node [right] {\small \ninofour};
\node [above] at (0.6,2.9) {\tiny \textbf{wino}} ;
\draw [dashed,line width=0.5mm] (0,2.9) -- ++ (1.2,0) node [right] {\small \ninothree,\chinotwopm};
\node [above] at (0.6,0.2) {\tiny \textbf{higgsino}} ;
\draw [dashed,line width=0.5mm] (0,0.2) -- ++ (1.2,0) node [right] {\small \ninoone,\ninotwo,\chinoonepm};

\node [above] at (4.6,3.8) {\tiny \textbf{higgsino}} ;
\draw [dashed,line width=0.5mm] (4,3.8) -- ++ (1.2,0) node [right] {\small \ninothree,\ninofour,\chinotwopm};
\node [above] at (4.6,2.9) {\tiny \textbf{bino}} ;
\draw [dashed,line width=0.5mm] (4,2.9) -- ++ (1.2,0) node [right] {\small \ninotwo};
\node [above] at (4.6,0.2) {\tiny \textbf{wino}} ;
\draw [dashed,line width=0.5mm] (4,0.2) -- ++ (1.2,0) node [right] {\small \ninoone,\chinoonepm};

\end{tikzpicture}

%% file: tex/exp_techniques.tex
Searches for Supersymmetry need to cover a very large range of experimental signatures and use all standard types of ``physics objects'' that can be reconstructed in a general purpose LHC detector: hadronic jets, including those identified as originating from the fragmentation of $b$ or $c$ quarks, electrons and muons, hadronically decaying $\tau$ leptons, and the transverse momentum imbalance, \ptmiss, with magnitude \etmiss.
The latter is itself obtained from a combination of the other objects, and a key signature for all RPC SUSY searches.

New challenges in LHC Run~2 arose from the increase in ``pileup'', the number of additional $\Pp\Pp$ collisions in the same LHC bunch crossing, which passed from an average of 21 interactions in the last year of Run~1 to almost 40 interactions in the two final years of Run~2.
At the same time, the experiments intensified the exploration of unconventional signatures that had not been a primary goal in the design of the experiment and the reconstruction algorithms: soft leptons and jets, as well as disappearing and emerging tracks and jets, and out-of-time energy deposits from long-lived particles.

Methods of background estimation evolved, with simulations based on higher-order calculations becoming the standard for many SM processes, and inputs from the experiments' own measurements of inclusive and differential cross sections of increasingly rare processes such as associated production of vector bosons with top-quark pairs, or multiboson production.
The sensitivity of the searches was improved by employing more complex algorithms for the reconstruction of the event kinematics, and the use of machine learning (ML).

The basic concepts used for the statistical analysis of data and their interpretation in terms of exclusion or significance stayed the same as in Run~1.
They are based on a likelihood method, with a likelihood function, which uses the observed and expected yields from several signal- and background-enriched regions.
Systematic uncertainties are included as nuisance parameters and profiled.
Exclusions are established based on the CL$_S$ criterion~\cite{Junk:1999kv,Read:2002hq}.
In order to reduce the computational cost related to the scan of a two- or higher-dimensional SUSY parameter space, an asymptotic approximation of the test statistics~\cite{Cowan:2010js} is frequently employed.

\subsection{Reconstruction and identification of leptons, jets, and \etmiss}\label{sec:expTechReco}

In most searches for SUSY, standard reconstruction and identification techniques are used, with some conceptual differences between the experiments.

Photons are reconstructed from energy deposits in the electromagnetic calorimeters~\cite{Aad:2019tso,Aaboud:2019ynx,CMS:2020uim} without matching signals in the inner tracking systems.
The reconstruction of electrons is also initiated in the electromagnetic calorimeters, followed by a combination with signals from the inner tracking detectors~\cite{Aad:2019tso,Aaboud:2019ynx,CMS:2020uim}.
In CMS, the efficiency for low-momentum electrons is increased by complementing this procedure with a tracker-based seeding, followed by a match with energy deposits from the original electron, and possible radiation, in the calorimeter.
Muons are reconstructed from a simultaneous fit to trajectory segments in the muon detectors and the inner tracking system~\cite{Aad:2016jkr,Sirunyan:2018fpa,Aad:2020gmm,Sirunyan:2019yvv}.

In addition to standard identification criteria designed to reduce the number of candidates that do not match a genuine lepton, ``prompt'' leptons produced in the electroweak decays of SM gauge bosons or sparticles have to be distinguished from leptons produced in association with jets, from hadron decays or photon conversions.
This discrimination is mainly based on isolation, a measure of the activity close to the candidate:
particle energies (projected on the transverse plane), or transverse momenta, are summed in a cone \dr~\footnote{Particle momenta are expressed in terms of the transverse momentum, \pt, the azimuthal angle, $\phi$, and the pseudo-rapidity, $\eta$. The distance \dr is defined as $\sqrt{\Delta\phi^2+\Delta\eta^2}$.} around the candidate, with a typical value of 0.2\,.
Deposits directly related to the candidate, e.g., from radiated photons, are excluded, the sum is corrected for the estimated contribution from pileup, and the final discrimination is based on the ratio between the sum and the lepton \pt.
The ratio should not exceed values that are typically in the range 10--20\%.
Analyses that critically depend on the efficiency of lepton reconstruction and identification can use refined criteria that take into account additional effects as the collimation of radiation from the candidate, the reduced separation of decay products of highly boosted particles~\cite{Rehermann:2010vq}, and variables such as the distance from the closest jet.
All of these variables, and even more detailed information on all reconstructed particles close to the candidate, are also used in ML-based discriminators.
Finally, the purity and identification efficiency can be optimised by a suitable disambiguation algorithm that is designed to avoid double counting between various objects. 
In the following, reconstructed leptons that are either misidentified, i.e., not associated to a genuine lepton, or not "prompt" according to the definition above, will be called "misidentified-or-non-prompt" (\FNP).

Hadrons and jets are obtained from signals in the calorimeters, combined with information from the tracking system.
In CMS, a global description of each event is systematically used, based on a particle flow (``PF'') algorithm that aims to reconstruct and identify all individual charged and neutral particles using a combination of all relevant detector systems~\cite{Sirunyan:2017ulk}.
In ATLAS, two approaches are used, depending on the analysis.
Some analyses employ PF jets~\cite{Aaboud:2017aca}, similar to what is done in CMS.
For others, jet reconstruction is mainly based on the energy deposits in the calorimeter system, complemented with tracking information for pileup rejection and for improved resolution for low-momentum jets.
In both experiments, the inputs (energy deposits or PF candidates) are clustered using the anti-\kt algorithm as implemented in the FastJet package~\cite{Cacciari:2011ma}, with a standard choice for the radius parameter of $R=0.4$.
Jet energies are corrected for detector effects~\cite{ATLAS:2020cli,Khachatryan:2016kdb} and contributions from pileup~\cite{Cacciari:2007fd,Aad:2015ina}.

Jets produced in the decay of highly boosted particles will tend to merge into a single, large-$R$ jet.
Reconstruction of these jets, frequently followed by an analysis of their mass and / or substructure, is a powerful tool to identify the presence of high-\pt Higgs, $W$, or $Z$ bosons, and top quarks.
Several different approaches have been followed: a second, independent clustering of all particles using the anti-\kt algorithm with a radius parameter of $R=0.8$, or by reclustering the standard-radius jets obtained in the first step with a radius parameter of $R=1.0$.
The latter approach is also used in a recursive procedure, starting at $R=3.0$, with a successive reduction of the parameter.

Jets that include long-lived $b$ or $c$ hadrons ($b$ or $c$ jets) are tagged using multivariate algorithms~\cite{Aad:2019aic,Sirunyan:2017ezt}, with the most recent versions using detailed information about individual jet constituents by means of deep neural networks.
Hadronically decaying $\tau$ leptons are reconstructed as narrow jets; as in the case of heavy flavour jets, the identification uses multivariate algorithms~\cite{CMS:2016gvn,ATLAS:2017mpa,ATLAS:2019uhp,Aad:2014rga}.

In ATLAS, the \ptmiss\ vector is determined as the negative vector sum of electrons, photons, muons and jets, and low momentum tracks associated to the selected primary $\Pp\Pp$ interaction vertex~\cite{ATLAS:2018ghb,Aaboud:2018tkc}.
In CMS, the negative vector sum of all PF candidates is used~\cite{Sirunyan:2019kia}.
Jet energy corrections are propagated to the estimate and events with anomalous contributions to \etmiss\ from detector noise or non-collision backgrounds are rejected.
Several versions of the \etmiss\ significance~\cite{Chatrchyan:2011tn,Khachatryan:2014gga,Aaboud:2017hdf,ATLAS:2018uid} --- estimates of the probability that \etmiss is compatible with zero --- are used in some of the analyses. 
The resolution of \etmiss\ approximately scales with the square root of the visible energy in the event, following the evolution of fluctuations in the calorimeters.
Hence, simplified versions of the significance use \metsigHT or equivalent quantities.
Alternatively, a $p$-value can be calculated based on the estimated resolutions of the individual input objects in each event.

Analyses targeting events with long-lived or low-\pt objects often need specific approaches.
Charged particle tracks originating from or ending at a displaced decay vertex are either covered by specific steps in standard reconstruction, or by custom algorithms used for individual analyses.
The reconstruction of secondary vertices at distances from the interaction point that are larger than what is expected from heavy flavour decays is typically performed within the analysis workflow, using an optimised preselection of tracks.
Timing information can be extracted from signals in the calorimeters or in the muon systems.

\subsection{Modelling of backgrounds\label{sec:expTechBkgs}}

Robust and precise estimates of the contribution of SM backgrounds to the signal-enriched regions are a critical element in searches for BSM processes.
Techniques for background estimation have evolved since LHC Run~1:
\begin{itemize}
\item a tremendous amount of work has been invested by the theory community in an increased precision of the theoretical predictions, both for the calculation of inclusive cross sections and for event generation;
\item the simulation software of the experiments was shown to model detector and beam conditions, and the detector response, to high accuracy;
\item precision measurements of inclusive and differential cross sections for some of the dominant background processes, such as production of top quark pairs, or electroweak gauge bosons in association with jets, have become available and can either be used directly in background estimates, or for the validation and tuning of simulation;
  \item evidence for, and first measurements of, a large number of rare SM processes, such as multi-boson production, associated production of a top quark pair with a vector or Higgs boson, or double parton scattering have become available.
\end{itemize}

As a consequence, many estimates for leading SM backgrounds are now based on a combination of MC predictions and control regions in data, providing an in-situ normalization of backgrounds in the final, global fit to the observed yields in control and signal regions.
Control regions are designed to be dominated by a specific background process, but as close as possible to the signal regions.
Typically they are binned in a similar way as the signal regions and use modifications or inversions of a small subset of the selection criteria.
The background estimates in the signal regions are obtained with MC-based transfer factors, implying the use of shapes in the selection variables predicted by simulation.
Signal contamination in the control regions is a concern, in particular due to the wide range of signatures and kinematics spanned by different SUSY models.
Control regions are designed to suppress signal contributions for the signal models tested in an analysis, and small residual levels of contamination are handled by the global fit.
However, care has to be taken when applying the results to other models, in particular, when deriving constraints on full SUSY models from results obtained for simplified models.
Another set of background-dominated regions, different from both control and signal regions, can be used to validate the prediction procedure, or to estimate systematic effects.

Some categories of background cannot be reliably modeled with simulation, or have an extremely low selection efficiency, making the cost of computation prohibitively high.
Typical examples are soft multijet production, or contributions from \FNP leptons.
These processes are therefore entirely estimated from data.
Examples are extrapolations from a set of three control regions to a signal region, frequently called ``ABCD'' method, or the determination of the contribution of misidentified objects by measurement of the misidentification probability in data.

The first case can be used for global estimates of background components that are difficult to model, such as SM multijet production in jets + \etmiss\ topologies.
It relies on three control regions that are distinguished from the corresponding signal regions by two sets of selection criteria, frequently  by inversion of some criteria used to define the signal region.
The two sets of criteria are chosen such that the control regions are enriched in the background under study, and that - by prior knowledge or previous studies - are known as being statistically independent.
In this case, the selection probabilities factorise and the yield in the signal region can be predicted by a ratio of yields in the control regions.
The observed yields in the control regions need to be corrected for the contribution of other backgrounds, and for signal contamination, if relevant.

In the second approach, in what follows called ``fake-ratio'' method~\footnote{In ATLAS and CMS publications, the method is frequently called "fake-factor", "fake-rate", or "matrix" method.}, background contributions from misidentified objects, such as leptons, are computed at the object level.
The method uses two levels of object identification with low (``loose'') and high (``tight'') purity.
The probability that a loosely, but not tightly, identified object also passes the tight selection (the fake-ratio) is measured in a background-dominated control region.
Under the assumption that this number is identical for events passing  control or signal region selections, and independence in case of multiple objects, the probability for an event in the signal region to contain one or more misidentified objects can be computed.
In practice, the ratio can depend on several variables.
It is typically measured as a function of \pt\ and $\eta$, but it can be necessary to take into account correlations with other variables, such as the estimated momentum of the originating hadron in the case of \FNP leptons, in order to make it applicable to the signal region.
In some cases, the ratio needs to be determined separately for different background components.
As in the first approach, a correction for a contamination of the control region by other background processes has to be done, in particular for those that constitute sources of genuine objects.

Many variations of these approaches exist as details of the estimation procedure are always tuned in the context of a specific analyses.
In addition, several other procedures to estimate backgrounds from data have been developed and are discussed in the next sections, where relevant.

%% file: tex/simplified_model_results.tex
Starting from the collaborations' summary plots, this section provides a global overview of the analyses performed by ATLAS and CMS to search for the production of supersymmetric particles. The final states considered, the analysis techniques and the sensitivities obtained in terms of mass exclusion limits in simplified models are discussed. Unless differently stated, all limits provided are given at 95\% confidence level.    

\subsection{Strong production}
\input{tex/strong}

\clearpage
\subsection{Electroweak production}
\input{tex/eweak}

%% file: tex/strong.tex
We start our review of the Run 2 results of the LHC collaborations by focusing on those obtained by analyses designed to look for the production of SUSY particles via the strong interaction. This means we focus on the search for the production of squarks and gluinos. Generally speaking, the fact that the couplings involved in the production are proportional to the strong coupling constant $\alpha_{\mathrm{s}}$ makes the cross sections corresponding to these production modes larger than those corresponding to SUSY particle production via electroweak interaction for the same SUSY particle masses. As an example, the production cross section for a pair of gluinos with mass $m= 1.5\ \TeV$, assuming decoupled squarks, is 15.7 fb, yielding more than two thousand gluino pairs produced in Run 2 - a signal which is relatively easy to identify, unless the final state topology and kinematics are challenging. Under analogous conditions (gluinos decoupled and same mass), and assuming a five-flavour mass degeneracy ($u,d,c,s,b$), the squark-antisquark pair production cross section is 2.6 fb, corresponding to more than 350 signal events in the Run 2 data - still a yield that is identifiable, depending on the following decay chain. This simple considerations already lead us to the conclusion that the squark and gluino mass scale to which the Run 2 LHC data are sensitive is about $2\ \TeV$.

\subsubsection{Gluino pair production}
\label{sec:gqq}

\begin{figure}[tb]
\begin{center}
\subfigure[]{
\includegraphics[width=0.4\textwidth]{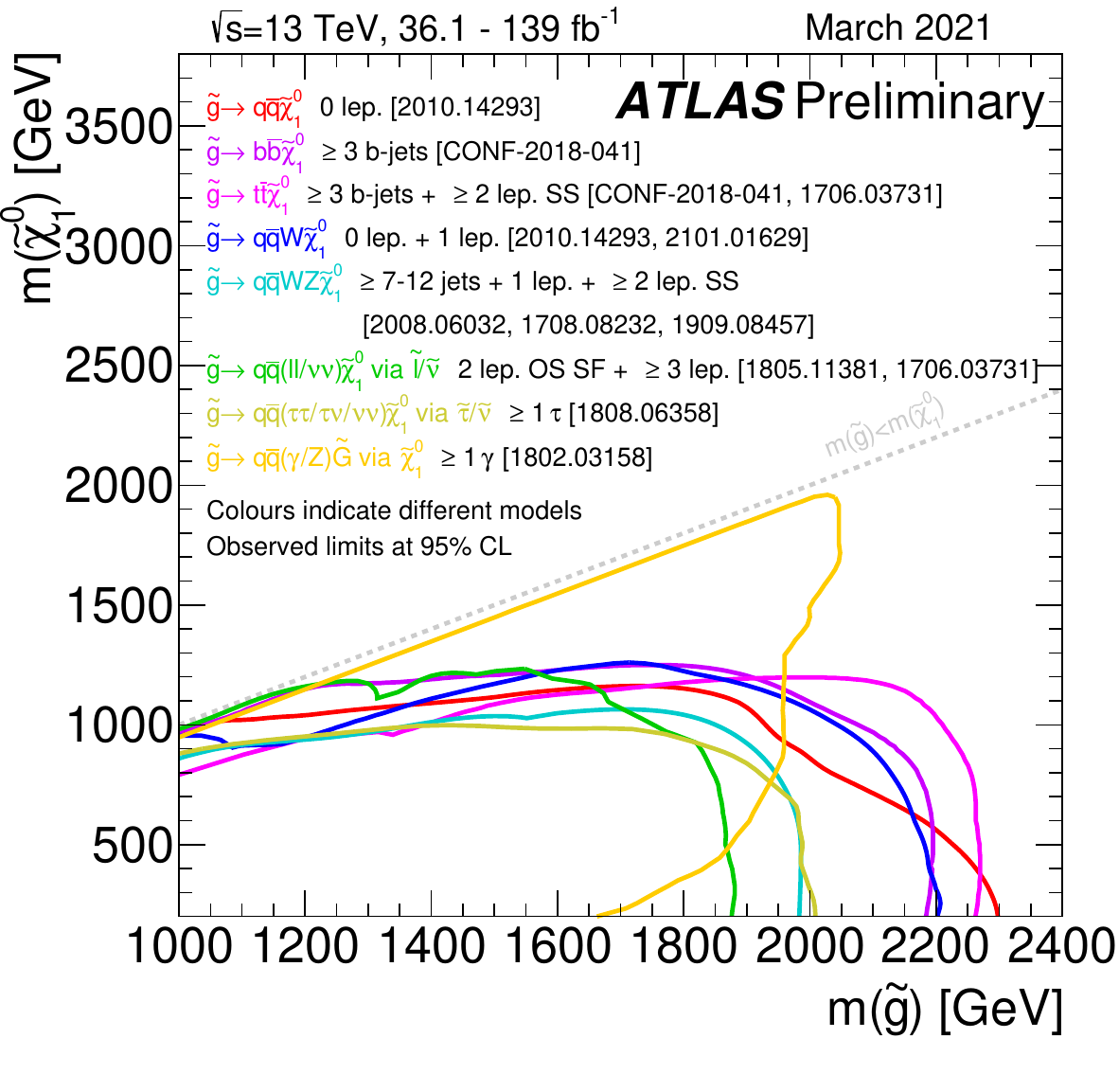}
}
\subfigure[]{
\includegraphics[width=0.405\textwidth]{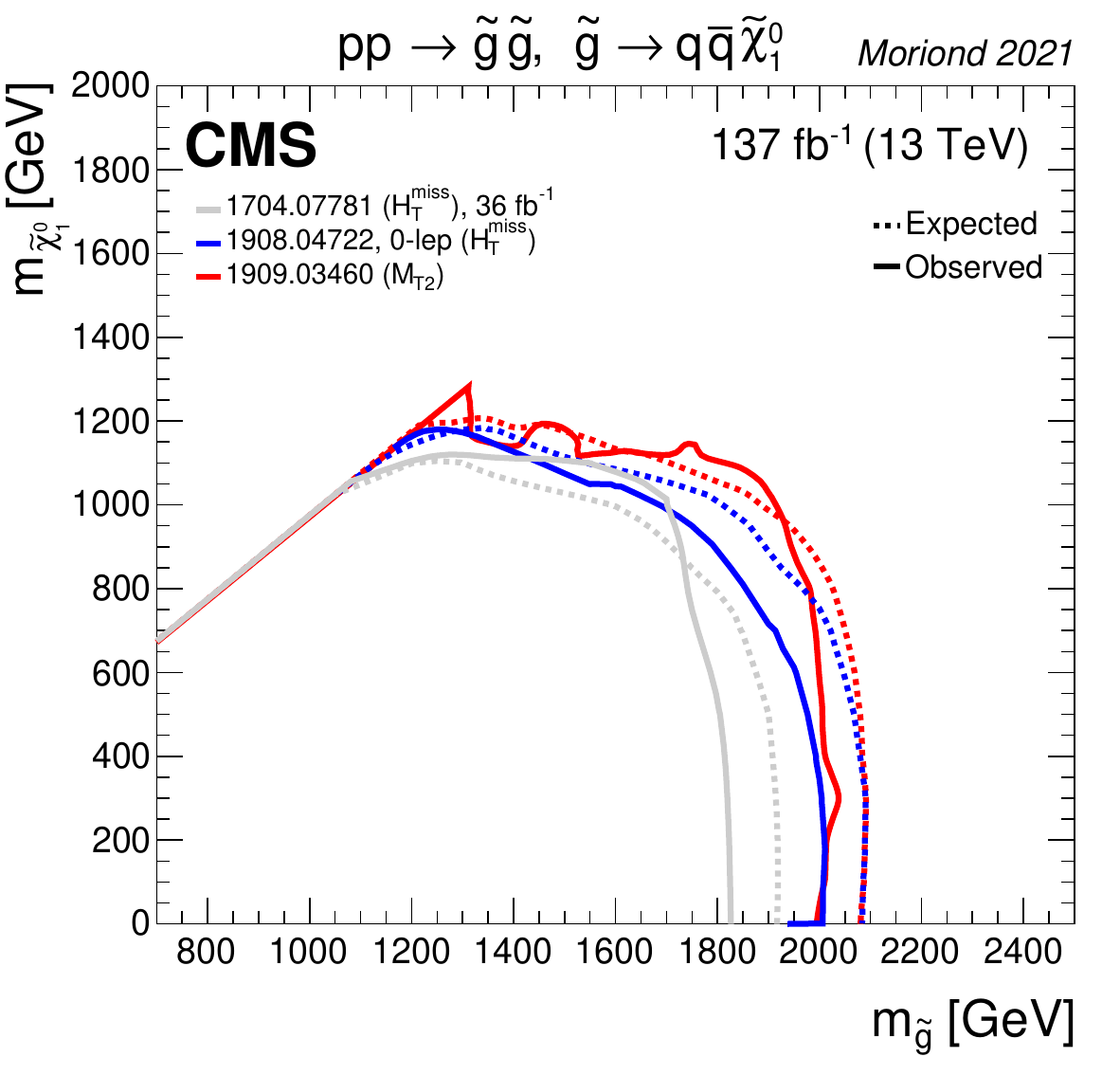}
}
\subfigure[]{
\includegraphics[width=0.45\textwidth]{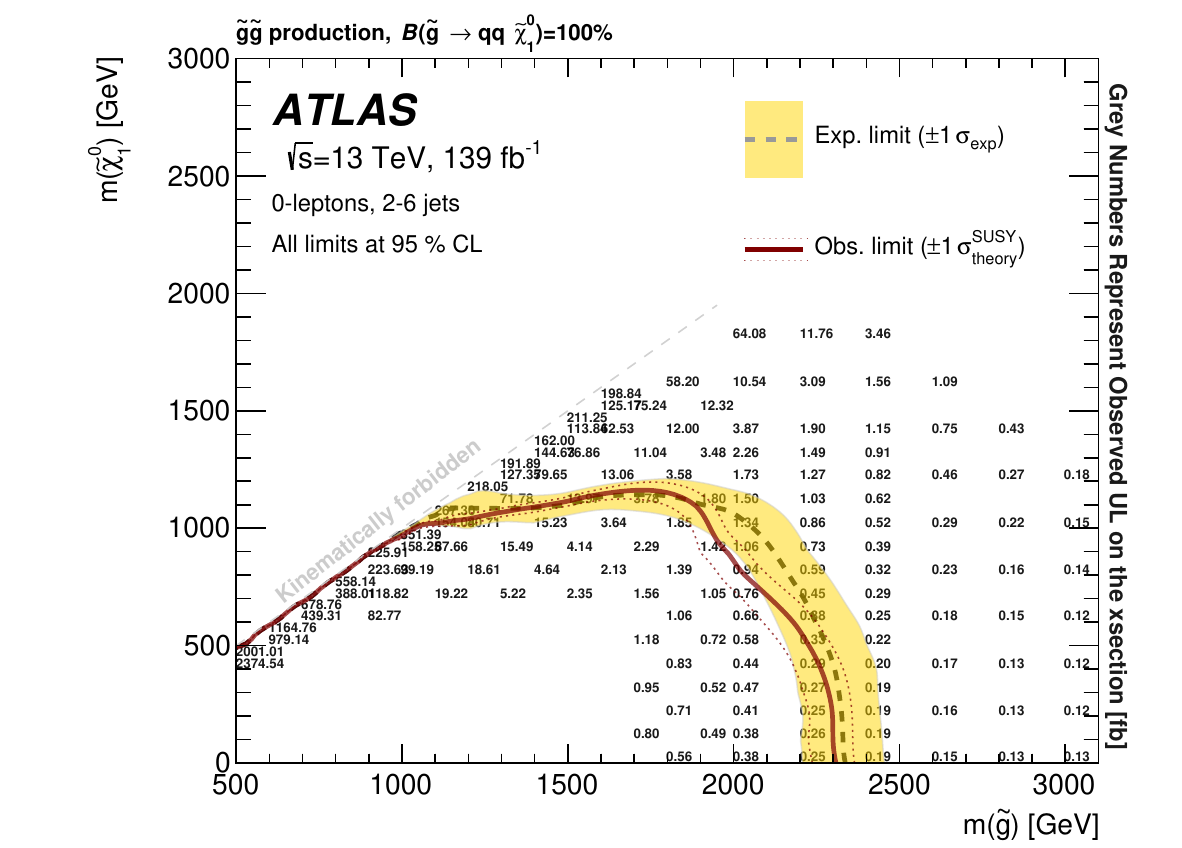}
}
\subfigure[]{
\includegraphics[width=0.35\textwidth]{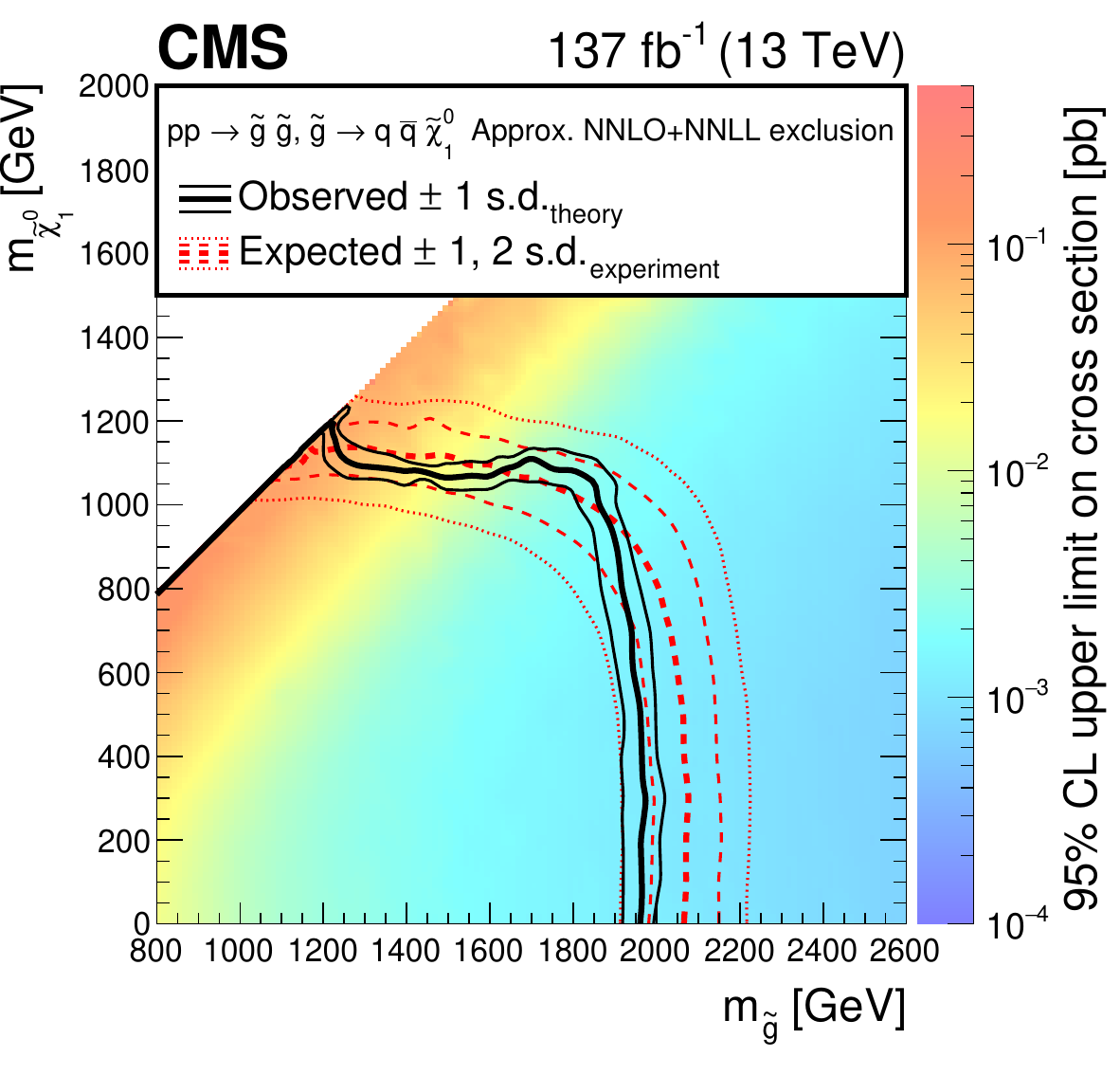}
}

\end{center}
\caption{Gluino pair production summary plot of 95\% CL mass exclusion limits in the gluino-neutralino plane in (a) ATLAS\cite{ATL-PHYS-PUB-2021-019} and (b) CMS\cite{CMS-SUSY-Summaries}, also including 95\% CL upper limits on cross sections for gluino pair production followed by $\gluino\rightarrow  q\bar{q}\ninoone$ for (c) ATLAS\cite{Aad:2020aze} and (d) CMS\cite{CMS:2019ybf}. For (a), each curve refers to a different decay mode of the gluino according to the legend.}
\label{fig:summaryPlot_gg}
\end{figure}

A summary of the mass exclusion limits of ATLAS and CMS are available in Figs.~\ref{fig:summaryPlot_gg} and \ref{fig:summaryPlot_gtt}. A set of simplified models has been used to derive the limits shown. In all cases, it is assumed that the only SUSY particle production process taking place is the production of pairs of gluinos. It is furthermore assumed that no diagram including any other SUSY particle contributes to the production process. This second assumption makes the production cross section dependent only on the gluino mass. The corresponding cross sections are shown in Fig.~\ref{fig:xsec}. In all cases, the limits are presented as a function of the gluino and lightest neutralino mass. Different curves represent either different assumptions on the gluino decay chain, or different analysis results. Glancing at the plots, one immediately realises that the rough prediction we made based on the cross section and Run 2 integrated luminosity is broadly speaking satisfied, although there are clearly some regions of these planes where the limits are weaker than naively expected. 

As discussed in Sec.~\ref{sec:models}, a widely used gluino pair-production simplified model is one where the gluinos decays to a pair of quarks and the LSP ($\gluino\rightarrow  q\bar{q}\ninoone$). The model is inspired by a bino-like LSP scenario, where the wino and higgsino mass parameters are too high to play any significant role for the LHC phenomenology. The gluino decay happens through an intermediate virtual squark state ($\gluino\rightarrow \squark^{*} \ninoone\rightarrow  q\bar{q} \ninoone$). The decay is assumed to be prompt. The limits set by the collaborations in this scenario are summarised in Fig.~\ref{fig:summaryPlot_gg}(a) (red line) and in Fig.~\ref{fig:summaryPlot_gg} (b). 

The ATLAS result is determined by an analysis\cite{Aad:2020aze}, which targets mainly direct gluino and direct squark pair production. The final state of interest contains no electrons or muons, and significant missing transverse momentum which is exploited for triggering purposes. The analysis targets many strong production scenarios: the offline selection is based on the jet multiplicity in the event (\njet), the significance of the missing transverse momentum \metsigHT, and the $\meff = \HT + \etmiss$. Here \HT denotes the scalar sum of the \pt of the jets in the event. Two analyses are considered: one using an explicit selection on these and other variables (multi-bin search), and one using a Boosted Decision Tree (BDT) trained using similar variables as input. A dominant background process in the considered signal regions is $\Zjets$ followed by $Z\rightarrow \nu\nu$. This is estimated using simulated events corrected with auxiliary measurements of the $\gamjets$ and $\Zjets\ (Z\rightarrow\ell\ell)$ processes in a similar phase space region. Other relevant background processes are $\Wjets$ and $\ttbar$ production, with a $W\rightarrow\ell \nu$ decay to satisfy the missing transverse momentum requirement. For the latter class of processes, the lepton fails identification, or it is a $\tauhad$. These are estimated using the MC simulation, whose normalisation is adjusted thanks to auxiliary measurements in one-lepton control regions. A small contribution from multijet production is estimated with a jet smearing method, where simulated events are used as seed for the generation of the background prediction after the jet response is corrected using dedicated samples of real data to make sure response tails are correctly taken into account~\cite{Aad:2012fqa}. 

The CMS analysis described in Ref.~\cite{Sirunyan:2019ctn} follows a similar strategy. The signal phase space is explored in bins of \HT, \HTmiss (the magnitude of the negative vector sum of the jet transverse momenta), \njet and the $b$-jet multiplicity, \nbjet. Bins with $\nbjet > 0$ are in particular sensitive to the production of heavy-flavour squarks, either directly in pairs, or in the decay chain of gluinos. The bin where $\nbjet = 0$ is specifically sensitive to $\gluino\rightarrow  q\bar{q}\ninoone$. The SM background composition is similar to the ATLAS case. The philosophy behind the background estimation is also similar (relying on one-lepton control samples for the \Wjets and  \ttbar  production, on \gamjets and $Z\rightarrow \ell \ell$ samples for the \Zjets production, and on data-driven techniques for the small multijet background), although the specific techniques differ slightly from those used by ATLAS. 

Competitive exclusion limits are obtained in CMS by the analysis shown in red in Fig.~\ref{fig:summaryPlot_gg} (b)~\cite{CMS:2019ybf}. The approach of this zero-lepton analysis is to cluster all  jets in the event starting from the two that provide the highest invariant mass pair, until the whole event is clustered into only two pseudo-jets. At this point, the stransverse mass \mttwo \cite{Lester:1999tx} of the two pseudo-jets and the missing transverse momentum vector is computed. The stransverse mass is defined as

\begin{eqnarray}
\mttwo = \min_{{\ptmiss}^1 + {\ptmiss}^2 = \ptmiss} \left(\max\left(\mt^1,\mt^2\right)\right).
\end{eqnarray} 

The computation of \mttwo requires to find the decomposition directions of the \ptmiss vector into two vectors ${\ptmiss}^1$ and ${\ptmiss}^2$, such as to minimise the value of the maximum transverse masses of each pseudo-jet and ${\ptmiss}^1$ and ${\ptmiss}^2$. The \mttwo variable has interesting properties~\cite{CMS:2019ybf}, that make it assume larger values for the signal than for the SM background. A trigger selecting events with moderate values of both $\HT$ and $\etmiss$ is used. The events are then categorised based on \HT, \mttwo, \njet, \nbjet. 

Figure~\ref{fig:summaryPlot_gg} shows that each of the experiments excludes gluino masses of about 2.2 TeV for massless $\ninoone$. The sensitivity is significantly reduced if the gap in mass between the pair-produced gluino and the LSP is small. This is a consequence of the lower momentum of the gluino decay products. For example, there is no sensitivity to gluinos with masses above about 1.3 TeV if the mass of the gluino and that of the LSP are similar.  

It should never be forgotten that these limits are obtained  assuming gluino pair production (in a regime with decoupled squarks), followed by the decay $\gluino\rightarrow  q\bar{q} \ninoone$ with a branching ratio of 100\%. While this is a useful benchmark channel, the mass sensitivity reached in this case is not necessarily representative of the experiment sensitivity to more general SUSY scenarios. For example, a not completely decoupled set of squark states may make $t$-channel squark exchange diagrams relevant when computing the gluino pair-production cross section: they would interfere destructively with the $t$-channel quark and $s$-channel gluon exchange, effectively lowering the production cross section. At the same time, depending on the squark masses, additional squark-pair, or squark-gluino production diagrams, would contribute to the total SUSY particle production cross section. These different effects push the sensitivity in different directions, and the global sensitivity is affected in a non-trivial way~\cite{Aad:2020aze}. 
 Moreover, a richer electroweakino spectrum, where multiple eigenmass states are available for the gluino decay, forces an immediate relaxation of the hypothesis on the branching ratio of 100\% for the $\gluino\rightarrow  q\bar{q} \ninoone$ decay. In presence of multiple electroweakino states, different decay channels will be competing, making the mass exclusion limits of Fig.~\ref{fig:summaryPlot_gg} not necessarily applicable. 

Useful information about the dependence of the mass limit on the assumption on the branching ratio can be gathered by looking at the limits set on the cross section of a given model, as shown in Fig.~\ref{fig:summaryPlot_gg} (c) and (d). The yield of the $\gluino\rightarrow  q\bar{q} \ninoone$ signal scales with the square of the branching ratio. Typically the analysis acceptance and efficiency are nearly constant at fixed values of the mass gap between the gluino and the LSP. A guess about the location of the mass limit at a given branching ratio, assuming no sensitivity to the complementing decay channels, can be obtained by combining this information.\footnote{Both collaborations publish extensive auxiliary information about the analysis results on \texttt{hepdata}\cite{hepdata}.}. 

Other benchmark gluino decays have been considered by the collaborations. Inspired by configurations where either the wino or the higgsino mass are larger than the bino mass, but still light enough to play a role in the LHC phenomenology, models with longer gluino decay chains have been extensively taken into account in the analysis design phase. If vector bosons are produced in the gluino decay chain, then this may open up the possibility of having leptons in the final state~\cite{Aad:2021zyy}, possibly same-sign leptons~\cite{Aad:2019ftg}. Likewise, longer gluino decay chains with hadronic vector boson decays may also give rise to final states with higher jet multiplicity~\cite{Aad:2020aze,Aad:2020nyj,Sirunyan:2019ctn,CMS:2019ybf} or hadronic resonances. A CMS analysis\cite{Sirunyan:2020zzv}, for example, targets full hadronic decays of the $Z$ boson produced in gluino pair production followed by $\gluino\rightarrow  q\bar{q} \ninotwo$, followed by $\ninotwo\rightarrow Z \ninoone$: assuming a $\ninotwo$ mass similar to that of the pair-produced gluino, and a small $\ninoone$,  the $Z$ boson has a large boost, and its decay products can be collected in large-$R$ jets. The event selection is characterised by the requirement of large $\etmiss$ and large $\HT$, followed by the requirement of two high-\pt jets with $R=0.8$, with mass compatible with that of the $Z$ boson. 
The dominant SM background process is $\Zjets$. The SM background contribution is estimated by a method exploiting the sidebands of the jet mass distributions. The analysis sets a limit at $\mass{\gluino} =1920$ GeV assuming  $\mass{\gluino}-\mass{\ninotwo} = 50$ GeV and $\mass{\ninoone} = 1$ GeV.

Also considered by the collaborations are decay chains in GGM-inspired scenarios~\cite{Cheung:2007es,Meade:2008wd}, where the LSP is often a light (with mass of the order of 1 GeV or less) gravitino. The phenomenology is determined by which particle is the NLSP. Scenarios where the NLSP is a neutralino that decays to the LSP with the emission of photons, Higgs and $Z$ bosons give rise to topologies that are not normally considered when the neutralino is assumed to be the LSP. A recent ATLAS analysis\cite{ATLAS-CONF-2021-028} studies the case where the NLSP is a bino-higgsino admixture, producing final states containing $\gamma Z$ and $\gamma h$ from $\gluino\rightarrow  q\bar{q}\ninoone$ followed by $\ninoone\rightarrow \gamma/Z/h \tilde{G}$. At the time of the writing, the analysis has been released as a preliminary result. The analysis requires the presence of a high-\pt isolated photon in events with relatively large jet multiplicities (three to five, depending on the signal region), large \etmiss and \HT. The dominant backgrounds arise from $W\gamma$, $\ttbar \gamma$ and $\gamma+\mathrm{jets}$ production, which is estimated with the MC normalised in dedicated control regions. Limits are extracted in the $\mass{\gluino}$-$\mass{\ninoone}$ mass plane. Gluino masses above 2.1 TeV are excluded for any  $\mass{\ninoone} < \mass{\gluino}$.

One can gather some understanding of how much the gluino mass limits depend on the gluino decay chain from Fig.~\ref{fig:summaryPlot_gg} (a), where the limits for different options are reported.

\subsubsection{Gluino pair production with decays into third generation quarks}
\label{sec:gtt}

\begin{figure}[tb]
\begin{center}
\subfigure[]{
\includegraphics[width=0.4\textwidth]{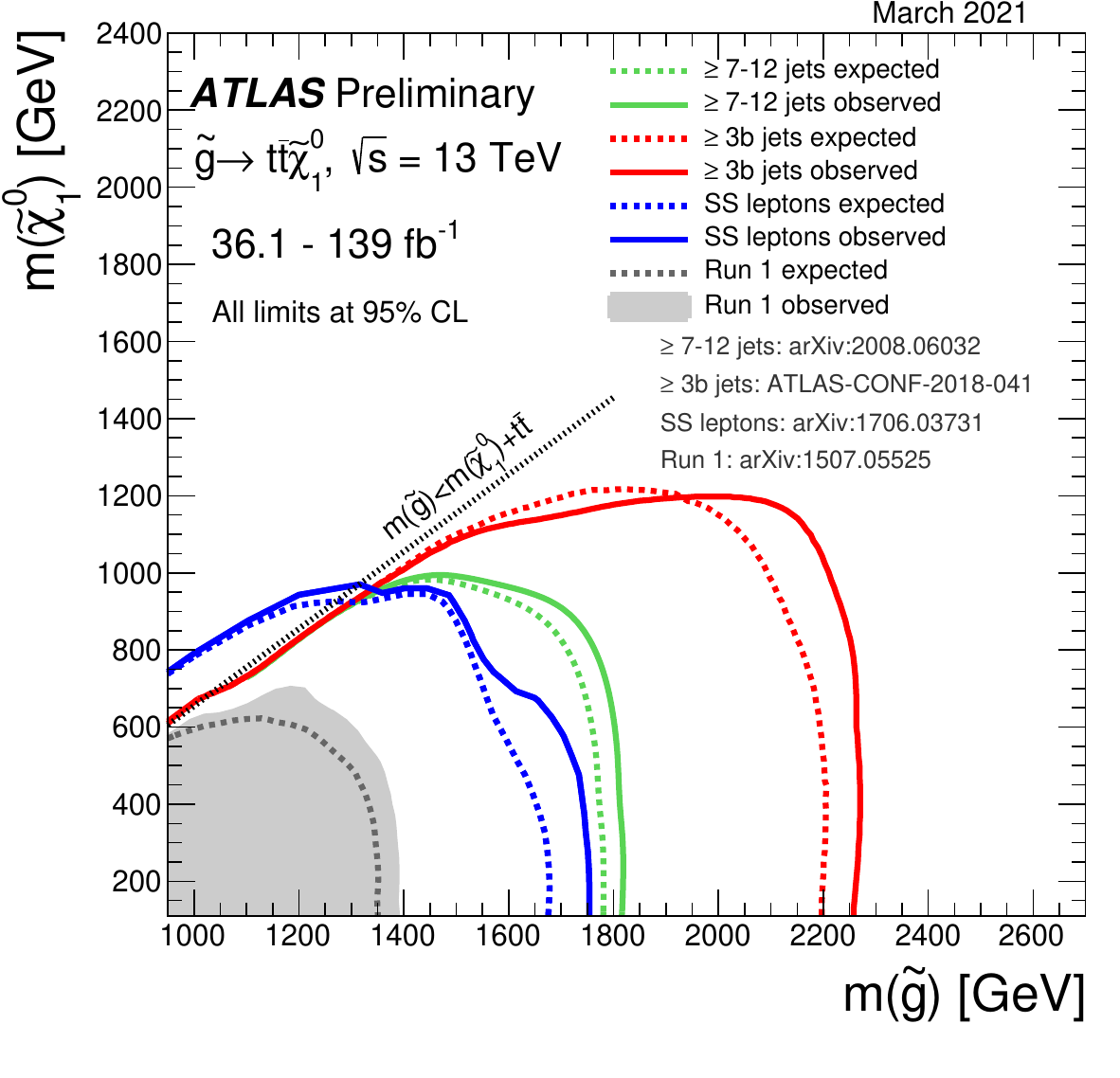}
}
\subfigure[]{
\includegraphics[width=0.41\textwidth]{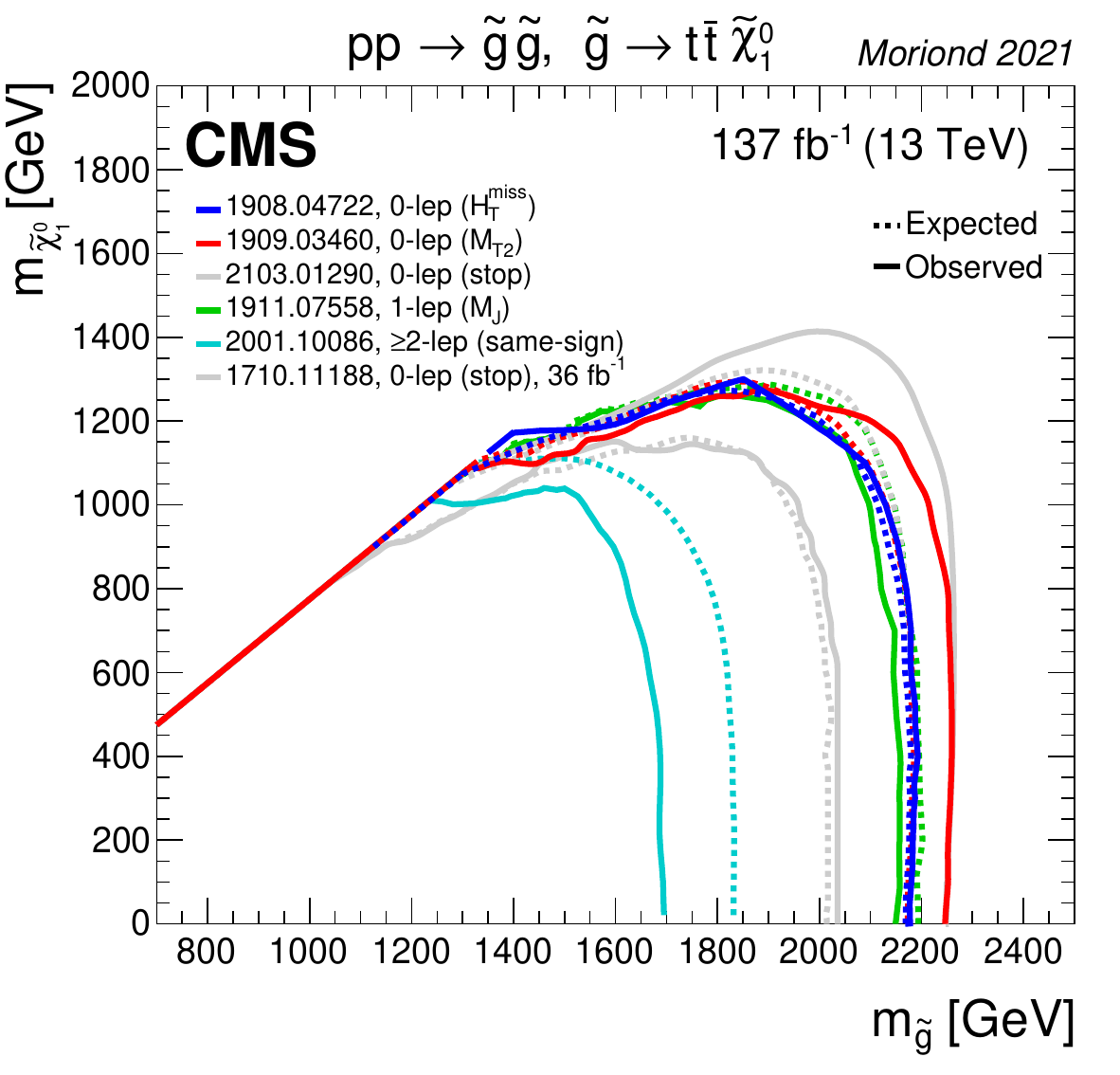}
}
\end{center}
\caption{Gluino pair production summary plot of the (a) ATLAS\cite{ATL-PHYS-PUB-2021-019} and (b) CMS\cite{CMS-SUSY-Summaries} collaborations, where the gluino is assumed to decay to a pair of top quarks and a neutralino.}
\label{fig:summaryPlot_gtt}
\end{figure}

As pointed out in Sec.~\ref{sec:models}, a particularly important simplified model studied by the experimental collaborations is the decay of pair-produced gluinos into third generation quarks. Similarly to the $\gluino\rightarrow  q\bar{q}\ninoone$ case, these final states arise from the decay of the gluino through an intermediate sbottom or stop. Naturalness requirements suggest that the third generation squarks (in particular the stop) are lighter than the other squarks, therefore favouring gluino decays into third generation squarks. A great experimental effort has been devoted to the search of $\gluino\rightarrow  t\bar{t}\ninoone$, leading to a striking final state containing four top quarks and missing transverse momentum for gluino pair production. Figure~\ref{fig:summaryPlot_gtt} shows the ATLAS and CMS summary plots for these scenarios. Some of the analyses driving the sensitivity to this scenario in CMS have been already discussed in the context of the $\gluino\rightarrow  q\bar{q} \ninoone$ scenario\cite{Sirunyan:2019ctn,CMS:2019ybf}: these zero-lepton analyses implement a categorisation of the events based on the number of the identified $b$-jets. The categories with large $b$-jet multiplicities are those that provide the highest sensitivity to the $\gluino\rightarrow  t\bar{t}\ninoone$ case. However, the richness of possible topologies that can arise from a final state containing four top quarks offers multiple ways to target this scenario. It is interesting to note, for example, how a similar sensitivity is obtained by an independent analysis selecting one-lepton final states\cite{CMS:2019tlp} (in green in Fig.~\ref{fig:summaryPlot_gtt} (b)).  In this case, signal events are characterised by the presence of a high-\pt lepton and large \etmiss. The transverse mass \mt between the lepton and the \etmiss is a variable which presents an endpoint at the $W$-boson mass for high-cross section SM background processes such as \Wjets and semileptonic decays of a \ttbar pair. The analysis exploits the approximate independence of a second mass variable (\mj), the sum of the masses of jets reconstructed using the anti-\kt algorithm with a $R$ parameter of $R=1.4$. The events are further categorised based on \etmiss, \njet, \nbjet. Another CMS analysis which provides excellent sensitivity will be better described in the context of direct stop pair production\cite{CMS:2021beq}.

The ATLAS sensitivity is driven by an analysis\cite{ATLAS-CONF-2018-041} targeting final states containing either zero or one leptons, and exploiting the presence of multiple $b$-jets in the final state (at least three) in signal regions characterised by large jet multiplicities. The rest of the selection exploits the fact that relatively large values of \meff, \mj, \etmiss are expected for the signal. At the time of the writing, the analysis has been released as a preliminary result using a partial Run 2 dataset. 

Finally, final states containing four tops can also be targeted with analyses looking for same-sign leptons, or three leptons\cite{Sirunyan:2020ztc,Aad:2019ftg}: the use of leptons to extract the signal from the Standard Model background leads to increased sensitivity to models with small mass gaps between the pair-produced gluinos and the LSP. It is worth noting that such analyses profit from a relatively low SM background, offering sensitivity to a variety of possible supersymmetric and generic BSM scenarios. 

Other scenarios of gluino pair production and decay into third generation quarks have been considered by the collaborations. It is worth mentioning the experimental effort targeting the decay of the gluino into a pair of $b$-quarks and the LSP, $\gluino\rightarrow  b\bar{b} \ninoone$, leading to a final state with four $b$-jets and significant missing transverse momentum, or that targeting the decay of a gluino into a top quark, a bottom quark and the LSP, $\gluino\rightarrow  t\bar{b}\ninoone + \mathrm{soft\ particles}$. The last final state is obtained, for example, via an intermediate virtual stop decaying into a bottom quark and a chargino, in a wino-like LSP scenario, with nearly degenerate $\chinoonepm$ and $\ninoone$, such that the additional particles emitted in the $\chinoonepm \rightarrow \ninoone$ transition have low \pt and are not reconstructed. The ATLAS and CMS analyses already mentioned in the context of the effort for the search of $\gluino\rightarrow  t\bar{t}\ninoone$ typically have signal regions which offer excellent sensitivities to these scenarios as well.

The explicit request of the presence of $b$-jets in the final state in the case of gluino decays mediated by intermediate stops and sbottoms causes significant changes in the typical background composition of the relevant signal regions. Depending on the specific final state targeted, relevant SM background processes arise from the production of top pairs, often in association with extra heavy flavour quarks, or vector bosons. The quality of the modelling of these low cross section processes is in general validated with a relatively low precision: estimates of these background processes relying on the yield normalisation in control regions, or on purely data-driven techniques are a key element of many of the mentioned analyses.

\subsubsection{Squark pair production} 

\begin{figure}[tb]
\begin{center}
\subfigure[]{
\includegraphics[width=0.4\textwidth]{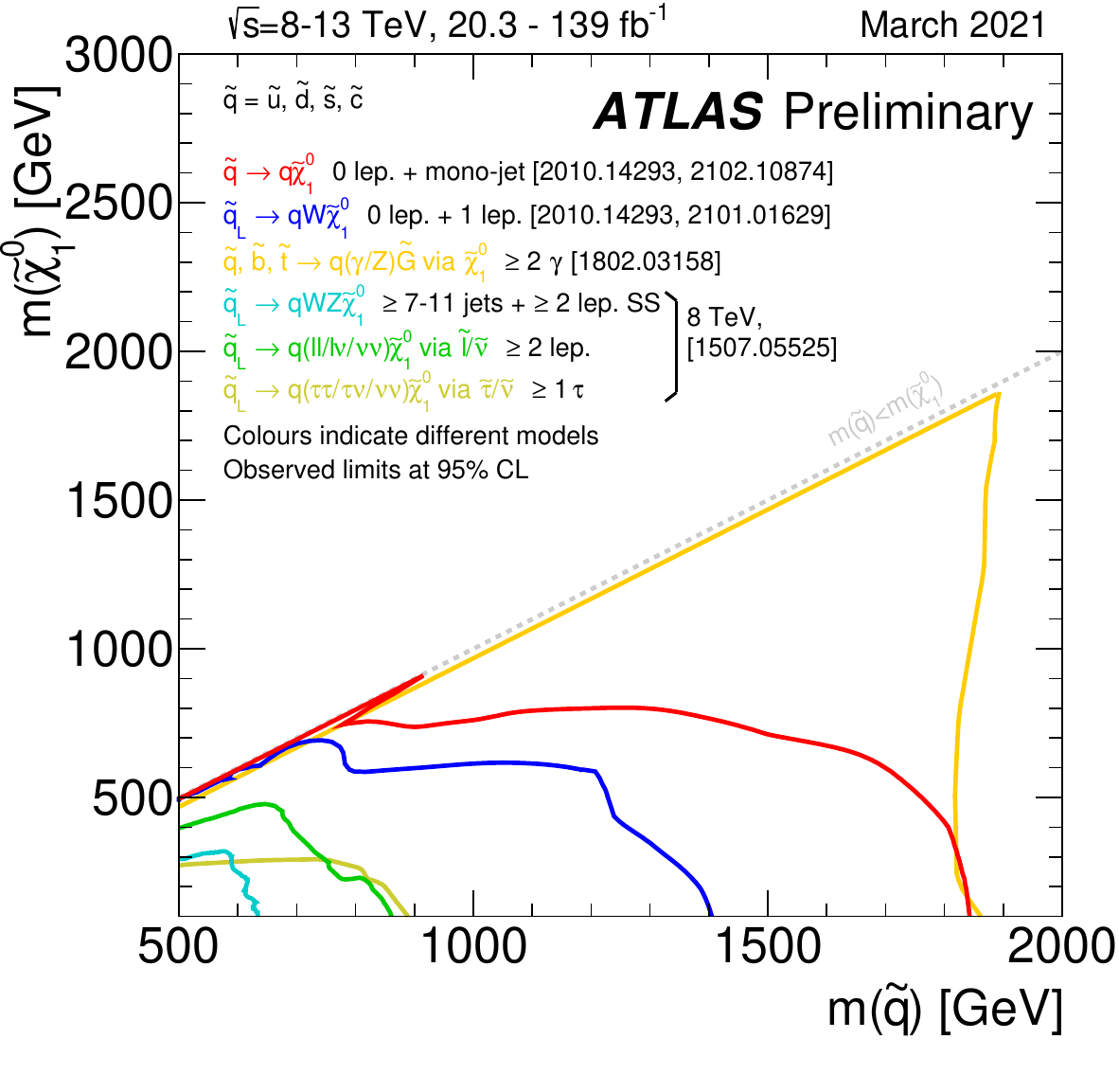}
}
\subfigure[]{
\includegraphics[width=0.4\textwidth]{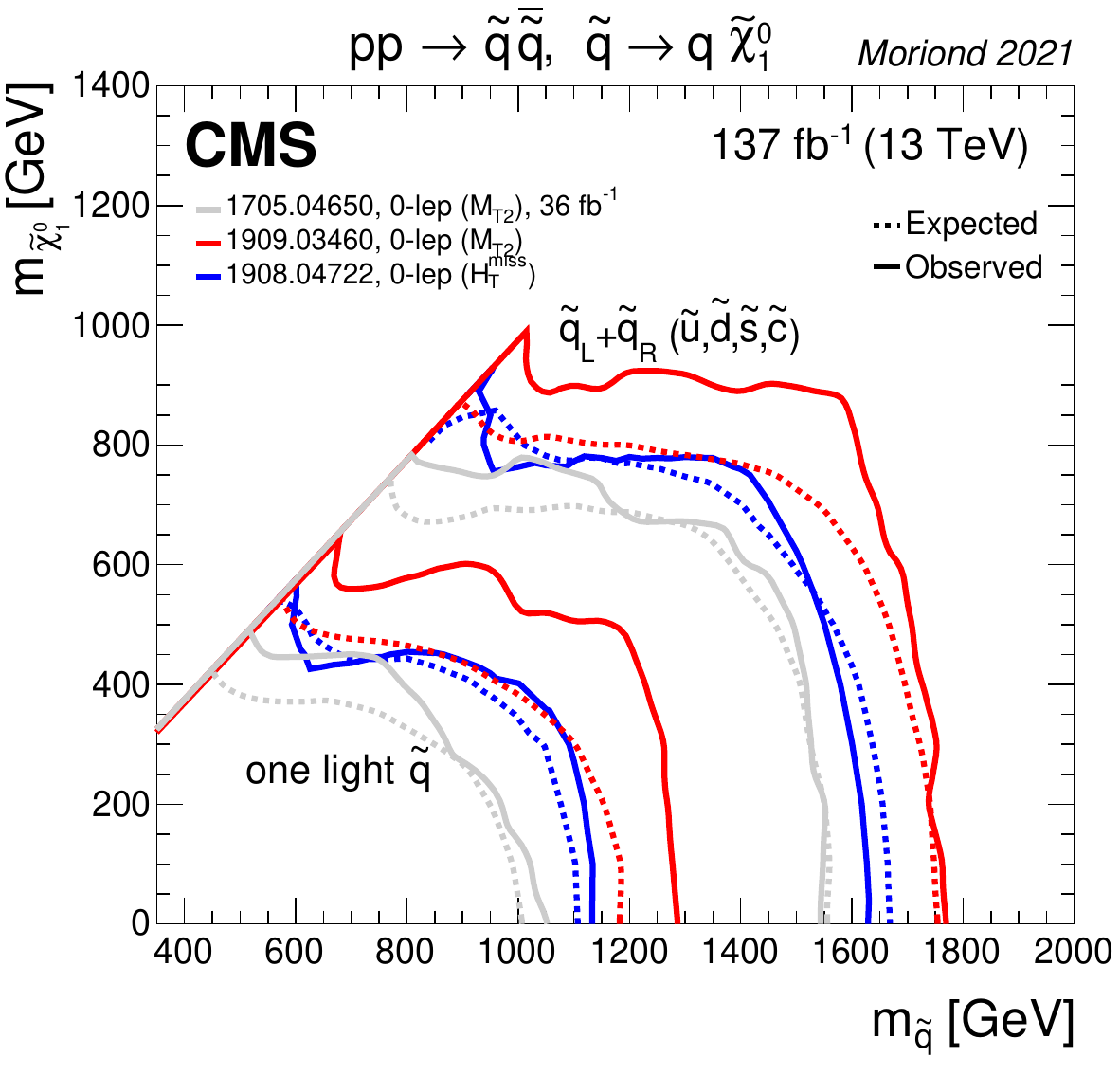}
}
\end{center}
\caption{Squark pair production summary plot of the (a) ATLAS\cite{ATL-PHYS-PUB-2021-019} and (b) CMS collaborations\cite{CMS-SUSY-Summaries}.}
\label{fig:summaryPlot_qq}
\end{figure}

Closely connected with the search for gluino pair production, the search for squark pair production relies on similar experimental strategies. Figure~\ref{fig:summaryPlot_qq} shows a summary of the mass exclusion limits obtained by ATLAS and CMS. The analyses providing the best sensitivity are the same already discussed in Sec.~\ref{sec:gqq} in the context of gluino pair production. The decay chain of squarks is typically producing a lower jet multiplicity than that for gluinos: assuming a stable neutralino as LSP, the shortest decay chain is $\squark\rightarrow q\ninoone$, yielding two expected jets in the final state. The search regions which are relevant for these searches are therefore those targeting lower values of \njet.

It is important to point out that the limits of Fig.~\ref{fig:summaryPlot_qq} (a) and part of those in Fig.~\ref{fig:summaryPlot_qq} (b) assume a four-flavour, two-chirality degeneracy of the squarks ($\tilde{q}_\mathrm{L}$ and $\tilde{q}_\mathrm{R}$ at the same mass for each of the four light flavours). Fig.~\ref{fig:summaryPlot_qq} (b) also shows the limit corresponding to the assumption of a single accessible chirality state of one flavour, differing in cross section by a factor eight.

An impressive set of dedicated searches have been developed by ATLAS and CMS for pair production of third-generation squarks (\stop and \sbottom). As already mentioned, the mass of the stop is bound below or at the TeV scale by some naturalness arguments. The \sbottomL  belongs to the same weak isospin multiplet as \stopL, therefore sharing a common SUSY breaking mass term, making also the sbottom a good candidate for discovery. The keyword for third-generation squark searches is $b$-jets: unless flavour violation is assumed, $b$-quarks will be produced as part of the decay chain, giving a very clear experimental handle to these searches. 

\begin{figure}[tb]
\begin{center}
\subfigure[]{
\includegraphics[width=0.55\textwidth]{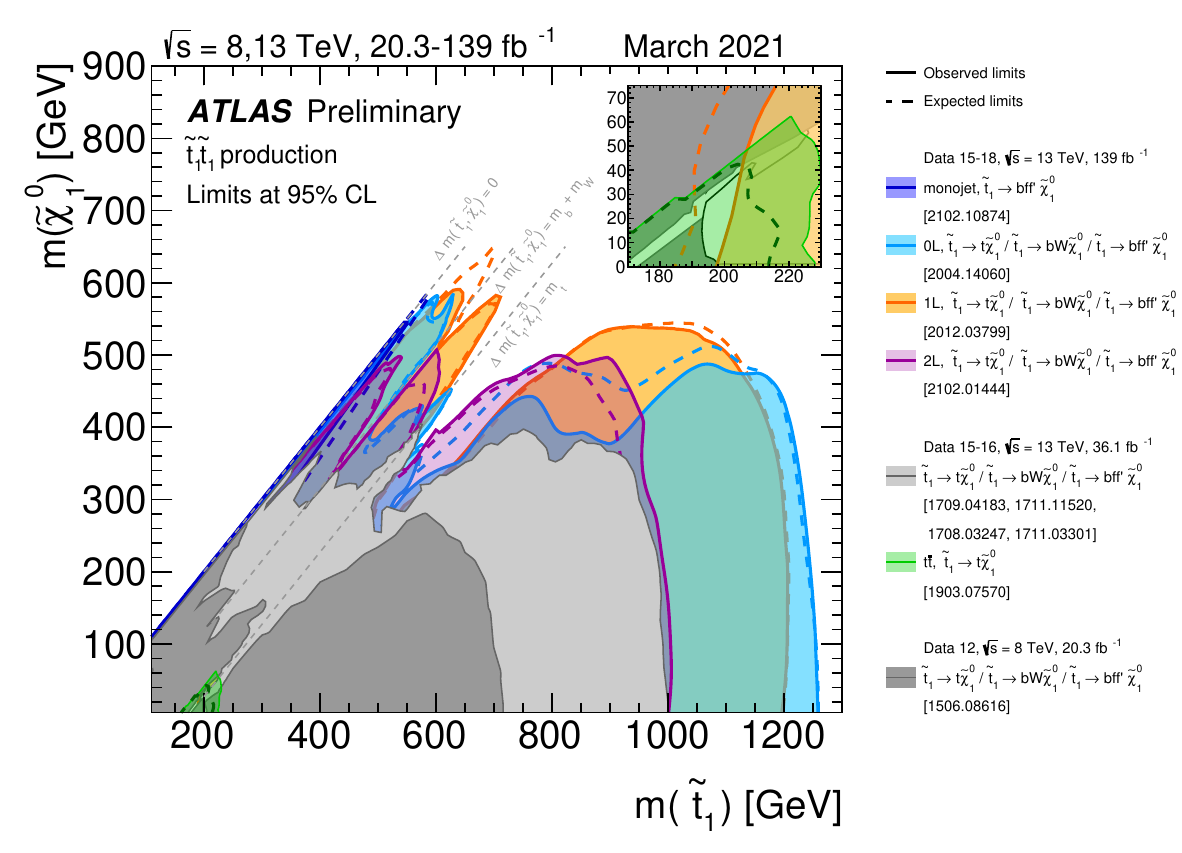}
}
\subfigure[]{
\includegraphics[width=0.4\textwidth]{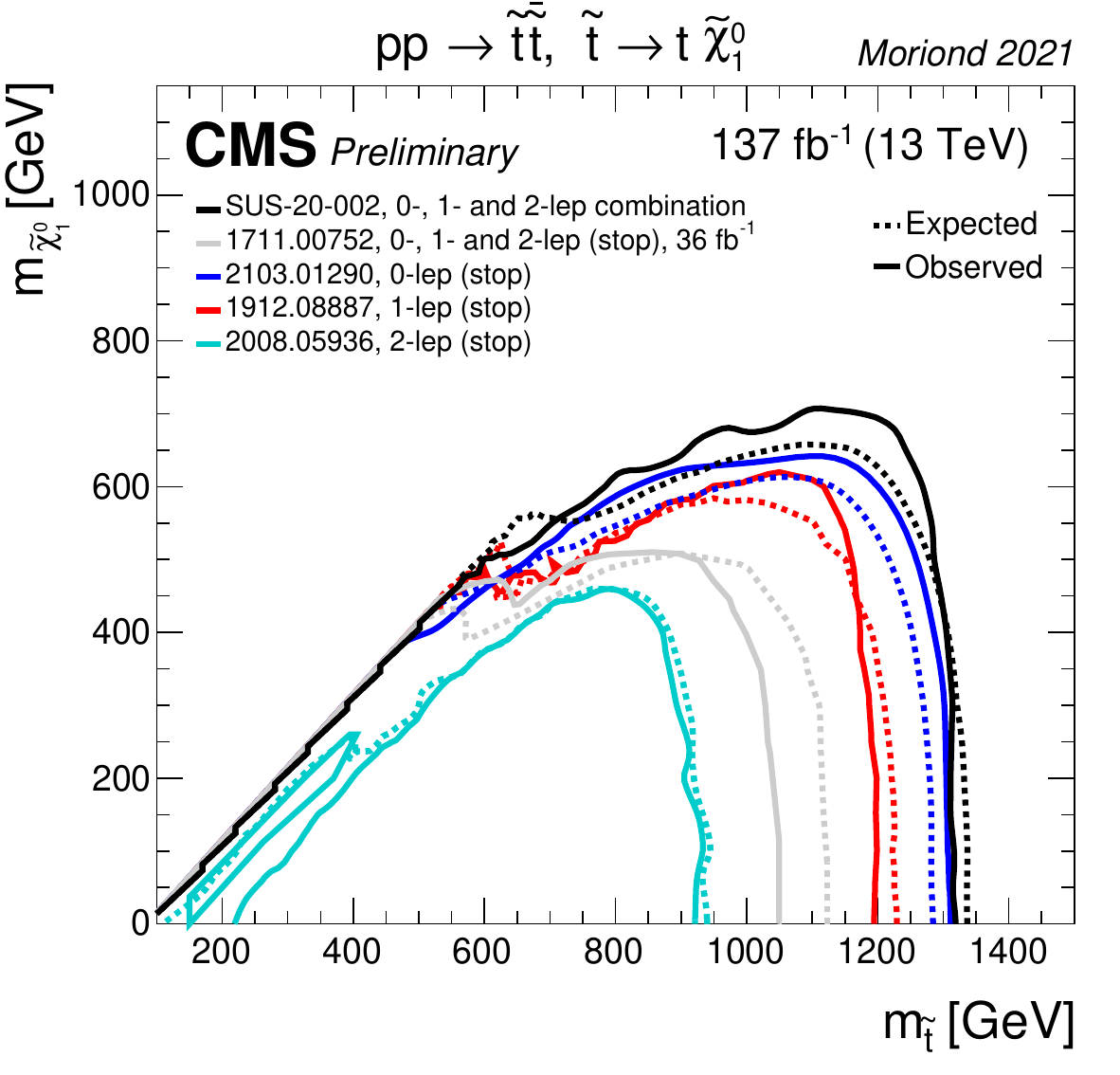}
}
\end{center}
\caption{Stop pair production summary plot of the (a) ATLAS\cite{ATL-PHYS-PUB-2021-019} and (b) CMS\cite{CMS-SUSY-Summaries} collaborations, where the stop is assumed to decay via $\stopone\rightarrow t^{(*)}\ninoone$.}
\label{fig:summaryPlot_tt}
\end{figure}

Even in its simplest possible decay mode in models with a neutralino LSP, $\stopone \rightarrow t^{(*)} \ninoone$, the strategy for stop pair production search is relatively complex: because of the large top-quark mass, and depending on the mass splitting between the \stopone and the \ninoone, on-shell top quarks may or may not be present in the final state. Figure~\ref{fig:summaryPlot_tt}
shows the summary plots of the collaborations, assuming that the branching ratio of $\stopone \rightarrow t^{(*)} \ninoone$ is 100\%. Different regions are clearly visible (and explicitly highlighted in Fig.~\ref{fig:summaryPlot_tt}(a) ) for $\Delta m\left(\stopone,\ninoone\right) > \mtop$ (often referred to as two-body stop decay), $\mass{W} + \mass{b} < \Delta m\left(\stopone,\ninoone\right) < \mtop$  (three-body stop decay) or $\Delta m\left(\stopone,\ninoone\right) < \mass{W} + \mass{b}$ (four-body stop decay). Limits for the four-body stop decay region are not shown in  Fig.~\ref{fig:summaryPlot_tt}(b).
 
Clearly, the analyses dominating the sensitivity in the two-body regions are those requiring either zero\cite{Aad:2020sgw,CMS:2021beq} or one\cite{Aad:2020aob,Sirunyan:2019glc} lepton from the top-quark decay in the final state. The all-hadronic analysis from CMS\cite{CMS:2021beq}, which is sensitive also to gluino pair production followed by the gluino decay into top quarks, exploits both boosted and resolved neural-network-based  top- and $W$-tagging algorithms, depending on their expected \pt range: at high \pt, the decay products of tops and $W$ can be collected in large-$R$ jets that can be tagged exploiting their mass and substructure; at low \pt, top and $W$ can be identified from combinations of three and two jets: the challenge is typically to define a suitable algorithm to solve the combinatorics. The final selection defines 123 bins (high $\Delta m$ selection in the original publication) based on the minimum transverse mass between the $b$-jets and \etmiss ($m_{\mathrm{T}}^b$), \etmiss, \HT, \njet, \nbjet, the number of boosted and resolved top quarks, and the number of identified $W$ bosons.  

The analysis defines a second set of 52 search bins (low $\Delta m$ selection) targeting mainly the three- and four-body stop decays. The general strategy (used in many other analyses looking for small mass differences between the pair-produced particle and the LSP) is to deploy a selection requiring the pair-produced sparticle system to recoil against a high-\pt ISR jet. This ISR-like selection also exploits a soft $b$-quark selection, where secondary vertices associated to $b$-hadron decays are tagged independently of the presence of a jet associated to it. Selections on \njet and \nbjet complete the signal regions definition.  

Depending on the search bin considered, the background is dominated either by the production of \Zjets in association with heavy-flavour quarks, or $\ttbar$ and single top  (and to a lesser extent $\Wjets$) events where one of the two top quarks yields a lepton in the final state, which is either not reconstructed or is a \tauhad. They are estimated with the help of one-lepton control region selections (for \ttbar, single top and \Wjets) and two-lepton and $\gamma+$ jets selections for \Zjets. Other low cross section processes of top pair production in association with vector bosons (in particular $\ttbar Z$) are relevant - they are estimated with the simulation relying on measured total cross section. 

\begin{figure}[tb]
\begin{center}
\includegraphics[width=0.55\textwidth]{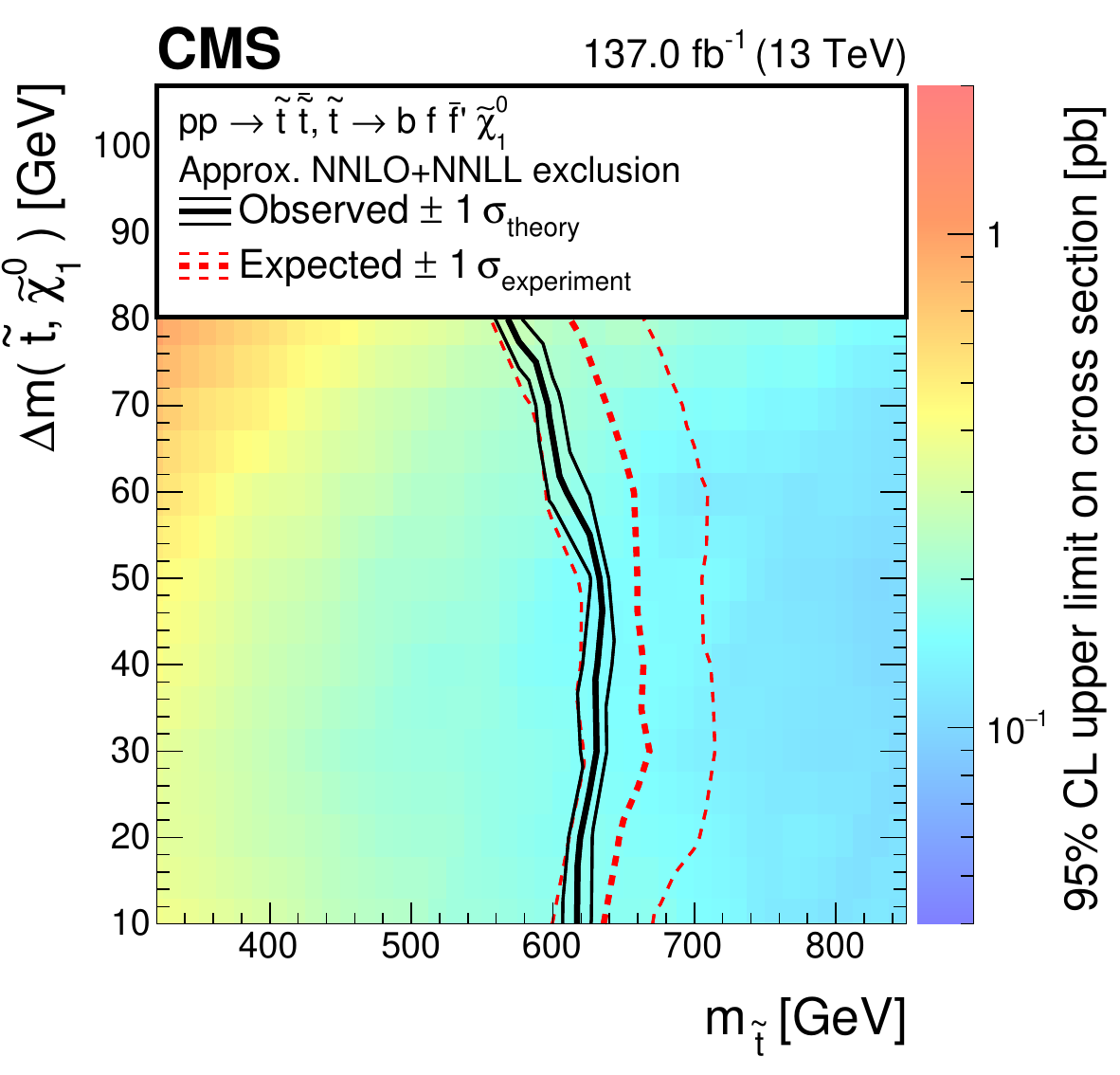}
\end{center}
\caption{Exclusion limits in mass and cross section for the four-body decay of the stop\cite{CMS:2021beq}.}
\label{fig:compressedCMStt}
\end{figure}

The ATLAS all-hadronic analysis, and the CMS and ATLAS one-lepton analyses follow similar strategies. A challenging background process for the one-lepton analyses is \ttbar production where both top quarks decay leptonically, and one lepton is lost, or is a \tauhad. Dedicated strategies are adopted to reject this background: the ATLAS analysis\cite{Aad:2020aob} makes use of a variable called {\em topness}, which yields an estimator of how much a given event with one lepton in the final state resembles a dileptonic \ttbar decay based on expected detector resolutions and mass constraints present in such events.

A challenging mass hierarchy is the one where $\Delta m\left(\stopone,\ninoone\right) \approx \mtop$: the final state and kinematics of $\stopone\stopone\rightarrow  t\bar{t}\ninoone\ninoone$ closely resembles that for \ttbar production, especially if $\mass{\ninoone} \approx 0$. In this case, precision measurements of the SM \ttbar production can come into rescue, as indicated in the inset plot of Fig.~\ref{fig:summaryPlot_tt} (a): in this case, a precise measurement of the spin correlations for \ttbar events\cite{ATLAS:2019zrq} allows to constrain the additional production of a pair of scalar particles.    

Under the assumption of a 100\% branching ratio of $\stopone \rightarrow t^{(*)} \ninoone$, Fig.~\ref{fig:summaryPlot_tt} shows that stop pair production is excluded for stop masses up to 1.3 TeV and neutralino masses up to about 0.6 TeV. The limit in the four-body region (shown in Fig.~\ref{fig:compressedCMStt} for the zero-lepton CMS analysis discussed), show limits on the stop mass at about 600 GeV even for  very small values of $\Delta m\left(\stopone,\ninoone\right)$.

Of course, the assumption of a 100\% branching ratio for $\stopone \rightarrow t^{(*)} \ninoone$ is a strong condition: one should not make the mistake of taking Fig.~\ref{fig:summaryPlot_tt} as a constraint on the stop mass tout-court. If a richer electroweakino spectrum exists below the top mass scale, for example, the possibility of $\stopone\rightarrow b \chinoonep$ and $\stopone \rightarrow t \ninotwo$ may open up, and even be dominant, depending on the electroweakino composition and on the stop chirality. The analyses already discussed have good sensitivities to these scenarios. Final states from the decay $\stopone\rightarrow b \chinoonep$ are explicitly targeted also by analyses requiring two leptons\cite{Aad:2021hjy,Sirunyan:2020tyy} (which also perform well in the more compressed regions of the  $\stopone\rightarrow t^{(*)}\ninoone$ scenario already discussed). The decay $\stopone \rightarrow t \ninotwo$ may give rise to signatures containing $Z$ or $h$ bosons, and these final states are targeted by an analysis looking for multilepton or explicit $b\bar{b}$ resonances in the final state\cite{Aad:2020qwe}. Finally, flavour-changing decays of the stop $\stopone\rightarrow c\ninoone$ can compete with the four-body (and, to some extent, even with the three body) stop decay depending on the $\ninoone$ composition, stop chirality, $\Delta m\left(\stopone,\ninoone\right)$ and flavour structure, and they are targeted as part of the effort for compressed final states.

\begin{figure}[tb]
\begin{center}
\subfigure[]{
\includegraphics[width=0.4\textwidth]{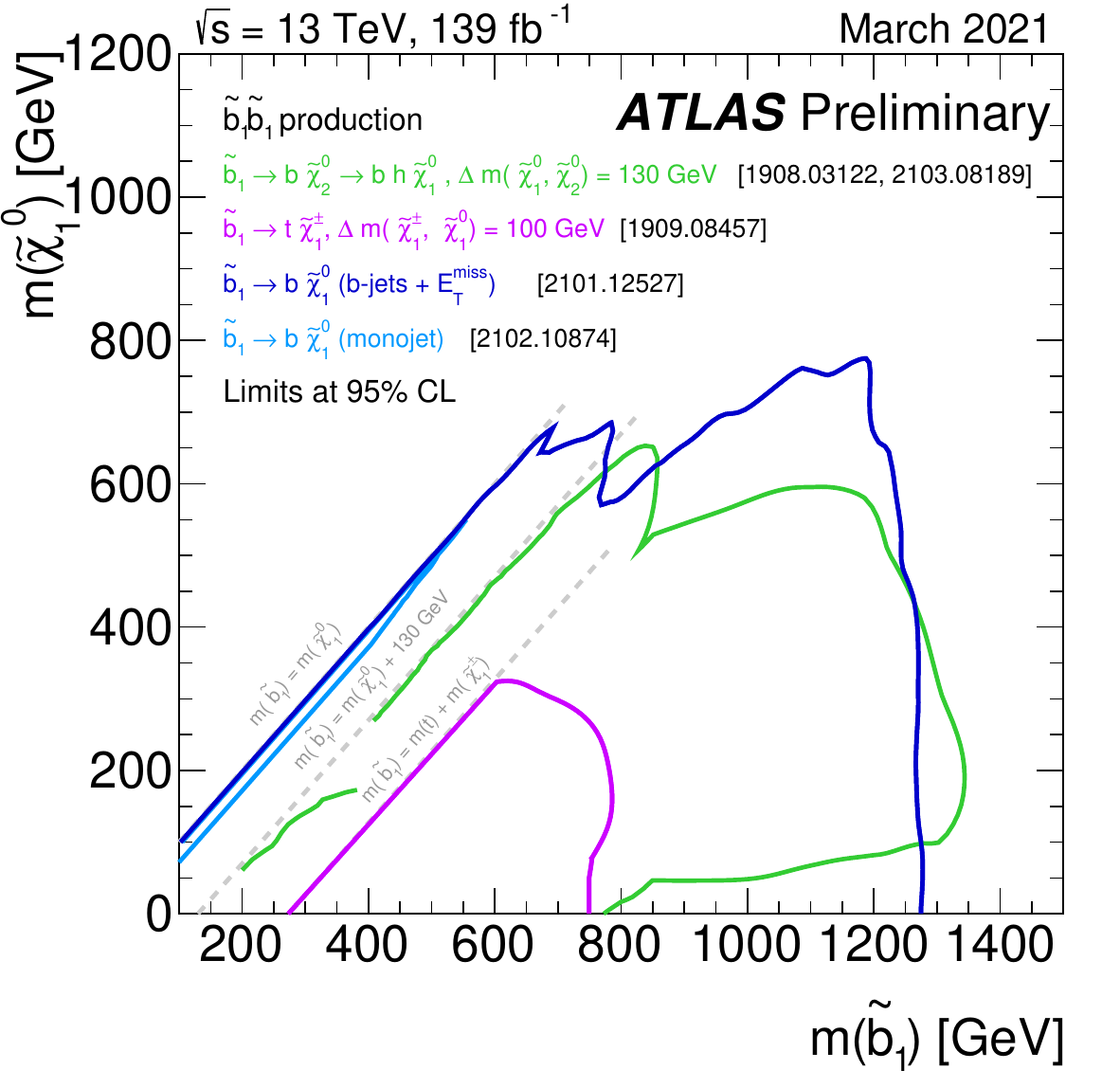}
}
\subfigure[]{
\includegraphics[width=0.4\textwidth]{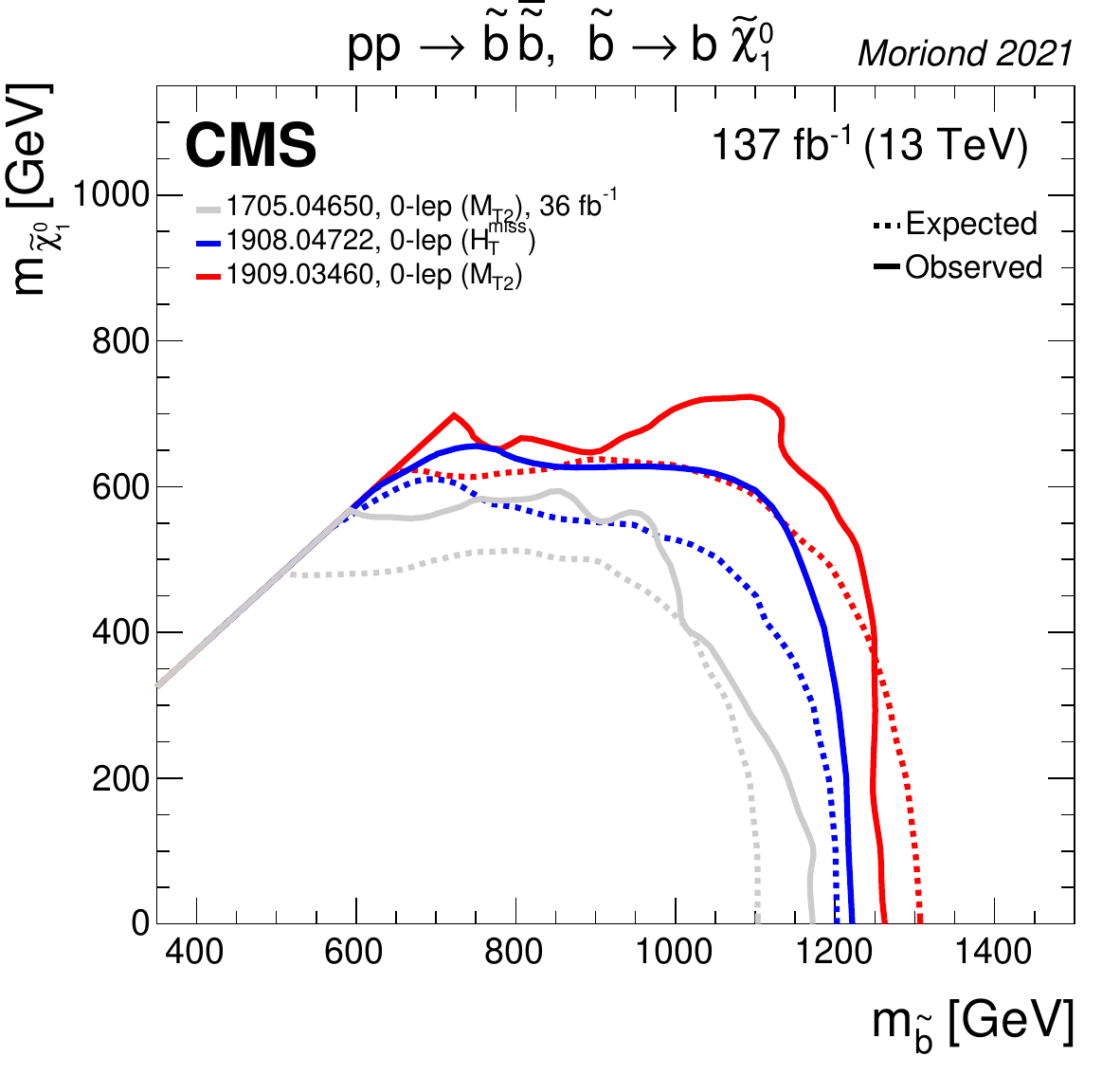}
}
\end{center}
\caption{Sbottom pair production summary plot of the (a) ATLAS\cite{ATL-PHYS-PUB-2021-019} and (b) CMS collaborations\cite{CMS-SUSY-Summaries}.}
\label{fig:summaryPlot_bb}
\end{figure}

As already mentioned, closely connected to the search for stop pair production is that for sbottom pair production. Figure~\ref{fig:summaryPlot_bb} shows a summary of the 95\% CL mass exclusion limits obtained by the two collaborations, assuming a neutralino LSP. Figure~\ref{fig:summaryPlot_bb} (b) focuses on the sbottom decay into a $b$ quark and a neutralino, $\sbottomone \rightarrow b \ninoone$. The analyses that determine the sensitivity there have been already discussed in the context of the gluino or stop pair production. Figure~\ref{fig:summaryPlot_bb} (a) is instead summarising the results for a few different possible decays of the sbottom. If the sbottom mass eigenstate is dominated by its left chirality component, and in presence of a bino-like LSP and a wino-like doublet, the sbottom decay may preferentially be into the wino state $\sbottomone\rightarrow b \ninotwo$. This decay was targeted in final states containing  a Higgs boson from $\ninotwo\rightarrow h \ninoone$ both in the $h\rightarrow b\bar{b}$\cite{Aad:2019pfy} and $h\rightarrow\tau^+ \tau^-$\cite{Aad:2021oos} channels. The   $\sbottomone \rightarrow b \ninoone$ decay is instead explicitly targeted by a dedicated analysis\cite{Aad:2021jmg} with different sets of signal regions for the large, intermediate and small $\Delta m\left(\sbottomone,\ninoone\right)$ regions. For the large $\Delta m\left(\sbottomone,\ninoone\right)$ region, a selection based on \etmiss, \meff is defined, after exploiting the end-point of the contransverse mass\cite{contransverse} variable (similar to the stransverse mass already discussed) to reject \ttbar production. The intermediate $\Delta m\left(\sbottomone,\ninoone\right)$ selection uses a BDT exploiting jet-level kinematic quantities on top of higher-level variables (\etmiss, \meff). The small $\Delta m\left(\sbottomone,\ninoone\right)$ defines an ISR-like selection and exploits a soft $b$-hadron selection through a secondary vertexing algorithm. The mass limits set by these analyses, each assuming 100\% decay branching ratio into different final states, are similar to those obtained on the stop.

\subsubsection{Strong production in RPV and non-prompt scenarios}
\label{sec:strongRPV}

We have so far focused on scenarios where each step in the decay of the gluino or the squark is assumed to be prompt, and the LSP is assumed to be stable. An interesting phenomenology arises when these assumptions are dropped. 

RPV SUSY scenarios open an almost endless list of potential final state topologies. The review of the motivation and theory behind RPV scenarios goes well beyond the scope of this paper - we refer the interested reader to excellent existing publications~\cite{herbiDreiner}. From the phenomenological point of view, the paradigm shift in the experimental search strategy for RPV SUSY is the lack of missing transverse momentum associated to the presence of a stable, weakly interacting LSP. At the same time, the decay of on-shell SUSY particles into fully measurable decay products may allow the reconstruction of resonances. 

\begin{figure}[tb]
\begin{center}
\subfigure[]{
\includegraphics[width=0.4\textwidth]{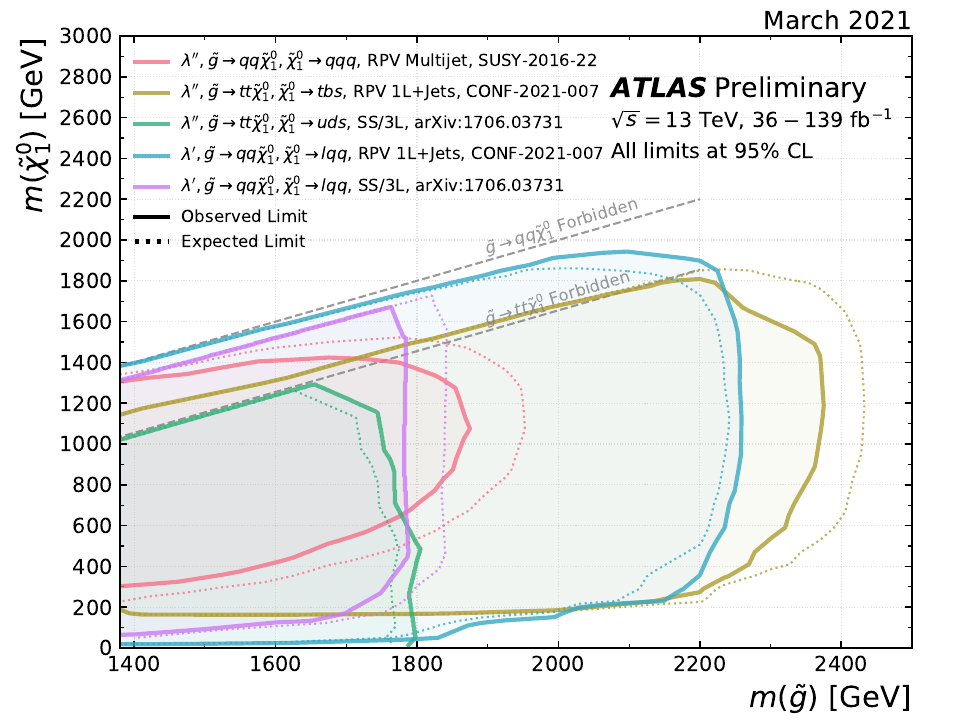}
}
\subfigure[]{
\includegraphics[width=0.43\textwidth]{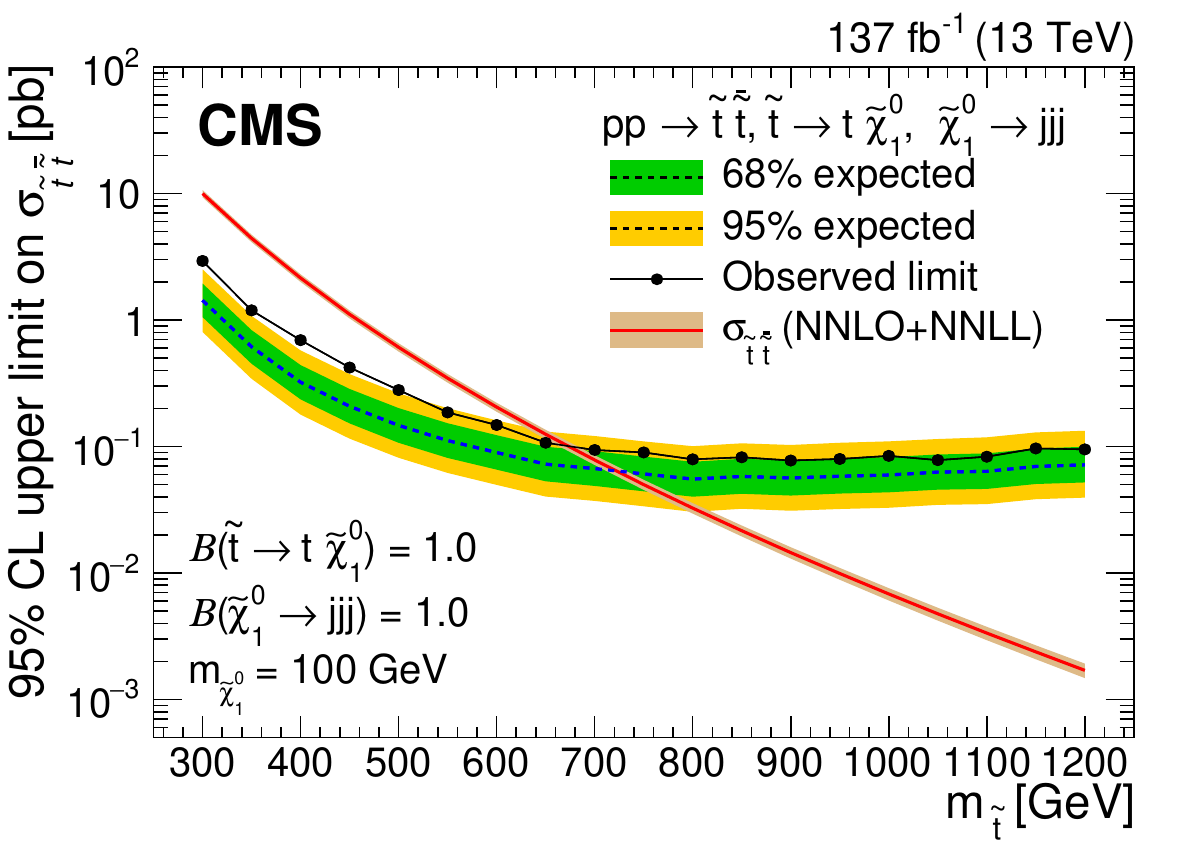}
}
\end{center}
\caption{Limits on RPV gluino and neutralino decays: (a) summary of ATLAS\cite{ATL-PHYS-PUB-2021-019} mass exclusion limits in the  $\left(\mass{\gluino},\mass{\ninoone}\right)$ mass plane for different RPV decays of the neutralino. Each shaded curve corresponds to a different gluino and neutralino decay mode, or to a different analysis. (b) Cross-section limits as a function of the stop mass set by a CMS analysis\cite{CMS:2021knz} assuming stop pair production followed by the $\stopone\rightarrow t \ninoone$ (with branching ratio 100\%) and by the RPV decay $\ninoone\rightarrow jjj$.  }
\label{fig:summaryPlot_gg_RPV}
\end{figure}

 Depending on which RPV coupling is assumed to be different from zero, and depending on the size of the coupling, different topologies of interest for the gluino decay should be considered. Figure~\ref{fig:summaryPlot_gg_RPV} (a)  shows a summary of the ATLAS sensitivity to gluino pair production, followed by decay chains concluding with an RPV decay of the LSP. The decays of the gluino itself are the same as those considered in Sec.\ref{sec:gqq} and \ref{sec:gtt}. However, this time the neutralino LSP is allowed to decay promptly with a decay into quarks and leptons ($\lambda' \neq 0$) or into quarks only ($\lambda'' \neq 0$). We will concentrate on the results of an ATLAS analysis focusing on final states containing at least one lepton and multiple jets\cite{ATLAS:2021fbt}. 
 
 The analysis considers classes of models with different hypotheses for the electroweakino mass hierarchy (pure bino, pure wino or pure higgsino, with reference to Fig.~\ref{fig:ewikino_spectrum}), and different production and decay modes of SUSY particles (gluino, stop, and electroweakino pair production). Following the assumption of minimal flavour violation~\cite{DAMBROSIO2002155,Csaki:2011ge}, the only $\lambda''$ coupling which is not heavily constrained is $\lambda''_{323}$. The analysis focuses therefore in scenarios with $\lambda''_{323} \neq 0$, causing chargino and neutralino transitions to triplets of second- and third-generation quarks. Depending on the size of $\lambda''_{323}$, RPV stop decays may become relevant. The analysis also considers the possibility of $\lambda' \neq 0$, introducing the possibility of a neutralino decay involving leptons. 

Two categories of signal regions are defined, depending whether events contain a single lepton, or a pair of same-sign leptons. Events are further categorised based on the jet multiplicity (defined based on multiple jet \pt thresholds) and on the $b$-jet multiplicity. The sensitivity to the signal due to a $\lambda''_{323}$ coupling is primarily obtained with the signal regions with high jet and $b$-jet multiplicities. The dominant background production is \Wjets and \Zjets in the selections with no $b$-jets, and \ttbar production otherwise. The modelling of all these processes for selections with high jet and $b$-jet multiplicities is expected to be poor, therefore a detailed data driven estimation strategy is used for the background: the background yields at high jet and $b$-jet multiplicities are extrapolated from those at lower multiplicities assuming a quasi-staircase-scaling\cite{BERENDS1989237} behaviour of \njet, and a template for \nbjet extracted at low \njet and evolved assuming an almost fixed probability that any additional jet is a $b$-jet. 

The limit shown in Fig.~\ref{fig:summaryPlot_gg_RPV} for $\gluino\rightarrow \ttbar \ninoone\rightarrow \ttbar tbs$ refers to the case in which the LSP is assumed to be bino-like. The results for a higgsino-like or wino-like LSP show a similar trend in the $\mass{\gluino}$-$\mass{\ninoone}$ plane, and are weaker by about 200 GeV in mass than those shown. The analysis sets limits also for $\gluino\rightarrow q\bar{q}\ninoone\rightarrow q\bar{q}q\bar{q}\ell/\nu$, $\gluino\rightarrow t\bar{t}\rightarrow tbs$, and for the decay of pair-produced stops into $t\ninoone/\ninotwo$ or $\chinoonepm$ followed by electroweakino decays into hadrons. Finally, limits are set on the production of pair-produced wino-like and higgsino-like multiplets, followed by RPV $\lambda''_{323}$-allowed decays of the electroweakinos. 

A similar strategy for the background estimation, involving a background extrapolation from low- to high-\nbjet, is used by a zero-lepton analysis\cite{Aad:2020uwr}, obtaining important constraints on the masses of stops in scenarios with electroweakino \Rparity violating decays. 

A CMS analysis\cite{CMS:2021knz} also targets stop pair production followed by $\stopone \rightarrow t \ninoone$, and by the $\lambda''$ neutralino decay into three quarks in final states containing one lepton. Signal events are characterised by large jet multiplicities. The analysis selects events containing at least seven jets, one of which is $b$-tagged, and high \HT. The invariant mass between the lepton and the $b$-jet is required to be compatible with that of a top quark decay. The discriminant of a neural network trained to separate the signal from the dominant $\ttbar$ background based on the spatial distribution of jets and decay kinematic distributions, is the variable on which the event categorisation is based. The obtained cross section limits as a function of the stop mass are compared to the theoretical stop pair production cross section, excluding stop masses of $\mass{\stopone} < 660$~GeV. The analysis also has interpretations in models of Stealth SUSY. 

Further important constraints on many gluino RPV decays come from analyses searching for an excess of events with same-sign leptons, or with three leptons\cite{Sirunyan:2020ztc, Aad:2019ftg}. These analyses, already discussed in the context of RPC gluino pair production followed by the decay into long electroweakino decay chains, or third-generation quarks, also offer sensitivity to, e.g., RPV neutralino decays through $\lambda$ or $\lambda'$ couplings, immediately yielding leptons in the final state, or RPV gluino decays through $\lambda''$ couplings, yielding top quarks and then leptons in the final state.   

Significant constraints to gluino and squark RPV decay scenarios, where it is the pair produced particle that features an RPV decay, come from resonant searches. Depending on which RPV decay is allowed, one can have di-jet\cite{ATLAS:2017jnp}, or lepton-jet\cite{ATLAS:2017jvy} resonances, providing powerful signatures with two two-object resonances at the same mass in an event. 

The consequences of removing the ``prompt'' hypothesis for the sparticle decays are even more striking, in terms of phenomenology and strategy search. SUSY particles can be long-lived because of: 

\begin{enumerate} 
\item  \label{it:LL1} decay happening via intermediate particles with large mass; 
\item   \label{it:LL2} small phase space available in the decay; 
\item \label{it:LL3} small couplings in the decay.
\end{enumerate}

Before diving into the details of some of these analyses, it is worth mentioning some of the experimental challenges associated with them:
\begin{itemize} 
\item  In many cases, the need of signal events to pass the first, hardware based, trigger level imposes specific analysis choices. Given that the first level of trigger is unable to reconstruct, e.g., a displaced vertex, or a highly ionising track, often auxiliary final state objects are required (missing transverse momentum, or jets, or leptons). A lot of exciting work is being done by the collaborations to mitigate some of the limitations introduced by the trigger needs. 
\item The specificity of the experimental signature often imposes non-standard reconstruction streams for long-lived analyses. Analyses involving tracks with extreme impact parameters, for example, may need to optimise the reconstruction step for such tracks, requiring a non-standard reconstruction flow. 
\item There is hardly any SM process producing an irreducible background for massive, long-lived particles: the main background for long-lived analyses often comes from fake/mis-measured objects (for example displaced vertices from intersections of random combinations of tracks, or beam backgrounds). It is often impossible to rely on the simulation predictions for the background estimation. All analyses use background estimation strategies exploiting dedicated data-driven methods. 
\end{itemize}

\begin{figure}[tb]
\begin{center}
\subfigure[]{
\includegraphics[width=0.43\textwidth]{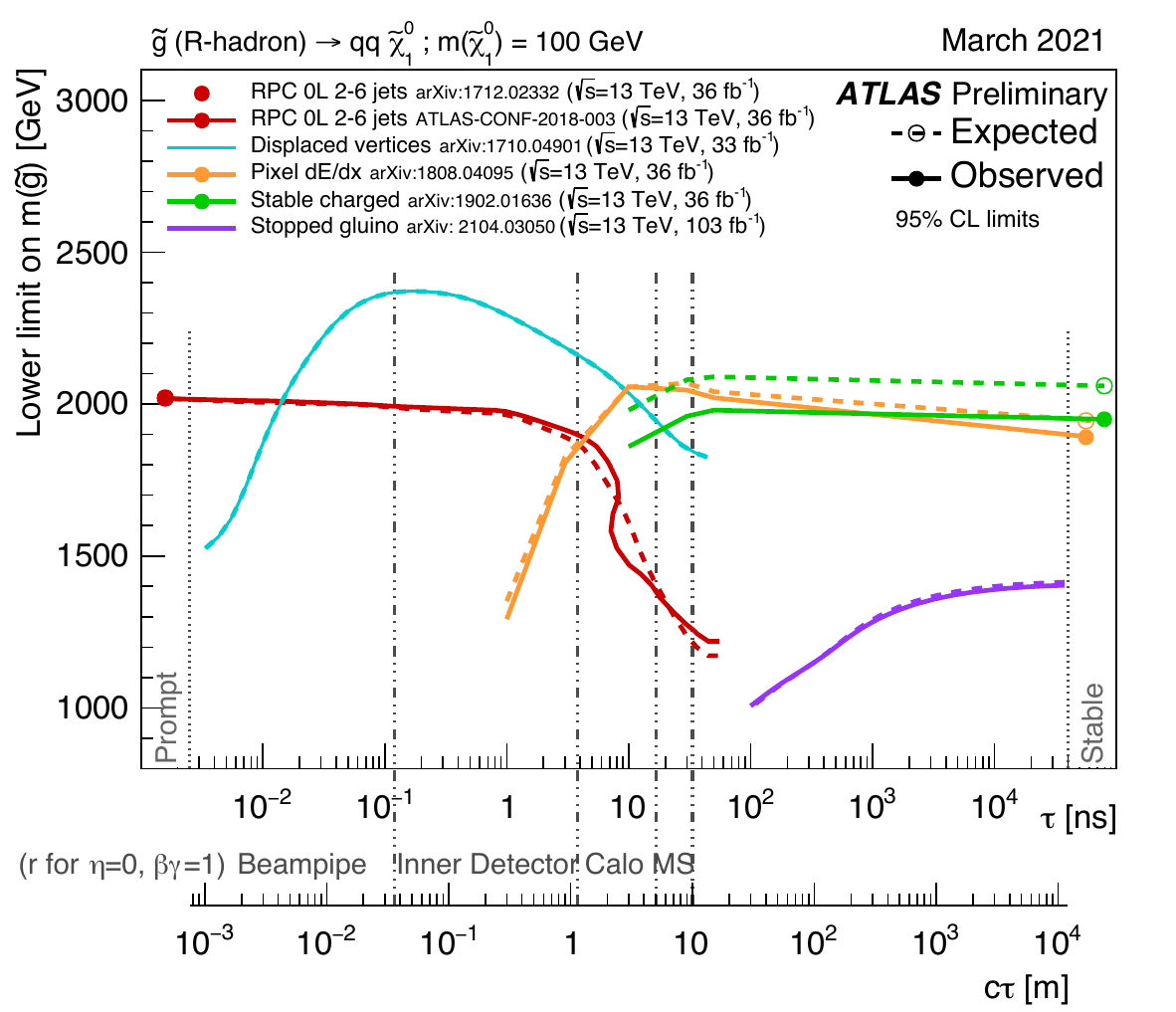}
}
\subfigure[]{
\includegraphics[width=0.4\textwidth]{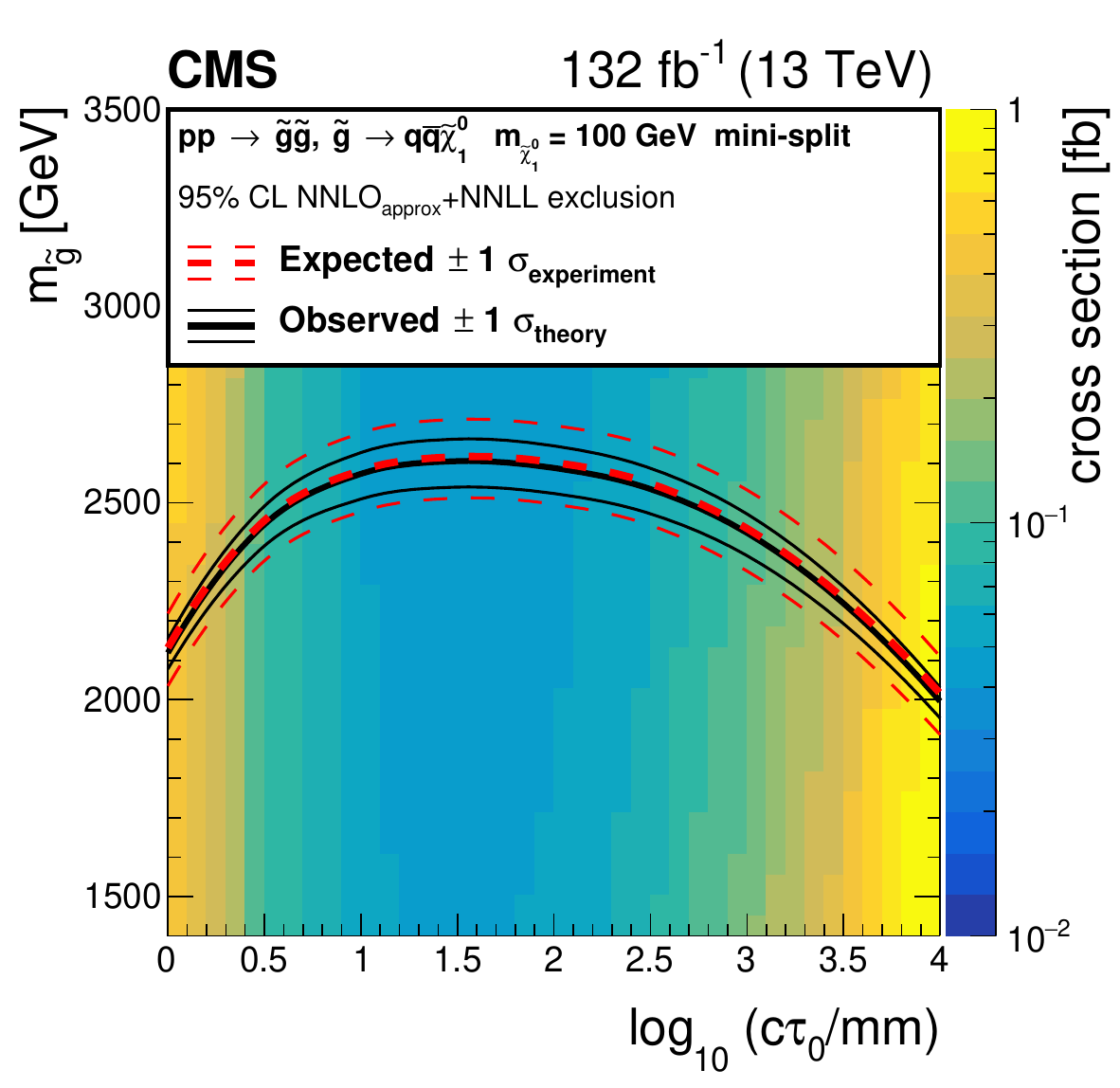}
}
\end{center}
\caption{Results of searches for long-lived particles: (a) ATLAS Long-lived gluino $R$-hadron summary plot\cite{ATL-PHYS-PUB-2021-019}. (b) Sensitivity of the CMS displaced jets analysis\cite{CMS:2020iwv} to long-lived $\gluino\rightarrow  q\bar{q}\ninoone$ decays.}
\label{fig:summaryPlot_LL_gg}
\end{figure}

An example of item (\ref{it:LL1}), that is, scenarios where the large decay mediator mass is responsible for the particles being long-lived  is split SUSY\cite{Giudice:2004tc}, where, generically speaking, SUSY scalar particles are predicted to have large masses, while SUSY fermions can have masses accessible by the LHC. In this case, the gluino decay $\gluino\rightarrow q\bar{q}\ninoone$, which happens through an intermediate $\gluino\rightarrow q\squark^{*}$, may actually become long-lived, producing stable hadrons ($R$-hadrons\cite{GATESJR2000216}), which can travel through the detector and decay at different radii, depending on their lifetime. A summary of these searches is shown in Fig.~\ref{fig:summaryPlot_LL_gg} (a): at low lifetimes of the $R$-hadron, standard RPC searches for gluino pair production are still sensitive, but their sensitivity decreases with the lifetime, and for $c\tau \approx 1$ cm, dedicated searches explicitly looking for a massive decay vertex in the inner detector region\cite{PhysRevD.97.052012} already dominate the sensitivity. The corresponding analysis selects events based on the presence of relatively large \etmiss due to the presence of the $\ninoone$ from the $R$-hadron decay. The analysis strategy relies on the identification of displaced vertices built from tracks reconstructed with relaxed requirement on their transverse and longitudinal impact parameters. The yield of events containing displaced vertices with at least five associated tracks and a mass $m_{\mathrm{DV}} > 10$ GeV, and large missing transverse momentum is the observable used to determine the presence of new phenomena. The background arises from intersection of random track crossings, and it is extrapolated from events containing displaced vertices with lower mass and/or lower track multiplicity.   

If the decay happens beyond the inner detector volume, searches looking for highly ionising tracks in the inner detector, with a possible continuation in the calorimeter\cite{PhysRevD.93.112015,2016647,CMS:2016kce} provide a very good sensitivity. The trigger selection requires again the presence of large missing transverse momentum, and the event selection relies on the measurement of charge collected in single hits of the silicon detector, and, depending on the analysis, on the timing measurement from the calorimeter and muon spectrometer. The background (arising from large fluctuations in the measurements of these quantities) is estimated relying on the independence of all these measurements for SM particles.  

Another technique used for long lifetimes is to look for a signal from the decay of a $R$-hadron stopped in the calorimeter\cite{ATLAS:2021mdj,CMS:2017kku}. 
Dedicated triggers are used to select events associated with empty bunches that feature significant calorimeter energy deposits. 
The analysis is highly non-standard and requires events without any reconstructed primary proton-proton vertex. 
The main backgrounds arise from cosmic rays and beam induced backgrounds.
They are suppressed using selections on the time of the signal and requirements on the presence of jets and missing transverse momentum, and the remaining contributions are estimated thanks to detailed data driven procedures.

A CMS analysis\cite{CMS:2020iwv} sets limits on a variety of split SUSY and RPV gluino and stop decay models, by looking for displaced jets. The analysis starts by selecting events by making use of either the \HT trigger, or of a dedicated displaced trigger, requiring lower values of \HT, but in addition to the presence of jets with associated tracks with a significant impact parameter. The offline selection aims at identifying di-jet events with a secondary vertex associated to them. The event selection exploits variables such as the fraction of jet-associated tracks arising from the secondary vertex, and the total track energy from the di-jet system associated to the primary vertex. After a preselection based on these quantities and on the SV quality and significance, a multivariate discriminant is built based on the vertex- and track- associated variables. The analysis sensitivity to split SUSY gluino decays is shown in Fig.~\ref{fig:summaryPlot_LL_gg} (b).    

Overall, the sensitivity to long-lived $R$-hadrons is at least on a par with that to prompt gluino decays. 

An example of item (\ref{it:LL2}) above (small phase space available in the decay) is the case where the LSP is almost a pure wino. The difference in mass between the LSP neutralino and the NLSP chargino is of the order of few hundreds MeV, reducing the decay width for, e.g., $\chinoonepm\rightarrow \pi^{\pm} \ninoone$, and making the $\chinoonepm$ long-lived with lifetimes $c\tau$ of $O(\mathrm{cm})$~\cite{Giudice:1998xp,Randall:1998uk}. The signature is therefore that of a disappearing track (there is no attempt to reconstruct the soft pion produced in the decay) and it is discussed in detail in Sec.~\ref{sec:eweak-ll} in the context of electroweak production, but it is also used explicitly\cite{CMS:2019ybf} to set limits for gluino pair production in presence of a wino-like LSP multiplet. 

Finally, an example of item (\ref{it:LL3}), where the long-lived nature of the particles is caused by small couplings for the decay process , is the case where SUSY particles can decay only through (small) RPV couplings. Typical signatures of these long-lived particle RPV decays are displaced production of jets and leptons, depending on which RPV coupling is active.   A CMS analysis\cite{CMS:2021tkn} looking for the presence of displaced vertices in events selected using a $\HT$ trigger and containing multiple jets is sensitive to the presence of long-lived SUSY particles (gluinos and stops) decaying into quarks via $\lambda''$ RPV couplings with lifetimes $0.1\ \mathrm{mm} < c\tau < 100\ \mathrm{mm}$. Similarly to the ATLAS analysis looking for displaced vertices in presence of \etmiss, the background estimation relies on the extrapolation of the background from events containing displaced vertices with low track multiplicities or low vertex mass. Limits are extracted  on the gluino and stop masses as a function of their lifetime, assuming $\lambda''$ RPV decays: gluino masses up to 2.5 TeV and stop masses up to 1.6 TeV for proper lifetimes in the range 0.6 to 90 mm (70 mm), respectively.

If the RPV coupling which is causing the pair-produced particle to be long-lived is instead $\lambda'$, then leptons can be produced at the displaced vertex associated with the sparticle decay. An ATLAS analysis\cite{Aad:2020srt} looks for events containing a muon with large transverse impact parameter with respect to the primary vertex and a displaced vertex. The muon selection is optimised to reject background from cosmic muons and heavy flavour production by requiring low activity in regions of the muon spectrometer opposite to the identified muon and by requiring the muon to be isolated. The background estimation strategy uses control regions defined by inverting the muon and the displaced vertex selection criteria to extrapolate the background to the signal region. The results are used to set limits in a scenario with the $\lambda'$ stop decay $\stopone\rightarrow \mu q$ obtaining limits on the stop mass above 1.3 TeV for lifetimes between 0.01 ns and 30 ns, and up to 1.7 TeV for a lifetime of 0.1 ns.

%% file: tex/eweak.tex
Searches for the production of supersymmetric particles in proton-proton collisions via electroweak interactions are challenging due to the low cross sections, as shown in Fig.~\ref{fig:xsec}.
The first part of the parameter space to become accessible in searches at the LHC corresponded to low sparticle masses, with the additional complication of relatively low-energy decay products.
The high integrated luminosity and, to a lesser extent, the higher collision energy of LHC Run~2 led to a spectacular increase of sensitivity to these processes and, in some cases, to the first limits set at the LHC.
The searches described below are targeting pair production of electroweakinos and sleptons, supersymmetric partners of electroweak gauge bosons and Higgs bosons, and leptons, respectively.
As for strong production, experimental analyses are signature-based and can be interpreted in the context of multiple models, even if the use of the multiplicity and kinematics of charged leptons in most of the studies tend to create a stronger link to specific benchmark models.

\subsubsection{Overview of searches for electroweakinos and sleptons}

\begin{figure}
  \begin{center}
    \subfigure[]{
      \includegraphics[height=.35\textwidth]{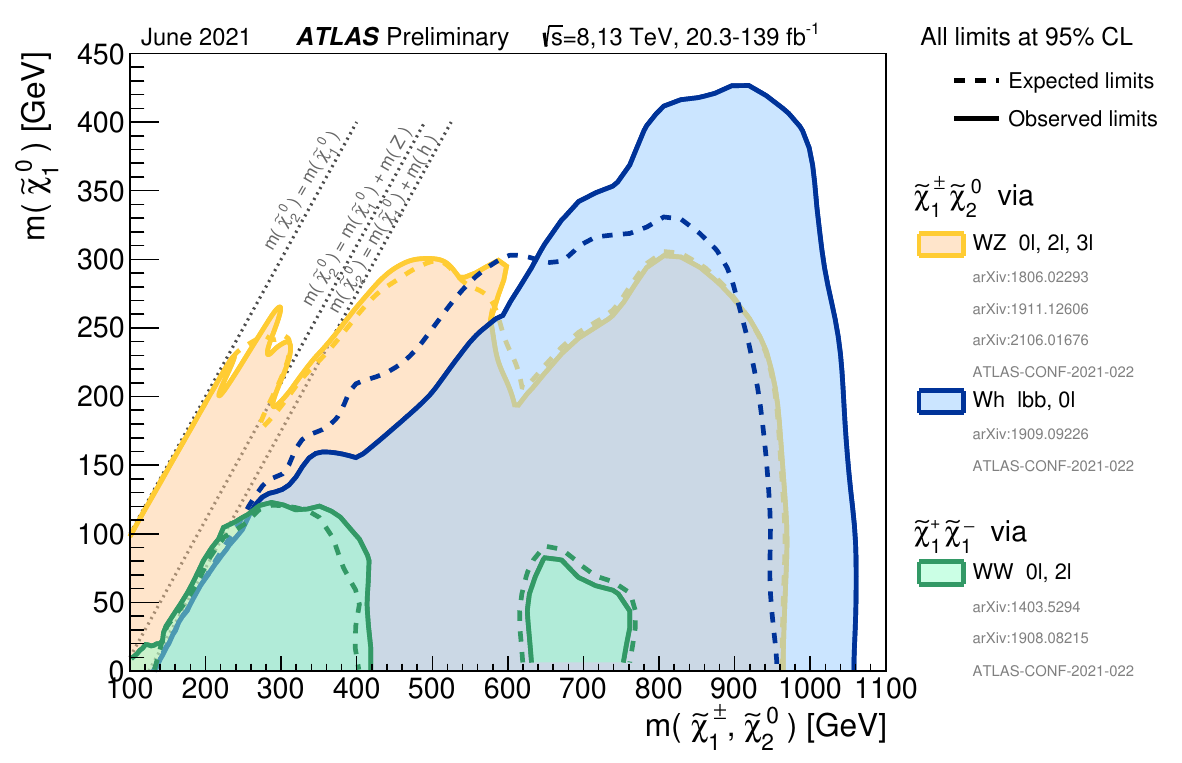}
      }\hfil
    \subfigure[]{
      \includegraphics[height=.35\textwidth]{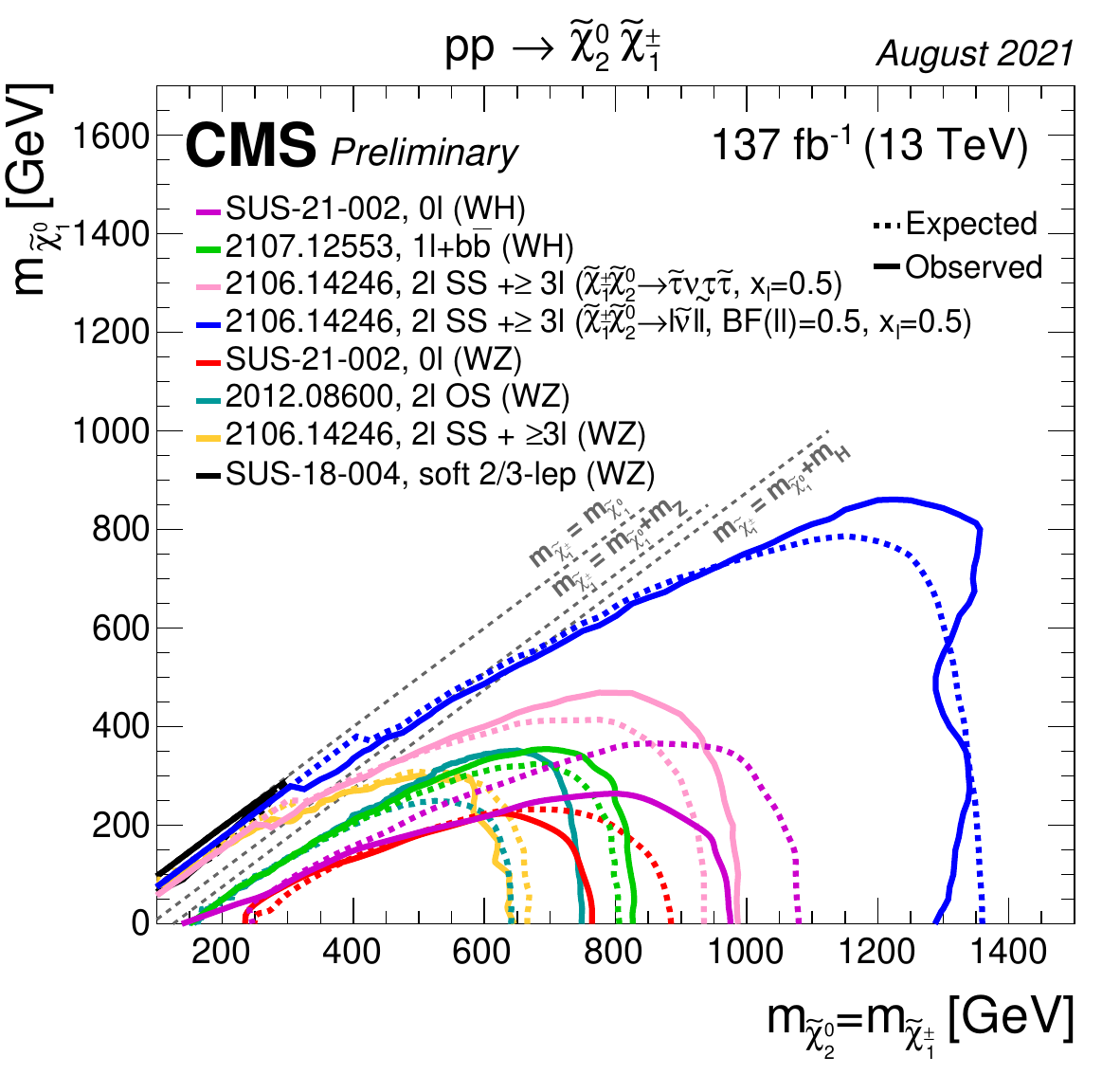}
      } \\
    \subfigure[]{
      \includegraphics[height=.35\textwidth]{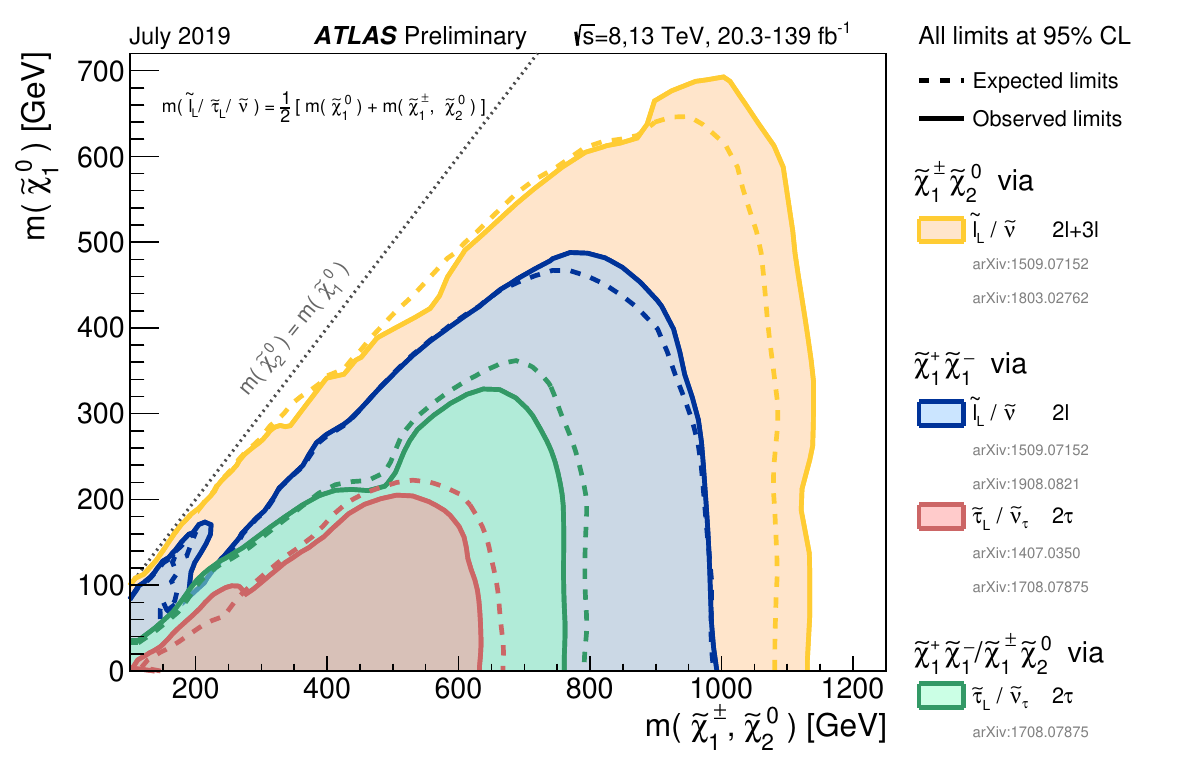}
    }
  \end{center}
  \caption{Limits on the production of pairs of charginos, and associated production of a chargino with the second-lightest neutralino, with ``wino'' cross sections, as a function the mass of the lightest neutralino, and the common mass of the lightest chargino and the second-lightest neutralino:
(a) unions of regions excluded by ATLAS analyses for decays via different SM bosons to the lightest neutralino~\cite{ATL-PHYS-PUB-2021-019};
(b) best limits on chargino-neutralino production obtained by CMS analyses, for both decay modes~\cite{CMS-SUSY-Summaries};
(c) unions of regions excluded by ATLAS analyses for decay chains via sleptons~\cite{ATL-PHYS-PUB-2021-019}.
Diagram (b) also includes limits on direct production of mass-degenerate partners of the left- and right-handed leptons of the first two generations. 
}
\label{fig:wino-summary}
\end{figure}

\begin{figure}
  \begin{center}
    \subfigure[]{
      \includegraphics[height=.35\textwidth]{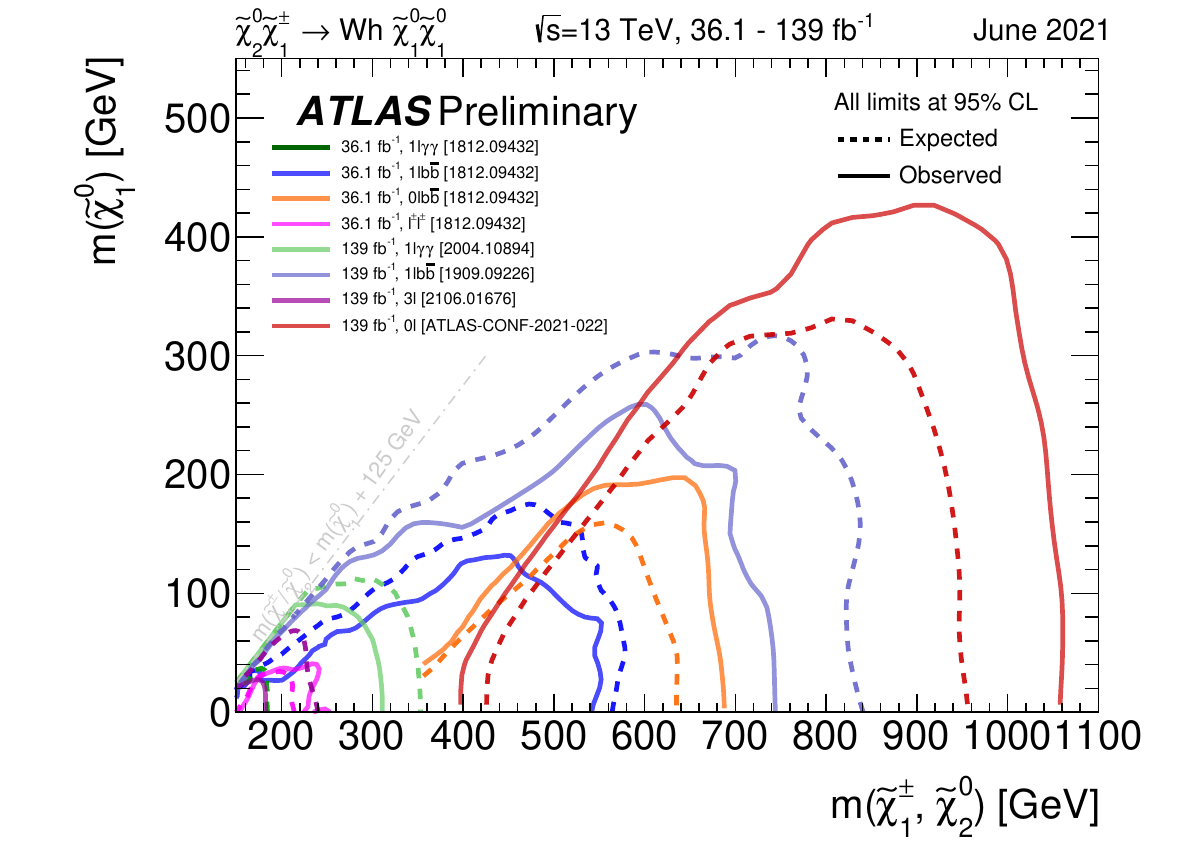}
      } \hfil
      \subfigure[]{
        \includegraphics[height=.35\textwidth]{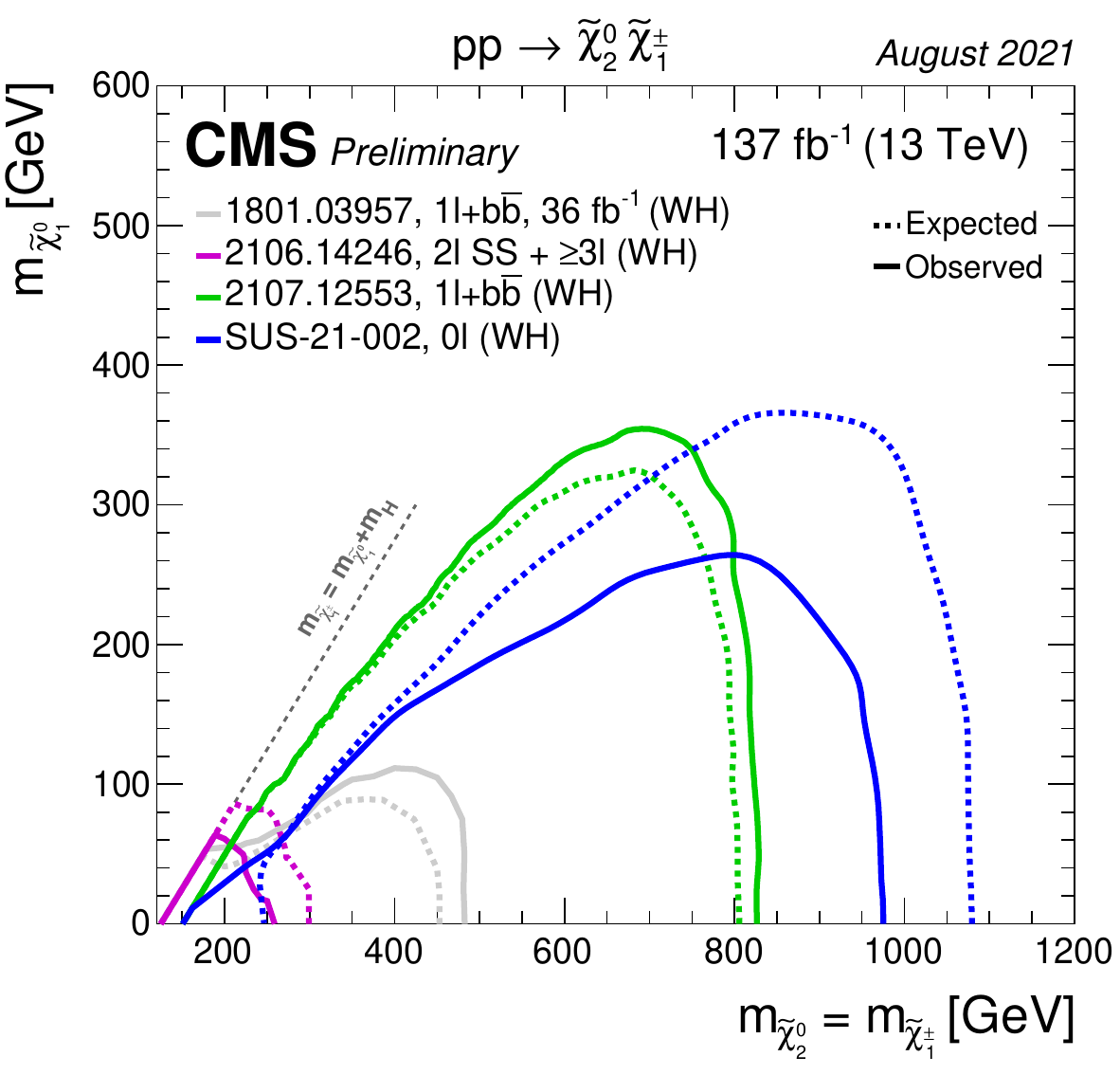}
        }
      \end{center}
      \caption{Limits on the associated production of a chargino with the second-lightest neutralino, with``wino'' cross sections and decays via W and h, respectively.
        ATLAS~\cite{ATL-PHYS-PUB-2021-019} and CMS~\cite{CMS-SUSY-Summaries} results are shown in diagrams (a) and (b), respectively.}	
      \label{fig:wh-summary}
\end{figure}

A large number of searches for electroweakinos are targeting models with a \ninoone LSP in one of the three scenarios shown in Fig.~\ref{fig:ewikino_spectrum}.
Many of these studies focus on the configuration with the highest cross section, the ``bino LSP'' scenario described in Sec.~\ref{sec:models-eweak}.
Under this assumption, the nearly mass-degenerate \chinoonepm and \ninotwo act as the NLSP.
Pairs of $\chinoonep\chinoonem$ and $\chinoonepm\ninotwo$ would be produced simultaneously with wino-like cross sections and decay via vector or Higgs bosons to the \ninoone LSP.
The signatures depend on the decay modes of the vector bosons, but all signal events feature \etmiss.
A summary of current limits in the plane of the LSP versus the NLSP mass is shown in Fig.~\ref{fig:wino-summary} for different decay modes.
Results for decays via an intermediate slepton are also shown.
The contributions from different event signatures for the specific case of $\chinoonepm\ninotwo \rightarrow W h \ninoone\ninoone$ are shown in Fig.~\ref{fig:wh-summary}.

The strongest constraints on $\chinoonepm\ninotwo$ production with direct decays to the LSP are currently set by an ATLAS study of the $W(\rightarrow q \bar{q}')h(\rightarrow \mathrm{b\bar{b}}$) decay channel and reach up to $1060\unit{GeV}$  in $m(\chinoonepm,\ninotwo)$~\cite{ATLAS:2021yqv}.
The corresponding numbers for the $W(\rightarrow q \bar{q}')Z(\rightarrow \mathrm{q\bar{q}})$ channel are $960\unit{GeV}$ and $300\unit{GeV}$.
The second-most sensitive signature for decays via $Wh$ is $1\ell + 2b + \etmiss$, with maximum limits in $m(\chinoonepm,\ninotwo)$ of about $820\unit{GeV}$ and $740\unit{GeV}$   from studies in CMS~\cite{CMS:2021few} and ATLAS~\cite{Aad:2019vvf}, respectively.

Because of the lower production cross section, limits on $\chinoonepm\chinoonemp$ pair production are weaker.
For light LSPs, chargino masses below $420\unit{GeV}$ and in the range $630$--$760\unit{GeV}$ are excluded by two ATLAS analyses~\cite{Aad:2019vnb,ATLAS:2021yqv}.
Sensitivity is increased in models with decay chains via charged or neutral sleptons as their decay products increase the fraction of signal events with charged leptons, and / or high \etmiss due to neutrinos in the final state.
The branching ratios into different sleptons depend on the assumptions on the slepton masses and on the nature of the next-to-lightest electroweakino states.
The highest reported mass limit on $\chinoonepm\ninotwo$ production under the assumption of equal branching ratios to left-handed sleptons is $1450\unit{GeV}$, set by a CMS analysis~\cite{CMS:2021cox}.
For $\chinoonepm\chinoonemp$ pair production, the limits set by an ATLAS study~\cite{Aad:2019vnb} reach up to $1\unit{TeV}$.

\begin{figure}
\begin{center}
  \subfigure[]{
    \includegraphics[height=.32\textwidth]{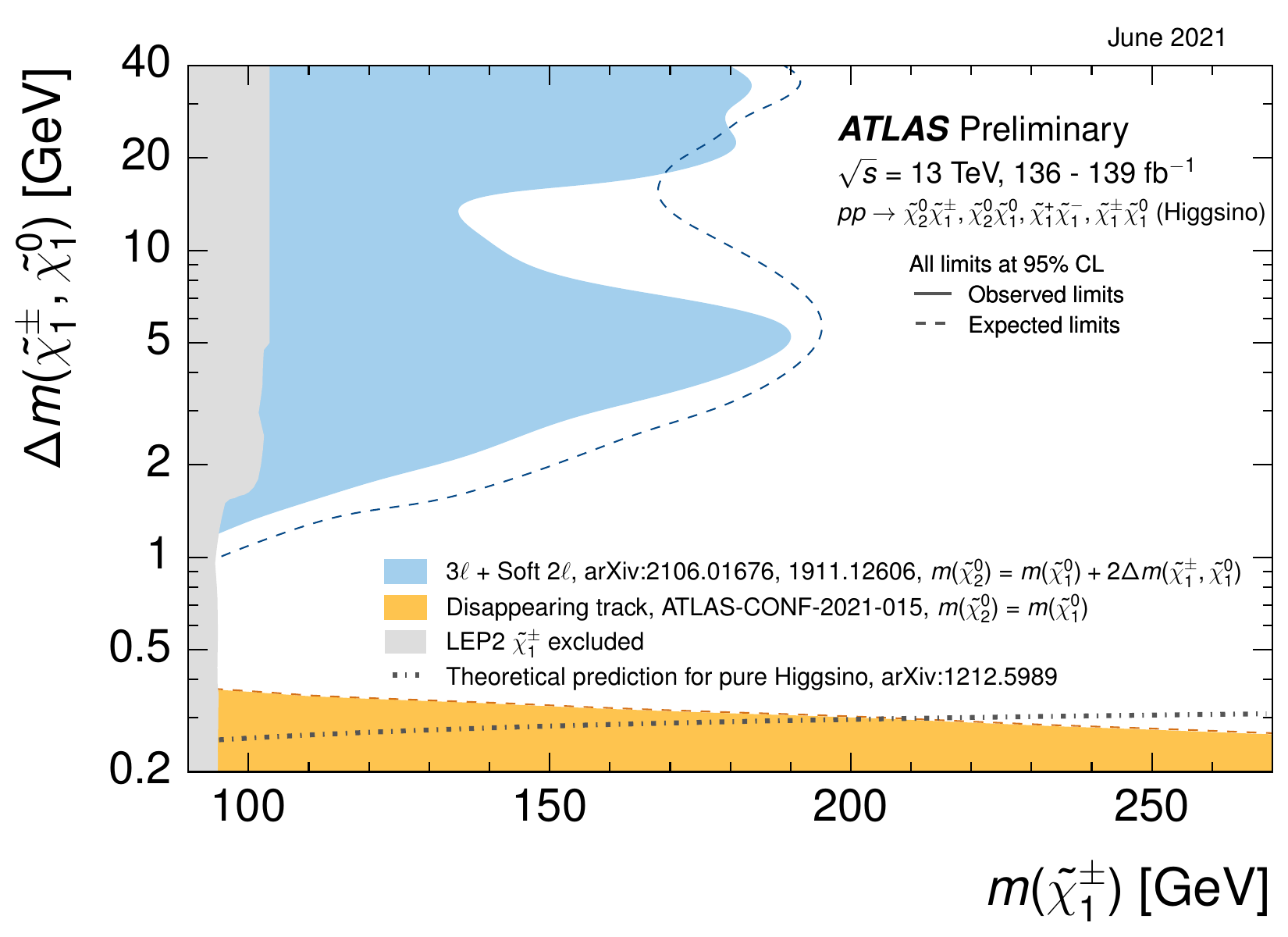}
  } \hfill
  \subfigure[]{
    \includegraphics[height=.32\textwidth]{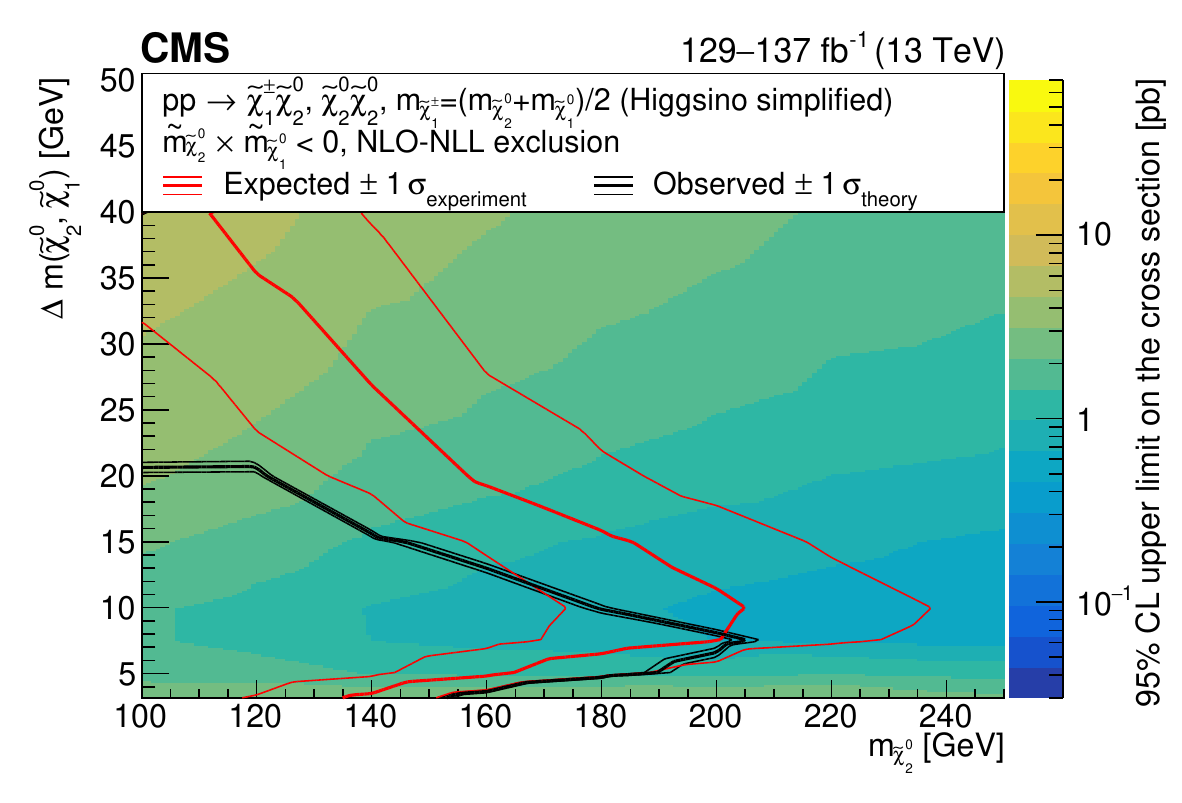}
    }
  \end{center}
  \caption{Limits on the production of pairs of charged and/or neutral electroweakinos in the ``higgsino LSP'' scenario from searches in events with two soft leptons as a function of the chargino mass and the mass difference.
  The mass difference between \ninotwo and \ninoone is set to twice the difference between \chinoonepm and \ninoone, and decays proceed via off-shell SM bosons.
  Diagram (a) shows ATLAS limits~\cite{ATL-PHYS-PUB-2021-019} as a function of the \chinoonepm - \ninoone mass difference and includes also results of a search for disappearing tracks~\cite{ATLAS:2021ttq}.
  Diagram (b) shows CMS limits~\cite{CMS:2021edw} as function of the \ninotwo - \ninoone mass difference.
}
\label{fig:higgsino-summary}
\end{figure}

Constraints on production of electroweakinos in the ``higgsino LSP'' scenario are considerably weaker:
production cross sections are smaller, and the models predict a small mass splitting between lightest electroweakinos, with transverse momenta of visible decay products at the limit of the acceptance.
Nevertheless, limits up to $\approx 200\unit{GeV}$ have been obtained using several production modes and under specific assumptions for the mass hierarchy of the lightest electroweakinos and a mass difference between \chinoonepm and \ninoone of $\approx 5\unit{GeV}$, as shown in Fig.~\ref{fig:higgsino-summary}.
Cross sections for the "wino LSP" and "higgsino LSP" scenarios, calculated under the assumption of a mass degenerate pair or triplet of states, respectively, are shown in Fig.~\ref{fig:xsec}.

In scenarios motived by gauge-mediated SUSY breaking, where decay chains end with a nearly massless gravitino, the acceptance can be substantially increased as the scale of the momenta of the decay products is now determined by the mass of the NLSP.
E.g., in the case of higgsino production, masses of an almost degenerate set of \chinoonepm, \ninotwo, and \ninoone are excluded up to $800\unit{GeV}$ and $380\unit{GeV}$ in case of a 100\% branching ratio for the decay $\ninoone \rightarrow Z \gravino$ by a CMS analysis~\cite{Sirunyan:2020eab} and $\ninoone \rightarrow h \gravino$ by an ATLAS analysis~\cite{Aad:2020qnn}, respectively.

In several SUSY scenarios, electroweakinos can acquire long lifetimes, either because of near mass degeneracy with the LSP, with mass differences below the \unit{GeV} scale, or small couplings to the LSP.
Examples are models motivated by anomaly-mediated or gauge-mediated Supersymmetry.
In the first case, as already discussed in Sec.~\ref{sec:strongRPV}, long-lived charginos give rise to a "disappearing" track signature as neither the LSP, nor the low-energy SM decay products are reconstructed in the tracking system.
In terms of lifetime, the sensitivity is limited at low values by the required minimum track length, and at high values by the smaller fraction of decays within the volume of the tracking system.
E.g., using reference cross sections with either a wino- or higgsino-like LSP, and varying $\tau(\chinoonepm)$, chargino masses up to $884\unit{GeV}$ ($750\unit{GeV}$) have been excluded for combined $\chinoonepm\chinoonepm$ and $\chinoonepm\ninoone$ ($\chinoonepm\chinoonepm$ and $\chinoonepm\ninoonetwo$) production under the assumption of 100\% branching ratio for the decay $\chinoonepm \rightarrow \pi^{\pm} \ninoone$ and for a wino (higgsino) LSP by a CMS analysis~\cite{CMS:2020atg}.
Similar values have been obtained by ATLAS~\cite{ATLAS:2021ttq}.
Using the lifetimes predicted by the models described in more detail in the references quoted above, of the order of a few 100\textsuperscript{th} (10\textsuperscript{th}) of a picosecond for the wino (higgsino) scenario, limits are $660\unit{GeV}$ ($210\unit{GeV}$)~\cite{ATLAS:2021ttq}.

At the same mass, production cross sections for pairs of sleptons~\cite{Beenakker:1999xh,Bozzi:2007qr,Fuks:2013vua,Fuks:2013lya,Fiaschi:2018xdm} are more than an order of magnitude smaller than the cross sections for $\chinoonepm\chinoonepm$ or $\chinoonepm\ninotwo$ production in the wino scenario as shown in Fig.~\ref{fig:xsec}..
For a slepton NLSP, and under the assumption of \Rparity conservation, the event signature is the same as the one expected for $\chinoonepm\chinoonepm$ production, followed by leptonic $W$ decays, under the same conditions:
a pair of oppositely charged leptons, and \etmiss.
Interpretations are done under different assumptions about the mass degeneracy of partners of left- and right-handed leptons, and of different flavours.

\begin{figure}
  \begin{center}
    \subfigure[]{
      \includegraphics[height=.35\textwidth]{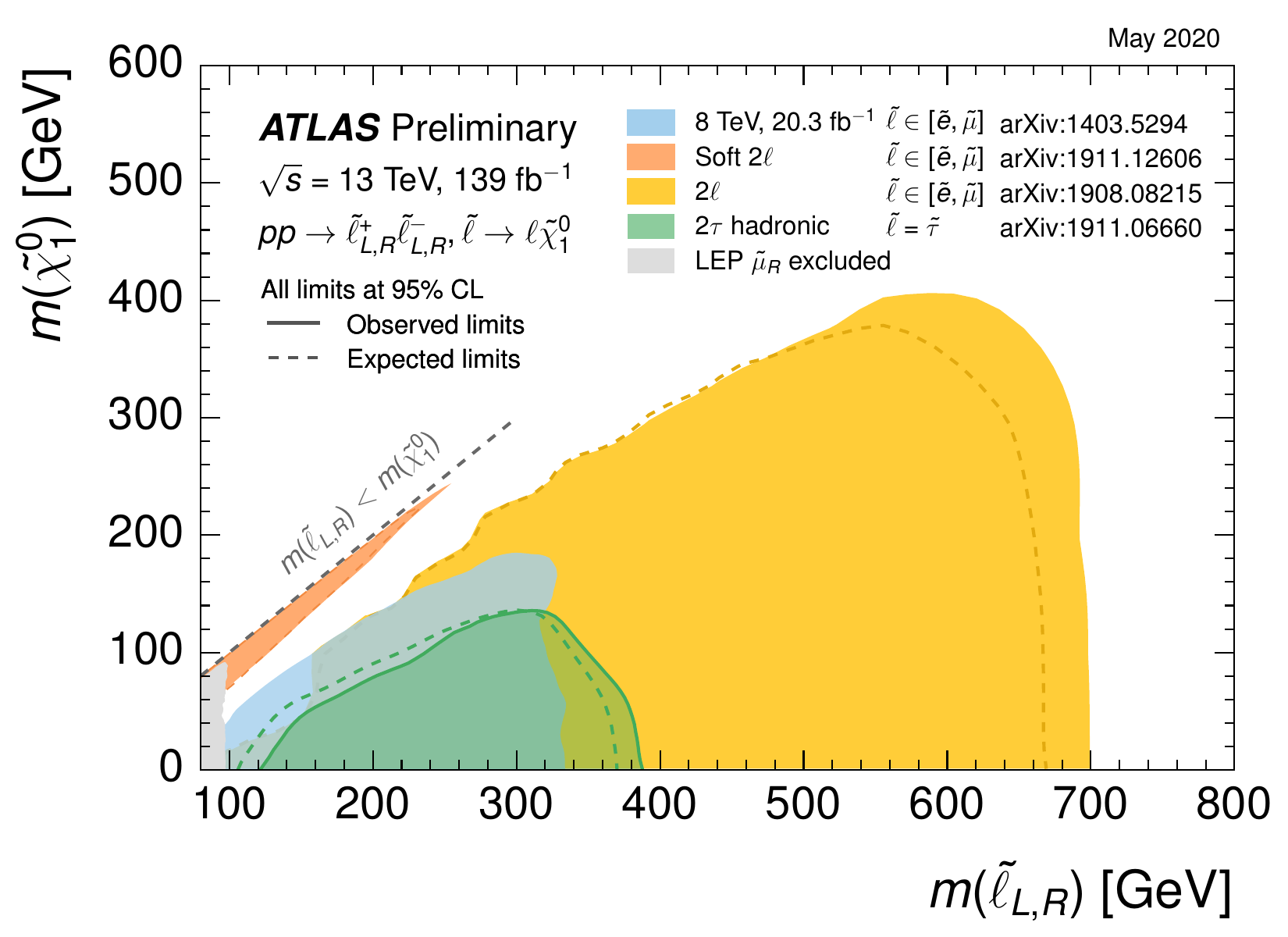}
    } \hfil
    \subfigure[]{
      \includegraphics[height=.35\textwidth]{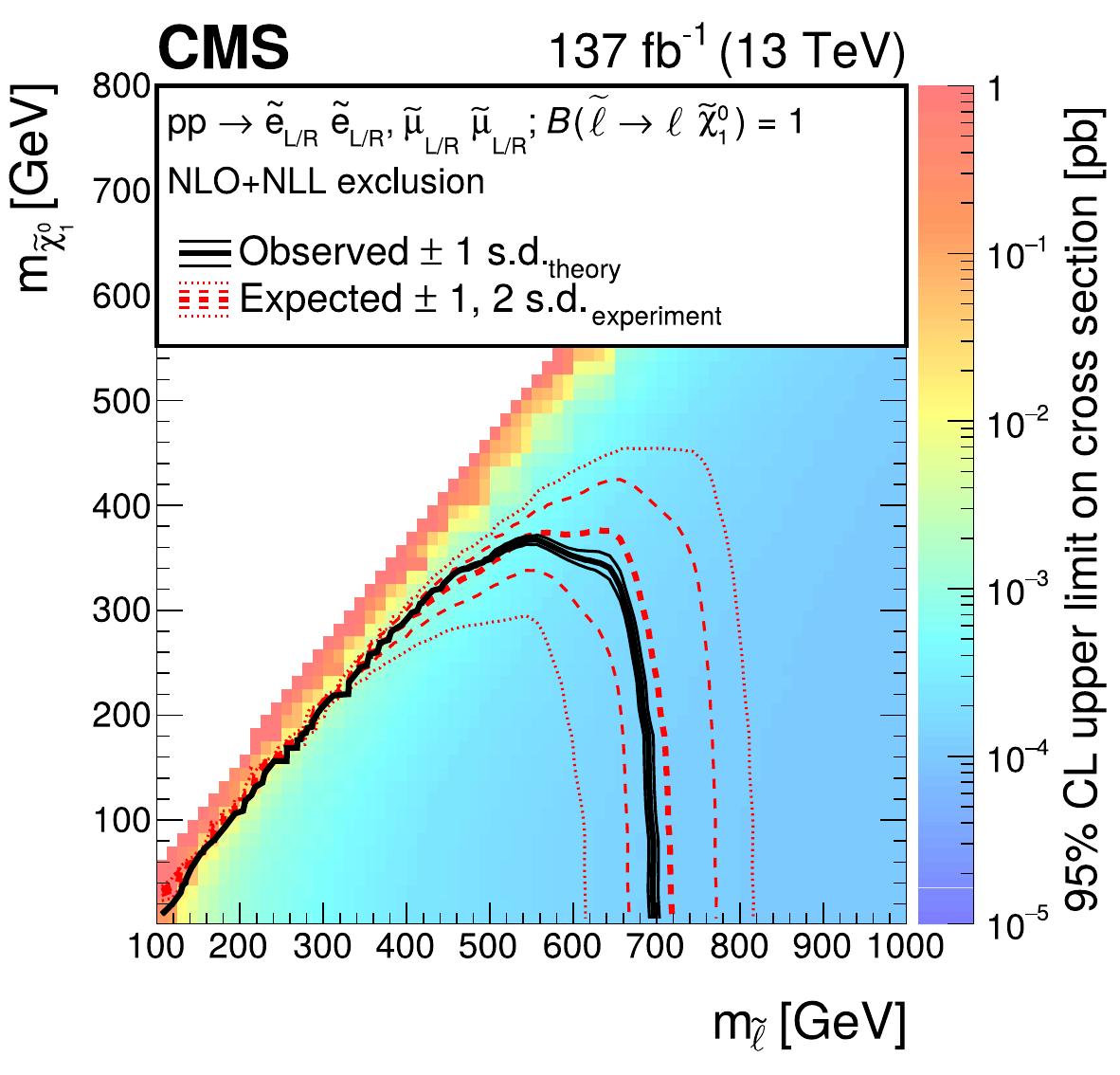}
    }
  \end{center}
  \caption{Limits on the production of charged slepton pairs, with decays proceeding via the corresponding SM lepton:
    (a) ATLAS limits~\cite{ATL-PHYS-PUB-2021-019} for production of degenerate \sleptonL and \sleptonR states of the first two generations, or of the third generation.
    (b) CMS limits~\cite{Sirunyan:2020eab} for production of degenerate \sleptonL and \sleptonR states of the first two generations.
  }
  \label{fig:sleptons}
\end{figure}

In both experiments, pairs of electrons or muons can be reconstructed and identified at low transverse momenta, both at the trigger and the analysis level.
For mass differences to the LSP of $\dm = m(\slepton)-m(\ninoone) \approx 60\unit{GeV}$ or above, mass limits obtained by ATLAS~\cite{Aad:2019vnb} and CMS~\cite{Sirunyan:2020eab} reach up to $700\unit{\unit{GeV}}$ under the assumption of four degenerate states of the first two generations (Fig.~\ref{fig:sleptons}).
Smaller mass differences need to be addressed with different analysis techniques using different online selections and focusing on production of soft leptons.
For the same mass hierarchy, mass limits of up to $251\unit{GeV}$ have been reached by an ATLAS analysis~\cite{Aad:2019qnd}.

The thresholds for hadronically decaying $\tau$ leptons are considerably higher due to an overwhelming SM background of jets, in particular in the online selection, resulting in considerably weaker limits for \stau production.
For a massless LSP, \stau masses between $120$ and $390\unit{GeV}$ have been excluded by an ATLAS analysis~\cite{Aad:2019byo}.

Several SUSY models predict long-lived sleptons, e.g., in gauge-mediated models with a weak coupling between a slepton NLSP and a gravitino LSP.
A search for the displaced lepton signature predicted by such a model was performed by the ATLAS collaboration~\cite{Aad:2020bay} and sets mass limits varying between $340\unit{GeV}$ and $820\unit{GeV}$ for a lifetime of $0.1\unit{ns}$, where the lower value corresponds to a \stau LSP, where $\stau_\mathrm{L}$ and $\stau_\mathrm{R}$ are degenerate in mass, and the higher value to a scenario where all six slepton states have the same mass.

In the following sections, some analyses will be discussed in more detail in order to illustrate state-of-the-art analysis techniques.
For this purpose, these sections are organised as function of the event signature.

\subsubsection{Searches in events with at least three charged leptons, or a same-charge lepton pair}\label{sec:eweak-multi-l}

Searches for Supersymmetry in events with three ore more charged leptons, or a pair of same-charge leptons, have been performed at previous accelerators and since the start of LHC operations. 
In this topology, signal events in RPC models can arise from decays of pairs of electroweakinos, either to $W$ or $Z$ bosons (directly, or via an SM-like Higgs boson), or from extended decay chains involving intermediate sleptons.
In other models, these events can be produced in the decay of electroweakinos or sleptons via \Rparity- and lepton-number-violating couplings.
Backgrounds are small and mainly composed of rare SM processes such as production of pairs of vector bosons or $t\bar{t}$ pairs in association with a vector boson, or events with leptons that are either misidentified or resulting from hadron decay.
The low background levels allow the use of relaxed criteria in the selection of leptons and for \etmiss, and hence higher signal efficiency, both at the trigger and at the analysis level.

A search for RPC SUSY in events with three leptons (electrons or muons) has been reported by the ATLAS collaboration~\cite{ATLAS:2021moa}.
The analysis targets pair production of electroweakinos with decays to an LSP pair and either $Wh$ or $WZ$, in the latter case with specific selections for on- and off-shell bosons.
The online selection relies on dilepton triggers that are complemented with single-, di- and trilepton triggers, and an \etmiss trigger, for the off-shell case.
The basic selection requires exactly three leptons (electrons or muons), with \pt thresholds as low as 4.5\unit{GeV} for electrons and 3\unit{GeV} for muons.
For the off-shell $WZ$ selection, the contribution from events where the lowest-\pt lepton is misidentified or non-prompt is further suppressed using a multivariate discriminator that combines eight variables related to lepton and track quantities, and the $b$-jet likeliness for tracks in a cone around the lepton.
Events are categorised according to whether or not an OSSF (opposite charge sign, same flavour) lepton pair is present.
In the latter case, a further binning is applied as function of the dilepton mass, \mll.
In the final selection, optimised \pt thresholds are applied to the three leptons, as well as a minimal \etmiss requirement of 50\unit{GeV}.
Additional selections are applied to suppress events from SM top-quark or $WZ$ production, low-mass resonances, and $Z \rightarrow \ell^{\pm}\ell^{\mp}\gamma^{(*)}$ decays.
Further discriminating variables are \mt, \njet, and the scalar sums of jet (\HT) and lepton ($\HT^\mathrm{\ell}$) transverse momenta.

The dominant background from SM $WZ$ production is estimated using MC simulation, with normalization in dedicated control regions similar to the signal regions.
Background from Drell-Yan production of dileptons with an additional \FNP lepton is estimated from data with a fake-ratio method as described in Sec.~\ref{sec:exp_techniques}.
Other backgrounds are predicted using MC simulation, and the estimations for all leading background sources are verified in validation regions.

No significant excess over the predictions is found. 
For degenerate, wino-like \chinoonepm and \ninotwo states with decays via $W$ and $Z$, the on- and off-shell $WZ$ regions are combined with results of a soft dilepton search~\cite{Aad:2019qnd} that decisively improves sensitivity to compressed spectra with mass gaps of $\dm \lessapprox 20\unit{GeV}$ (see also Sec.~\ref{sec:eweak-two-l-os}).
For equal signs of the mass parameters for the two lighest \nino states~\cite{Fuks:2017rio,Gunion:1984yn}, lower mass limits reach up to 640\unit{GeV} for a massless LSP.
For opposite signs, the highest mass limits are reached at $\dm \approx \mass{Z}$, with a value of 310\unit{GeV}. 
For the production of degenerate, higgsino-like \chinoonepm and \ninotwo particles decaying to the same final state, limits are set for values of \dm up to 60\unit{GeV}, reaching 210\unit{GeV} at the highest mass gaps.
For decays of wino-like  \chinoonepm and \ninotwo states to $Wh$, and under the hypothesis of equal signs of the mass parameters, mass limits reach up to 190\unit{GeV} for a massless LSP, weaker than the expectation of 240\unit{GeV} in absence of signal due to a slight excess in the signal region without OSSF pair, however, compatible with the background hypothesis within two standard deviations.

A search in events with three or four charged leptons, including up to two hadronically decaying $\tau$ leptons, or an SSSF (same charge sign, same flavour) pair of electrons or muons, has been performed by the CMS collaboration~\cite{CMS:2021cox}.
The study targets production and decay of electroweakinos in different scenarios:
\begin{enumerate}
\item \label{itm:SUS_19_012_1} production of a wino-like $\chinoonepm\ninotwo$ pair, degenerate in mass and decaying to a bino-like \ninoone LSP via $\chinoonepm \rightarrow W^{\pm} \ninoone$ and $\ninoone \rightarrow (Z \ninoone)$ or  $(h \ninoone)$, as in the ATLAS trilepton analysis discussed above, or with decay chains via intermediate sleptons, $\chinoonepm \rightarrow (\nu \slepton^{\pm}) \mathrm{or} (\ell^{\pm}\snu)$ and $\ninotwo \rightarrow \ell^{\pm} \slepton^{\mp}$;
\item pair production of nearly mass degenerate, higgsino-like \chinoonepm, \ninotwo, and \ninoone states in a model inspired by GMSB, where the decays of \chinoonepm and \ninotwo each produce a \ninoone NLSP and soft, undetected particles, and where \ninoone decays to $Z$ or $h$ and a nearly massless gravitino, \gravino.
\end{enumerate}
For the slepton-mediated decays in item (\ref{itm:SUS_19_012_1}), the lepton \pt spectrum depends strongly on the size of the mass gaps between \chinoonepm,\ninotwo and \slepton/\snu, and between \slepton/\snu and \ninoone.
The analysis has been optimised to cover a large range of the relative mass splitting $x = (m_{\slepton}-m_{\ninoone})/(m_{\ninotwo}-m_{\ninoone})$.
The online event selection is based on a combination of single, di- and trilepton triggers, where lepton stands for an electron or muon.
The primary classification of events in one of 12 exclusive categories is based on the numbers of light and $\tau$ leptons, and the number of opposite-charge pairs of the same (OSSF) or different flavour. 

The target event signatures include at least three charged leptons.
However, some signal leptons may be outside the acceptance or not pass the selection criteria, in particular in case of small mass gaps.
The same-sign dilepton category is used to increase the selection efficiency in these cases while retaining the advantage of low SM background.
The three-lepton category is particularly important for decays via sleptons to a $WZ$ pair, or for non-resonant lepton production via a Higgs boson decay.
In order to reduce the sizeable SM background in the category defined by three light leptons, including an OSSF pair, a novel approach using parametric, fully connected feed-forward neural networks~\cite{Baldi:2016fzo} has been deployed in order to achieve an optimal selection for different regions of parameter space.
As the kinematics of the signal events is largely determined by the mass difference between NLSPs and LSP, the networks are parametrised in this variable.

Depending on the category, the leading backgrounds are SM events matching the lepton requirements, additional leptons from photon conversions, \FNP leptons passing the selection, and, for the same-sign dilepton category, charge mismeasurement.
Several control regions are used to normalise predictions from MC simulation for leading backgrounds such as SM $WZ$ or $ZZ$ production, events with internal or external photon conversions, or to determine systematic uncertainties in the description of the tails of distributions related to \etmiss.
Mismeasurement of the sign of the electron charge is a source of background in di- and trilepton categories with same-sign lepton pairs.
The probability is determined from simulation, normalised to data in a Drell-Yan-enriched control region, and applied to the dominant opposite-sign events.
Contributions from events with \FNP leptons are estimated with a fake-ratio method.

No significant excess over the predicted background is observed, and exclusion limits are derived from  global fits to all categories relevant for a specific model.
Under the hypothesis of decays of wino-like $\chinoonepm\ninotwo$ pairs to $WZ$ ($Wh$) final states, the limits reach 650\unit{GeV} (190\unit{GeV}) for a massless LSP.
For slepton-mediated decays, three scenarios are distinguished: equal branching ratios to all three flavours, exclusive decays of \chinoonepm via $\tau$ leptons, or exclusive decays of both \chinoonepm and \ninotwo via $\tau$ leptons.
The two latter scenarios are motivated by assumptions on the couplings to partners of right-handed leptons, and the mass hierarchy of sleptons.
For each scenario, three values of the relative mass splitting $x$ are tested (0.05, 0.50, and 0.95).
For $x=0.50$, mass limits for \chinoonepm / \ninotwo reach up to 1450\unit{GeV}, 1150\unit{GeV}, and 970\unit{GeV} for the three decay scenarios mentioned above.
Finally, for higgsino production with \ninoone decays to $Z\gravino$ or $h\gravino$, mass limits vary from 200\unit{GeV} to 600\unit{GeV}, depending on the branching ratio.

A search using events with high lepton multiplicity has also been performed by the ATLAS collaboration~\cite{ATLAS:2021yyr}.
Here, only events with at least four charged leptons were considered.
Four categories were defined based on the number of identified light or $\tau$ leptons.
One difference with respect to the CMS analysis described above is the use of a region with at least five light leptons.
In addition to the \Rparity-conserving, GMSB inspired model with production of higgsino-like electroweakinos with decays $ \ninoone \rightarrow Z/h \gravino$ described above, several \Rparity-violating scenarios for the pair production of electroweakinos or sleptons are investigated, with decays to a \ninoone LSP, followed by its decay via lepton-number-violating RPV couplings to a pair of charged leptons and a neutrino.
The analysis uses an experimental approach similar to what was described for the CMS analysis, selecting events without $b$ jets, and exploiting \etmiss, the presence and mass of $Z$ candidates built from $e$ or $\mu$ pairs, or a veto on such pairs, and the effective mass built from the scalar sum of \etmiss and the lepton and jet transverse momenta.
No further requirements are imposed for the five-lepton signal region.
For higgsino production in the RPC model, mass limits for the NLSP are shown as a function of ${\cal B}(\ninoone \rightarrow Z\gravino)$ and reach up to 540\unit{GeV}.
For the RPV models, two coupling scenarios are studied where either $\lambda_{12k}$ or $\lambda_{i33}$ are non-vanishing, leading to different branching ratios for \ninoone with decays that include $\tau$ leptons in none, or all decays, respectively.
Limits are presented in the plane of the NLSP vs. LSP mass.
In the $\lambda_{12k}$ scenario, mass limits for \chinoonepm or \ninotwo (\sleptonL or $\tilde{\nu}_\mathrm{L}$) reach up to 1.6\unit{TeV} (1.2\unit{TeV}).
For the $\lambda_{i33}$ scenario, the corresponding numbers are 1.13\unit{TeV} (0.87\unit{TeV}).
Including signal regions with a $b$-jet requirement, the study also set limits on gluino pair production in the two RPV scenarios.

For many RPV models, decays can be fully reconstructed, leading to substantial improvements in background discrimination.
The ATLAS collaboration presented a search for trilepton resonances~\cite{Aad:2020cqu}, inspired by an extension of the MSSM with a spontaneously broken $B-L$ symmetry, leading to lepton number violation at the tree level.
The analysis targets $\chinoonepm\chinoonemp$ and $\chinoonepm\ninoone$ production, as wino-like \chinoonepm and \ninoone particles were identified as likely LSP candidates in a scan of MSSM parameter space.
Possible decay modes are $\chinoonepm \rightarrow Z\ell$, $h\ell$ or $W\nu$, and $\ninoone \rightarrow Z\nu$, $h\nu$ or $W\ell$, with branching ratios that depend on $\tan{\beta}$ and the neutrino mass hierarchy.
Events were recorded using single-electron or -muon triggers.
All signal regions target the trilepton $Z\ell$ resonance expected from chargino decay and require an OSSF lepton pair compatible with \mass{Z}.
Three regions target different decay modes of the second electroweakino: either fully-reconstructed decays, partially-reconstructed decays due to the presence of at least one neutrino, or decays without leptons.
Each region is subdivided in 16 bins in \mass{Z\ell}, where the mass resolution is improved by a correction using the known mass of the $Z$ boson.
In order to resolve ambiguities in the association of leptons with electroweakino decays, a matching procedure is applied that exploits the mass-degeneracy of the electroweakino pair.
Backgrounds from diboson and \ttbar production  are further reduced by region-specific requirements on dilepton masses, the relative mass difference between two fully reconstructed decays, \etmiss, \mt, or the separation between $b$ jets.
For the interpretation in scenarios with high branching ratios to either electrons or muons, versions of the three signal regions are defined with the requirement that the lepton(s) identified as direct decay product of the electroweakino(s) be either electron or muon.

Dedicated control regions are used for the normalisation of the MC simulation for leading background processes: $WZ$, $ZZ$, and $\ttbar Z$.
Contributions from events with \FNP leptons are estimated from data using a fake-ratio method.
No significant excess with respect to the predicted SM background is found, and a fit to the control regions and all signal region bins is performed to establish mass limits in specific models.
Limits are set in the plane defined by the branching ratio to $Z\ell$ versus the electroweakino mass, starting at a mass of 100\unit{GeV}, for four scenarios: exclusive decays to $Ze$, $Z\mu$, or $Z\tau$, or equal probabilities for the three decay channels.
They extend up to 1100\unit{GeV} (1050\unit{GeV}) for ${\cal B}(\chinoonepm,\ninotwo \rightarrow Z\ell) = 100\%$ and decays to electrons (muons) and are lower for the categories involving $\tau$ leptons, and for lower branching ratios.

\subsubsection{Searches in events with a pair of oppositely-charged leptons}\label{sec:eweak-two-l-os}

In the decay of supersymmetric particles, two oppositely-charged leptons can arise in several situations: 
in each decay branch of a pair of oppositely-charged SUSY particles, in the decay chain of a neutral SUSY particle via an intermediate slepton, or from the decay of a $Z$ boson.
In some of these cases, the two leptons will be of the same flavour.

The ATLAS collaboration has performed a generic search for events with two oppositely-charged light leptons in events with \etmiss~\cite{Aad:2019vnb}.
The analysis targets chargino pair production with decays involving $W$ bosons or sleptons, and slepton pair production.
Out of a sample of events recorded using dilepton triggers, only those with exactly two light leptons ($e$ or $\mu$) with $\pt>25\unit{GeV}$ and opposite charge are retained.
In signal events, the two leptons are produced in different decay chains, motivating a requirement of high dilepton mass, $\mll > 100\unit{GeV}$. 
Events with significant \etmiss are selected by imposing lower limits of 110\unit{GeV} for \etmiss and 10 for the \etmiss significance.
The background contribution from \ttbar production is strongly reduced by vetoing events with $b$ jets.
Events are required to have high values of \mttwo and are categorised according to whether the lepton pair is of same or different flavour.
In the first case, the lower limit on \mll is increased to reduce SM backgrounds with $Z$ bosons.
The events are further binned in the number of jets (zero or one), and \mttwo.
Dominant backgrounds with genuine, prompt leptons arise from SM diboson and top-quark production.
Their contribution is estimated by normalising MC simulation to data in control regions enriched in $WW$, $VZ$, or top-quark events.
The yield of backgrounds from events with \FNP leptons passing the selection is estimated with a fake-ratio method using the signal and a baseline definition for leptons.

Results are obtained from fits to the control and signal regions.
A background-only fit using only control regions shows good agreement of the predictions with the yields in the signal regions.
Exclusion limits for specific models are obtained from a fit to all control and signal region bins.
In a scenario of \chinoonepm pair production, with exclusive decays to $\Wpm \ninoone$, mass limits reach up to 420\unit{GeV}.
For slepton-mediated \chinoonepm decays, with equal branching ratios to the three flavours, and a slepton mass half-way between $m_{\chinoonepm}$ and $m_{\ninoone}$ , limits extend to 1\unit{TeV}.
Mass limits for mass-degenerate partners of left- or right-handed electrons or muons are also derived.
If the sleptons of the first two generations are assumed to be mass-degenerate, limits reach up to 700\unit{GeV}.
More inclusive signal regions are used to derive model-independent limits on the product of BSM cross section, acceptance, and efficiency.

An analysis of events with two oppositely-charged light leptons was also performed by the CMS collaboration~\cite{Sirunyan:2020eab}.
Here, only same-flavour pairs were considered.
One set of signal regions targets electroweak production leading to a $Z$ boson identified using the lepton pair, and either an additional, hadronically decaying vector boson ($W$ or $Z$) that can either be "resolved" and reconstructed as two jets, or "boosted" and identified as a large-$R$ jets, or a $h$ boson identified by its decay into two $b$ jets.
A second set of signal regions is optimised for slepton pair production.
In a model of $\chinoonepm\ninotwo$ production, with exclusive decays to $\Wpm \ninoone$ and $Z \ninoone$, respectively, mass limits reach up to 750\unit{GeV}.
In the case of nearly degenerate, higgsino-like \chinoonepm, \ninotwo, and \ninoone states, with \ninoone decays to a $Z$ or $h$ boson and a nearly massless gravitino, mass limits of 800 and 650\unit{GeV} are obtained for ${\cal B}(\ninoone \rightarrow Z \gravino)$ of 100\% and 50\%, respectively.
In all these cases, the observed limits exceed the expected ones by about one standard deviation because of a slight deficit in two high-\etmiss $VZ$ signal regions.
For slepton production, mass limits reach up to 700\unit{GeV} under the hypothesis of four degenerate states of the first two generations.
Furthermore, other signal regions target strong production of gluinos and squarks.

In RPC models with mass differences between the NLSP(s) and the LSP of the order of \unit{GeV}, the transverse momenta of decay products are small compared to typical thresholds, in particular those applied at the trigger level.
Here, leptonic decay modes play an essential role since, compared to jets, hadronic $\tau$ lepton decays, and \etmiss, electrons and muons can be reconstructed down to much smaller \pt.
An example of such an analysis is a CMS study based on events with two or three low-\pt ("soft") leptons and \etmiss~\cite{CMS:2021edw}.
For electroweak production, the target scenarios are production of wino-like $\chinoonepm\ninotwo$  with decays via $W^{\pm*}$ and $Z^{*}$, respectively, to a bino-like LSP, production of higgsino-like \chinoonepm and \ninotwo in the modes $\chinoonepm\ninotwo$ and $\ninotwo\ninoone$, with the same assumptions on the decay, and a higgsino model based on the pMSSM.
The analysis also covers strong production with chargino-mediated stop decays.
One of the challenges of the analysis was the online selection, since standard dilepton triggers at transverse momenta used by the analysis are not available due to their high rate.
The event sample was recorded using a combination of two triggers: a standard \etmiss selection, and a lower-threshold \etmiss trigger with the additional requirement of a dimuon pair with a selection on the its mass, and a minimum \pt of 3\unit{GeV} for each of the two muons, and the dimuon system.
In signal events, the sizeable amount of \etmiss is generated by the emission of one or more jets from initial state radiation,  resulting in a boost of the di-electroweakino system.
Events are selected for the analysis if they contain at least one OSSF lepton pair and categorised according to the number of leptons (two or three).
The $2\ell$ ($3\ell$) category is further divided into four (two) \etmiss bins.
Signal regions are further binned according to the mass of the OSSF lepton pair, \mll.
For signal events, this quantity is limited by $\mass{\ninotwo}-\mass{\ninoone}$.
Depending on the signal region, further criteria are applied.

The leading SM backgrounds are Drell-Yan, $WW$, and \ttbar production for the $2\ell$ signal regions, and $WZ$ production for the $3\ell$ regions.
For the background estimation, control regions are used that are enriched in Drell-Yan, \ttbar, and $WZ$ events.
The background contribution due to \FNP leptons is estimated with a fake-ratio method and further constrained using a control region with same-sign lepton pairs.
In this preliminary result, no significant excess over the SM prediction is observed, and cross section and mass limits are extracted using a fit to the bins of the control and signal regions.
In the bino LSP model, limits are derived for both options for the relative sign of the mass parameters for the two lightest neutralinos and reach maxima of up to 275\unit{GeV} in the region of low mass differences, $\dm \lessapprox 10\unit{GeV}$.
At higher values of \dm, the observed limits are weaker than the expected ones because of data exceeding the predictions in high \mass{\ell\ell} bins.
The maximum local significance in the \mass{\chinoonepm,\ninotwo}-\dm plane is found to be 2.4 standard deviations.
In the simplified higgsino model, the chargino mass is assumed to be half-way between the masses of the two lightest neutralinos.
Here, mass limits reach up to 205\unit{GeV} for $\dm=7.5\unit{GeV}$ and 150 GeV for $\dm=3\unit{GeV}$, where \dm is the \ninotwo-\ninoone mass difference.
For the pMSSM-based higgsino model, results depend mainly on the parameters $M_1$, $M_2$, $\mu$, and $\tan{\beta}$.
For the interpretation, the relation $M_2 = 2 M_1$ is used, motivated by a unification condition, and $\tan{\beta}$ is fixed to a high value ($10$). 
Results are presented in the plane $\mu$-$M_1$, with a maximum limit on $\mu$ of 190\unit{GeV} for $M_1 = 1000\unit{GeV}$.
Limits for two models of stop pair production are also provided.

A similar analysis~\cite{Aad:2019qnd} by the ATLAS collaboration focused on events with two light leptons of opposite charge and the same flavour.
Main differences with respect to the CMS analysis are the design of specific signal regions targeting electroweakino production via vector boson fusion, and slepton pair production.
The discrimination between signal and background is improved by the use of the recursive jigsaw reconstruction~\cite{Jackson:2017gcy}.
The interpretations use the bino LSP and higgsino simplified models described above.
In the bino LSP model, \chinoonepm/\ninotwo mass limits reach up to 240\unit{GeV} at $\dm = 7\unit{GeV}$.
In the simplified model of higgsino production, the corresponding values are 193\unit{GeV} and 9.3\unit{GeV}, respectively.
Limits are also set production of winos and higgsinos via vector boson fusion.
For sleptons, individual limits are presented for the production of supersymmetric partners of left- and right-handed electrons and muons.
Assuming degeneracy for the four states, mass limits reach up to 251\unit{GeV}.

A dedicated search for \stau pair production~\cite{CMS-PAS-SUS-21-001} was conducted by the CMS collaboration and released in preliminary form at the time of writing.
Light partners of $\tau$ leptons are motivated by the possible explanation of the observed dark matter relic density by stau-neutralino coannihilation.
The analysis also covers scenarios with long-lived \stau sleptons that can arise, e.g., in GMSB models.
Events with two hadronically-decaying $\tau$ leptons and significant \etmiss are studied.
The online selection relied on a combination of triggers requiring two $\tau$ leptons or high \etmiss.
All events are required to contain exactly two \tauhad of opposite sign and $\pt>40\unit{GeV}$, with a veto on additional $e$, $\mu$, or $\tau$ candidates, and $\etmiss>50\unit{GeV}$.
Background from \ttbar production is reduced with a veto on $b$ jets, and a minimum azimuthal separation between the two $\tau$ leptons is required in order to reduce the contribution from Drell-Yan production.
Two categories of signal regions are defined for prompt and displaced $\tau$ leptons, respectively.
In the first case, at least one \tauhad has to fail the identification as displaced \tauhad described below.
The sum of the \mt values calculated with the two $\tau$ leptons, $\sum{\mt}$, and \mttwo  are used for further discrimination, and the category is further binned according to the values of $\sum{\mt}$, \mttwo, \njet, and \pt of the leading $\tau$ lepton.
In the second category, both candidates have to be identified as displaced, based on minimal requirements on the significance of the impact parameter in the transverse plane, and the impact parameter in three dimensions.
Further criteria are based on $\sum{\mt}$, \mttwo, and the azimuthal separation between the two leptons, and two signal regions are defined based on the \pt of the trailing lepton.

The main backgrounds are due to \Wjets and multijet production with jets passing the $\tau$ selection, and events with genuine $\tau$ leptons from Drell-Yan, diboson, and top-quark production.
The contribution from misidentified \tauhad is estimated with a fake-ratio method.
The background from events with genuine $\tau$ leptons is estimated with an embedding technique~\cite{CMS:2019pkt}, where dimuon events are selected in data and the footprint of each muon is replaced by a simulated $\tau$ lepton decay.
Data yields in the 31 signal regions are compatible with the predictions.
For the case of prompt \stau decays, cross section limits are set for the hypotheses of production of partners of either left-  or right-handed $\tau$ leptons ($\stau_\mathrm{L}$ or $\stau_\mathrm{R}$), or for the case of two degenerate $\tau$ states ($\stau_\mathrm{L,R}$).
For $\stau_\mathrm{L}$, masses in the range 115--340\unit{GeV} are excluded.
For $\stau_\mathrm{L,R}$, the exclusion is stronger, while the $\stau_\mathrm{R}$ scenario is at the limit of experimental sensitivity.
For the long-lived scenarios, maximal mixing between $\stau_\mathrm{L}$ and $\stau_\mathrm{R}$ is assumed, with a cross section close to that of pure $\stau_\mathrm{R}$ pair production.
Results are derived as a function of the proper lifetime, for six values of $c\tau$ in the range 0.01--2.5mm, for a nearly massless LSP.
For $c\tau = 0.1$mm, masses between 150 and 220\unit{GeV} are excluded.

 A similar search for the production of promptly-decaying \stau pairs in events with two \tauhad by the ATLAS collaboration~\cite{Aad:2019byo} uses \mass{\tau\tau}, the separation between the two leptons in terms of \dr and $\Delta\phi$, and \mttwo as main discriminant variables and derives similar limits for $\stau_\mathrm{L,R}$ and $\stau_\mathrm{L}$.
 
 In models with \Rparity violation, a $\snu_{\tau}$ LSP could be produced resonantly via $\lambda'$ couplings, and decay either via $\lambda'$  to quarks, or via $\lambda$ to dilepton final states. 
 A study~\cite{CMS-PAS-EXO-19-014} performed by the CMS collaboration searches for such events in final states with two charged leptons of different flavour, $e\mu$, $e\tau$, and $\mu\tau$.
 Events were recorded based on the presence of a single, high-\pt electromagnetic or muon signature.
 In the $e\mu$ channel, the search is performed in the distribution of the mass of the $e\mu$ system. 
 In the other two channels, an identified \tauhad is required in addition to a light charged lepton, and events with a small transverse mass, defined using the light lepton and \ptmiss, are rejected to reduce background from misidentified $\tau$ lepton decays.
Here, the dilepton mass is reconstructed assuming collinearity between the neutrino momenta and the visible $\tau$ decay products.
Backgrounds from \ttbar and diboson production, as well as other, smaller contributions are predicted using simulation.
The yields of \Wjets and multijet events, which can enter the sample due to jets passing the lepton selection, are estimated from control samples.
Limits on the product of cross section and branching ratio are established using a Bayesian approach in this preliminary result.
Using a narrow-width approximation of the signal cross section, these limits are interpreted as excluded regions in the plane \mass{\snu_\tau}-$\lambda'$ for different values of $\lambda$.
For $\lambda = \lambda' = 0.1$, mass limits for individual channels range from 3.6 to 4.2\unit{TeV}.

Motivated by another RPV model, the ATLAS collaboration searched for an asymmetry in the cross sections for events with a $e^\pm \mu^\mp$ pair~\cite{ATLAS-CONF-2021-045}.
For proton-proton collisions, the SM predicts the yields for $e^+ \mu^-$ and $e^- \mu^+$ to be almost identical.
However, a non-vanishing $\lambda'_{231}$  coupling would give rise to associated production of a top quark and a smuon.
Assuming a neutralino LSP as the only other accessible sparticle, electron-muon pairs would then arise from the decays $t \rightarrow e^+ \nu_e b$, $\smu^- \rightarrow \mu^- \ninoone$ and their charge conjugates, with a strong enhancement of $\mu^-$ production due to the difference between the proton p.d.f.s for down quarks and anti-quarks.

The search was designed as a model-independent measurement of the ratio $\rho = (\sigma(e^+ \mu^- X) / (\sigma(e^- \mu^+ X)$.
Only well-reconstructed, isolated leptons are considered, and the impact of charge misidentification for electrons is reduced using a BDT discriminant~\cite{Aad:2019tso}.
Events with exactly one electron and one muon of opposite charge are considered.
All signal candidates are required to have a high value of the sum of the \mt values built with the electron and muon, respectively, $\sum{\mt}>200\unit{GeV}$.
Two generic signal regions are defined, as well as two model-specific regions.
For the RPV interpretation, events are required to have high values of the \etmiss significance and of \mttwo.
Potential biases that could lead to differences in the estimated cross sections for the two charges are carefully evaluated and corrected, if necessary.
A major source is due to a combination of the difference in the cross sections for $W^{+} + \mathrm{jets}$ and  $W^{-} + \mathrm{jets}$ production and the misidentification of jets as electrons.
Therefore, the contribution of events with misidentified leptons is estimated from data and subtracted before $\rho$ is calculated.

In this preliminary result, the global measurement of the ratio $\rho$ is found to be consistent with unity, and exclusions limits for the RPV model are obtained by a simultaneous fit to the signal regions of the two charges, and to two corresponding control regions.
For the maximum coupling considered, $\lambda'_{231}=1$, the excluded region for \mass{\smu} reaches up to 650\unit{GeV}, depending on the LSP mass.

\subsubsection{Searches in events with at most one electron or muon}\label{sec:eweak-one-l-had}

Events with a single electron or muon have been used by both collaborations in order to search for production of a pair of wino-like, mass-degenerate  \chinoonepm and \ninotwo, with decays via a $W$ and a $h$ boson that decay to leptons and $b$ quarks, respectively.
In the CMS analysis~\cite{CMS:2021few}, the online selection of events relied on \etmiss, or the presence of a single electron or muon.
The analysis targets events where the Higgs boson can be identified either as a system of two $b$ jets, or as a large-$R$ jet containing all its decay products for high $\pt(h)$.
Large-R jets are clustered with the anti-\kt algorithm with a size parameter $R=0.8$ and are associated with the decay $h \rightarrow  b\bar{b}$ using a dedicated tagging algorithm based on a deep neural network~\cite{Sirunyan:2020lcu}.

Events are required to contain exactly one isolated $e$ or $\mu$, 2--3 small-R jets, with exactly two tagged $b$ jets, no isolated tracks or \tauhad candidates, and to show high \etmiss.
Contributions from SM processes involving a single, leptonically-decaying $W$ boson are strongly suppressed by requiring high values of \mt.
Further selections are based on the contransverse mass\cite{Tovey:2008ui}

\begin{eqnarray}
\mct = \sqrt{2 \pt^{b_1} \pt^{b_2} (1 + \cos{\Delta\Phi(b_1 b_2)})} 
\end{eqnarray}

\noindent that features an endpoint for 
\ttbar events, a mass window around \mass{h} for the $b$-jet pair, and an upper limit for the non-$b$-tagged jet, if present.
The remaining events are assigned to one of 12 signal regions based on whether a $h$-tagged large-$R$ jet is present, and the value of \etmiss.
Backgrounds from SM top-quark production dominate at low \etmiss, while \Wjets and $WV$ production are the most important backgrounds at high \etmiss.
Both categories of background are estimated using dedicated control regions for the normalization of yields predicted from simulation.
Other, smaller background sources, including $Wh$ production, are directly estimated from simulation.
The observed yields in the signal regions are consistent with the predictions.
Mass limits on \chinoonepm, \ninotwo extend up to 820\unit{GeV} for a nearly massless LSP.

A very similar study~\cite{Aad:2019vvf} has been performed by the ATLAS collaboration, with slightly lower observed limits due to a modest excess in some signal regions.
Another search~\cite{ATLAS:2021fbt} by the ATLAS collaboration in final states with at least one light charged lepton, specifically designed for models with RPV decays to multijet final states, was already described in Sec.~\ref{sec:strongRPV}.
Under the assumption of RPV decays via $\lambda"_{323}$, the mass ranges 200--320\unit{GeV} and 197--365\unit{GeV} are excluded for higgsino- and wino-like electroweakinos, respectively.

A CMS search~\cite{CMS:2021pmn} for a heavy $W'$ gauge boson in the \mt spectrum of events with a single electron and muon,  and high \etmiss, is also interpreted in terms of the resonant production of a \stau via an \Rparity violating coupling $\lambda'_{3ij}$, with subsequent decays to $e\nu_e$ ($\mu\nu_{\mu}$) via a $\lambda_{231}$ ($\lambda_{132}$) coupling.
At the time of writing, the analysis is released in preliminary form and establishes model-independent limits on the product of cross section, branching ratio, acceptance, and efficiency as a function of the minimum \mt value.
These limits, together with the simulated \mt spectrum of the RPV signal, are used to set upper limits on $\lambda_{231}$ and $\lambda_{132}$ as a function of \mass{\stau} for the electron and muon channel, respectively, for different values of $\lambda'_{3ij}$ in the range 0.05--0.5 .

Typically, searches for electroweak production of supersymmetric particles have been performed in final states with at least one charged lepton, in order to reduce the overwhelming SM background to a level where sensitivity to the small signal could be achieved.
With increasing centre-of-mass energy and integrated luminosity, higher masses are probed, the expected yields of signal events with high-momentum decay products reach the detection level, and new analysis techniques give access to the identification of boosted, heavy particles.
An example of this development is a recent analysis by the ATLAS collaboration~\cite{ATLAS:2021yqv}, which selects events with hadronic decays of two boosted $W$, $Z$, or $h$ bosons, and \etmiss.
The efficient identification of large-$R$ jets containing the merged decay products of hadronic boson decays provides discrimination with respect to the large SM multijet background and increases signal efficiency as it gives access to higher branching ratios.
The analysis targets a generic class of \Rparity-conserving models with the production of a pair of nearly degenerate, heavy electroweakinos that each decay via a  $W$, $Z$, or $h$ boson into a lighter electroweakino.
The lighter state either terminates the decay chain, or decays via undetectable, soft particles into the LSP.

Two signal categories cover events with four light quarks from the decay of a pair of $W$ or $Z$ bosons, or two light and two $b$ quarks, where the latter originate from a $Z$ or $h$ decay.
The identification of hadronic boson decays is based on large-$R$ jets, clustered with the anti-\kt algorithm with a size parameter $R=1.0$.
The jets are trimmed~\cite{Krohn:2009th} to reduce the effects of soft radiation and pileup.
Subjets, clustered from the constituents of a large-$R$ jet with $R=0.2$, are removed if their \pt amounts to less than 5\% of the \pt of the large-$R$ jet.
The large-$R$ jets are required to have large transverse momentum, $\pt>200\unit{GeV}$, and $|\eta|<2$.
The jet mass, \mj, is  computed by a weighted sum of masses using calorimeter and tracking information~\cite{ATLAS:2016vmy} and required to exceed 40\unit{GeV}.
The $b$-jet multiplicity, $N_b$, is determined by applying $b$-jet identification to track jets inside the large-$R$ jet, clustered with a \pt-dependent $R$ parameter ranging from 0.02--0.4~.
Two types of boson tagging are applied for decays of $W$ or $Z$ bosons to light quarks ($N_b<2$), and for decays of $Z$ or $h$ bosons to $b$ quarks ($N_b$=2), respectively.
In the first case, optimised selections for $W$ and $Z$ decays are obtained using the number of tracks associated with the large-$R$ jet, and \pt-dependent requirements on \mj and an energy correlation function.
In the second case, \mj is corrected for the possible presence of a high-\pt muon, and a $Z$ ($h$) mass window is applied.
Correction factors are applied to the efficiencies determined with MC simulation~\cite{ATLAS:2020szu}.
Events considered in the analysis need to have at least two large-$R$ jets, no identified leptons, $\etmiss>200\unit{GeV}$ to match the trigger requirements, and fulfill further cleaning criteria.
Boson tagging is based on the leading two jets and determines the event category, with either two $W/Z \rightarrow  q\bar{q}^{(')}$ tags, or one $W/Z \rightarrow  q\bar{q}^{(')}$ and one $Z/h \rightarrow  b\bar{b}$ tag.
Each category is subdivided according to the expected composition of the boson pair by applying selections in the plane of the two jet masses.
Further selections are applied on the number of $b$ jets, \etmiss, and the effective mass computed from the two leading jets, the separation of small-R jets from the \ptmiss vector, and \mttwo.
Irreducible backgrounds, mainly from triple-boson and $\ttbar X$ production, are estimated from MC simulation.
Reducible backgrounds with a mistagged $W/Z \rightarrow  q\bar{q}^{(')}$ candidate are estimated by normalisation of the MC prediction in control regions where one of the jets fails the tagging requirement.
The procedure is validated in events with exactly one lepton or photon.

The statistical analysis follows several approaches, as described for the ATLAS three-lepton analysis in Sec.~\ref{sec:eweak-multi-l}.
No significant excess is found in any signal region.
Exclusion limits for specific models are set for a number of hypotheses, depending whether the heavier states or the lighter state are wino- ($\tilde{\mathbf{W}}$), bino- ($\tilde{B}$), or higgsino-like ($\tilde{\mathbf{H}}$).
For the ($\tilde{\mathbf{W}}$,$\tilde{B}$) and ($\tilde{\mathbf{H}}$,$\tilde{B}$) scenarios, the ${\cal B}(\chinoonepm \rightarrow W \ninoone)$ is set to one, while several values of ${\cal B}(\ninotwo \rightarrow Z \ninoone)$ are tested.
In the wino (higgsino) case, \chinoonepm/\ninotwo masses between 400 and 1060\unit{GeV} (\ninotwo masses between 450 and 900\unit{GeV}) are excluded for ${\cal B}(\ninotwo \rightarrow Z \ninoone) = 50\%$ and a light \ninoone.
For the ($\tilde{\mathbf{W}}$,$\tilde{\mathbf{H}}$) and ($\tilde{\mathbf{H}}$,$\tilde{\mathbf{W}}$) scenarios, the branching ratios are derived as a function of the MSSM parameters $M_2$, $\tan{\beta}$, and the sign of $\mu$.
Results are shown in two-dimensional slices of this parameter space.
As in other publications, limits are also derived for specific production modes of the ($\tilde{\mathbf{W}}$,$\tilde{B}$) scenario, setting the most stringent LHC limits to date.
Exclusions are also derived for a GMSB-motivated higgsino scenario, with nearly degenerate \chinoonepm, \ninotwo, and \ninoone, and decays of \ninoone to $Z \gravino$ or $h \gravino$, and for a model based on a SM extension with a QCD axion and with light higgsinos decaying to the superpartner of the axion, which constitutes the LSP.

The CMS collaboration~\cite{CMS-PAS-SUS-20-004} applied similar techniques in a search for decays of higgsino-like $\ninothree\ninotwo$ pairs, with both states (assumed degenerate) decaying to $h$ and a bino-like \ninoone, and for the GMSB-motivated higgsino signal mentioned above (for the specific case of $\ninoone \rightarrow h \gravino$ decays).
For what concerns the identification of decays of boosted $h$ bosons, the main differences with respect to the ATLAS study described above is the use of a specific, DNN-based double-$b$-tagging algorithm for large-$R$ jets, which are clustered with the a size parameter of $R=0.8$.
The identification of events with two boosted $h$ bosons reconstructed as double-$b$-tagged, large-$R$ jets, is complemented with a "resolved" category for lower \pt bosons, requiring 4--5 standard jets, with 3--4 of them identified as $b$-jets.
In one out of 16 signal regions of the resolved category, four events are observed for a prediction of $0.07^{+0.13}_{-0.05}$.
In this preliminary result, the global significance is estimated to be 1.9 standard deviations and the excess is attributed to a statistical fluctuation.
For the higgsino-bino scenario, there is no observed exclusion, while \ninothree, \ninotwo masses were expected to be excluded from 240--540\unit{GeV} for a massless \ninoone.
In the GMSB-inspired model, \ninoone masses from 175 to 1025\unit{GeV} are excluded.
The results are also interpreted in a model of \gluino pair production, each with a decay chain to $q q h \ninoone$.

\subsubsection{Searches in events with isolated photons}\label{sec:eweak-gamma}

Decay chains of neutralinos via an intermediate Higgs boson can be identified using $h \rightarrow \gamma\gamma$ decays.
The small branching ratio of this decay is compensated by lower background levels and combinatorics compared to hadronic decays, and a superior background discrimination due to the excellent mass resolution.
Furthermore, \pt thresholds for photons are higher than for leptons, but substantially lower than for jets or \etmiss.

The ATLAS collaboration used this approach, with techniques similar to those employed in studies of the SM Higgs boson, to search for the production of a pair wino-like $\chinoonepm\ninotwo$ with decays via $Wh$ to two neutralino LSPs, or of pairs of higgsino-like \chinoonepm, \ninotwo, and \ninoone particles, where the two former decay to soft, undetected particles and \ninoone, and with the subsequent decay $\ninoone \rightarrow h \gravino$, as motivated by GMSB models~\cite{Aad:2020qnn}.
In both cases, these electroweakinos are assumed to be almost degenerate in mass.
Events were recorded using a diphoton trigger.
In the event reconstruction, both converted and unconverted photons are used.
The identification of the primary $pp$ vertex is performed with a neural network as used in SM Higgs boson studies.
Events are preselected using properties of the pair of leading photons: the diphoton mass \mass{\gamma\gamma} and the ratios $\pt(\gamma) / \mass{\gamma\gamma}$ for both photons.
Twelve signal regions are then defined based on the number of leptons ($e$ or $\mu$), the number of jets, \njet, the mass of the pair of leading jets (for $\njet \geq 2$), and the \etmiss significance. 

The search is performed in the \mass{\gamma\gamma} distribution.
In this variable, signal and the SM Higgs boson background are described by independent Crystal Ball functions~\cite{Oreglia:1980cs}.
Dominant backgrounds include production of SM Higgs bosons, \gamjets, and di- or triboson production with one or two photons.
The non-resonant background is also estimated from the distribution in data. 
The shape of the \gamjets background is obtained from a control region.
Other backgrounds are directly obtained from simulation.
Results are obtained from simultaneous fits to the mass distributions in all signal regions.
The nominal values for nuisance parameters related to yields and shape parameters for SM Higgs boson production are obtained from theoretical predictions and simulation.
The observed spectrum is consistent with the predictions, with the largest excess in any region corresponding to about two standard deviations.
Limits are set on the visible cross section for each category.
For the wino model, mass limits are derived in the $\mass{\chinoonepm,\ninotwo}$-$\mass{\ninoone}$ plane and reach up to 310\unit{GeV} for a nearly massless LSP.
In the GMSB-inspired higgsino model, the gravitino is nearly massless, and higgsino masses below 380\unit{GeV} are excluded.

\subsubsection{Searches for long-lived particles}\label{sec:eweak-ll}

Searches for a pair of long-lived sleptons, motivated by GMSB models with NLSP sleptons and a nearly massless gravitino as LSP~\cite{Evans:2016zau}, have been performed by the ATLAS and CMS collaborations.
The experimental signature is a pair of leptons that can show displacement with respect to the primary $pp$ collision vertex, but do not form a common decay vertex.

For the ATLAS analysis~\cite{Aad:2020bay}, events were recorded using single- and dilepton triggers that do not use tracking information.
For electrons, this can be achieved with photon triggers that select a signature of electromagnetically interacting particles in the calorimeter.
For muons, only information from the muon system is used.
Thresholds are higher than for fully reconstructed electrons and muons, with a minimum \pt of 50\unit{GeV}, depending on flavour and multiplicity.
In offline reconstruction, tracks are associated with the lepton candidates.
Including a dedicated track reconstruction step, tracks with transverse impact parameters, $d_0$, with respect to the primary vertex of up to 300\unit{mm} are covered.

In order to reduce backgrounds, in particular from the production of prompt leptons, selected leptons are required to satisfy $3 < d_0 < 300\unit{mm}$.
Further selections are applied on \pt, reconstruction quality, and isolation with respect to tracks and calorimeter deposits.
The contribution from cosmic muons is reduced using the timing in the muon system.
At least two displaced leptons are required, and three signal regions are defined based on the flavour of the two leading leptons: $ee$, $\mu\mu$, and $e\mu$.
Standard model backgrounds are strongly reduced by the requirement of a displaced lepton pair.
A small set of further selections includes a minimum opening angle between the leptons in terms of \dr, and a veto on tagged cosmic muons.
Dominant backgrounds for the regions with at least one electron are due to misidentified leptons, or leptons from the decay of heavy-flavour hadrons.
They are estimated from the number of events in regions where the quality selections for one or both leptons are inverted.
In events with two muons, backgrounds originate mainly from cosmic muons failing the tagging requirement.
The contribution is estimated using events where the muon in the upper part of the detector has no cosmic tag, fails to fulfill the quality criteria, or both.

No events are observed in any of the signal regions, and limits on \mass{\slepton} are derived as a function of the lifetime, for different assumptions of the nature of the NLSP.
For the case of mass-degeneracy between the partners of a left- and right-handed lepton, and a lifetime of 0.1\unit{ns}, close to the maximum sensitivity, the lower limits range from 340\unit{GeV} for \stau production to 820\unit{GeV} for the case of equal masses for all flavours.
For the same lifetime, limits are also set for the exclusive production of partners of right-handed electrons and muons (at 580\unit{GeV} and 550\unit{GeV}, respectively), and of left-handed $\tau$ leptons (at 280\unit{GeV}).

The CMS analysis~\cite{CMS:2021kdm} follows a very similar strategy.
Electrons and muons are subject to selections on identification and isolation, and need to pass \pt thresholds varying from 35 to 75\unit{GeV}, depending on the lepton flavour and the year of datataking. 
Events are subject to preselection requirements similar to the ones in the ATLAS analysis and are categorised in $e\mu$, $ee$, and $e\mu$ channels. 
The transverse impact parameter is used to define for each category an inclusive signal region ($100\,\mu\mathrm{m} < d_0 < 10\unit{cm}$), with the upper limit of the latter determined by the requirement of at least one hit in the pixel system.
The signal regions are further divided in four bins based on whether the leptons used to define the category show a $d_0$ below or above $500\,\mu\mathrm{m}$.
Finally, the bins with two leptons with $d_0 < 500\,\mu\mathrm{m}$ are further split according to lepton \pt.
The main backgrounds are estimated using three control regions where the transverse impact parameter of one or both leptons is below $500\,\mu\mathrm{m}$ after taking account possible correlations between the two $d_0$ values.
The results are interpreted in terms of slepton pair production, but also for stop pair production, with \stop decays via RPV couplings to a quark and a lepton.
Compared to the analysis described above, the maximum reach of the mass limits obtained for sleptons are smaller, but the exclusion extends to lower lifetimes.
In terms of \stau pair production, the results are complemented by the dedicated analysis for production of \stau pairs described above in Sec.~\ref{sec:eweak-two-l-os}.

Decays of long-lived charginos to the LSP and soft decay products, which are often below the reconstruction thresholds, lead to a "disappearing track" signature - a charged particle trajectory originating from the primary collision vertex and ending within the tracking system.
The signature has been searched for by both the ATLAS and CMS collaborations.
Motivated by the need of an efficient online selection, the CMS analysis~\cite{CMS:2020atg} focuses on signal events for which a high-\pt ISR jet leads to a boost of the LSPs, and therefore high \etmiss. 
The data sample was collected in 2017 and 2018, after installation of an upgraded pixel detector.
At the first trigger level, which does not have access to tracking information, events were selected exclusively based on \etmiss.
In the high-level trigger, events were retained if the high \etmiss signature was confirmed, or by a dedicated selection combining a looser \etmiss criterion with the requirement of an isolated charged track.
At the analysis level, events are preselected by requiring \etmiss and at least one high-\pt jet.
High-\pt isolated tracks are selected if they are compatible with the reconstructed primary $pp$ vertex and pass a track-based isolation requirement, 
In order to retain only well-reconstructed tracks, no "missing hits" (tracker layers traversed by the trajectory without an associated hit) must be present between the primary vertex and the first hit, and between the first and last hit.
Finally, tracks must have at least three missing hits at the end of the track, and the energy sum in the calorimeters in a \dr cone around the track must be small.

One of the two dominant backgrounds arises from high-\pt leptons with missing hits in the outer tracking system, e.g., because of bremsstrahlung in the case of electrons, or nuclear interactions in the case of hadronically decaying $\tau$ leptons.
It is suppressed by vetoing tracks in a \dr cone around reconstructed leptons.
Furthermore, tracks in regions of lower lepton reconstruction efficiency, or low hit efficiency in the pixel detector, are rejected.
Three signal categories are then defined based on the number of tracker layers with hits, corresponding roughly to track lengths of $<20\unit{cm}$, $20$--$30\unit{cm}$, and $>30\unit{cm}$ for central tracks.
The remaining background from high-\pt leptons that pass the track selection is predicted by individually estimating the conditional probabilities to pass specific steps in the selection, in particular the lepton veto described above, and the \etmiss requirement.
The prediction is made separately for each lepton flavour and signal category and normalised using control samples enriched with leptons of a specific flavour.
A second major background contribution is due to spurious tracks, with an association of hits from different sources.
The probability for an event to contain such a track is estimated in control samples with $Z$ boson decays to pairs of electrons or muons.

Results are interpreted in two scenarios with a higgsino or wino LSP, based on AMSB models with $\tan{\beta}=5$ and $\mu>0$. 
While the mass difference between NLSP and LSP, and the NLSP lifetime can be calculated in the context of the model, a wide range of lifetimes is scanned in order to reduce model dependence.
In the higgsino case, chargino pair production with chargino decays to $\ninotwo X$ or $\ninoone X$, where $\ninotwo$ and $\ninoone$ are mass degenerate, is considered.
Decays to a pion are dominant, with smaller branching ratios to $e\nu$ and $\mu\nu$~\cite{Thomas:1998wy}.
For a lifetime of 3\unit{ns}, chargino masses below 750\unit{GeV} are excluded.
In the wino case, results are combined with the ones of a CMS analysis~\cite{CMS:2018rea} of data recorded in 2015 and 2016.
Production of $\chinoonepm\chinoonemp$ and $\chinoonepm\ninoone$ are considered, and  ${\cal B}(\chinoonepm \rightarrow \pi^{\pm} \ninoone)$ is set to 100\%.
Under these conditions, chargino masses below 884\unit{GeV} are excluded for a lifetime of 3\unit{ns}.

A very similar analysis~\cite{ATLAS:2021ttq} was performed by the ATLAS collaboration.
Here, triggers for the signal region are entirely based on \etmiss.
In a second-pass track reconstruction, pixel tracklets are reconstructed starting from seeds with four pixel hits.
Stringent quality criteria are applied on the tracklets, and the "disappearing" track signature is defined by a veto on hits in the strip semiconductor tracker and an upper limit on the energy deposited in the calorimeter in a narrow \dr cone around the tracklet.
Events with at least one tracklet are selected, with additional criteria on \etmiss, leading jet \pt, and a veto on electron or muon candidates.
The main background sources are the same as for the CMS analysis and are also estimated from data.
Results are reported for the higgsino and wino models mentioned above.
The analysis is also used to set limits on a model of gluino pair production, with charginos appearing in the gluino decay chain.
The same model of strong production with long-lived charginos is also covered by a CMS search for Supersymmetry with the \mttwo variable, already mentioned in Sec.~\ref{sec:gqq}.
At the time of writing, these results are released in preliminary form.

Decays of long-lived supersymmetric particles to quarks via RPV $\lambda''$ couplings can lead to displaced decay vertices with high track multiplicities.
An example are decays of a long-lived neutralino LSP to a top, a bottom, and a strange quark, motivated by an RPV model with minimal flavour violation~\cite{Csaki:2011ge}.
A CMS analysis~\cite{CMS:2021tkn}, already described in Sec.~\ref{sec:strongRPV}, searched for such displaced vertices and obtained limits on neutralino masses up to 1100\unit{\GeV} for decay lengths between 0.1 and 15\unit{mm}, under the assumption of pair production of mass-degenerate higgsino states.

%% file: tex/model_interpr.tex
\begin{figure}
  \begin{center}
    \subfigure[]{
      \includegraphics[width=.47\textwidth]{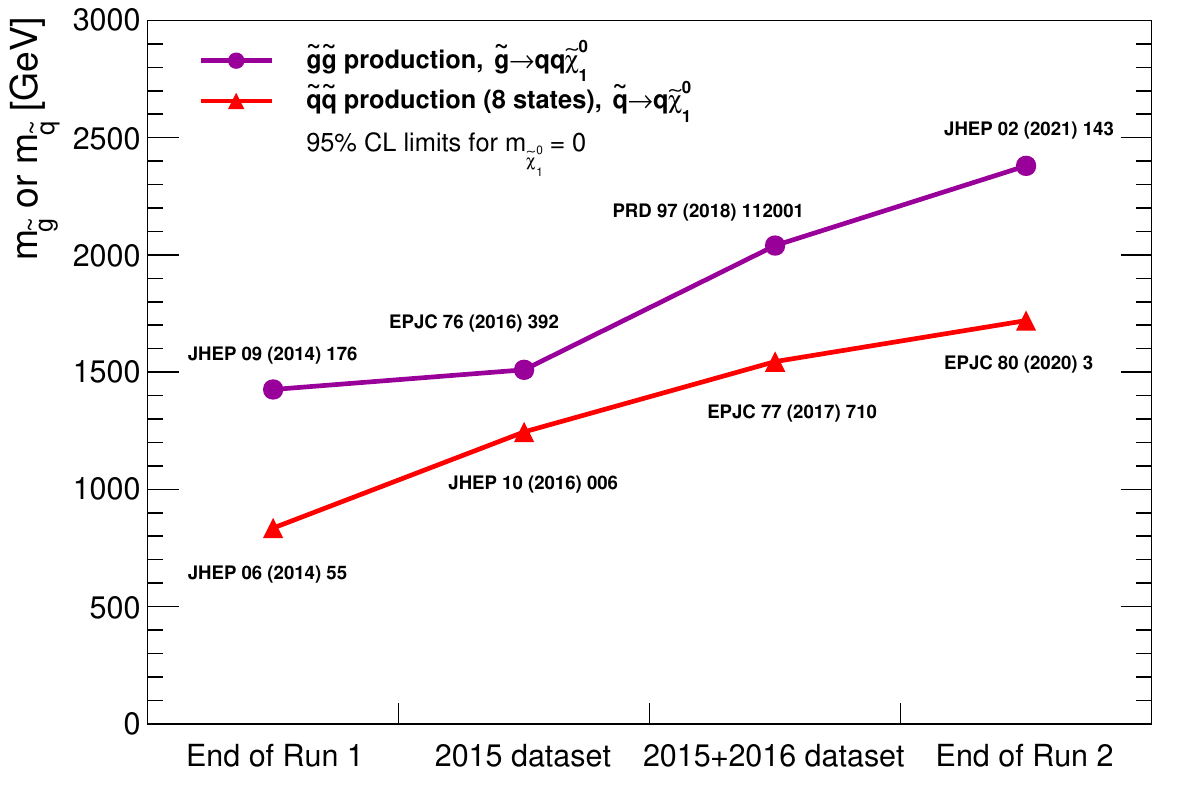}
      } \hfil
      \subfigure[]{
        \includegraphics[width=.47\textwidth]{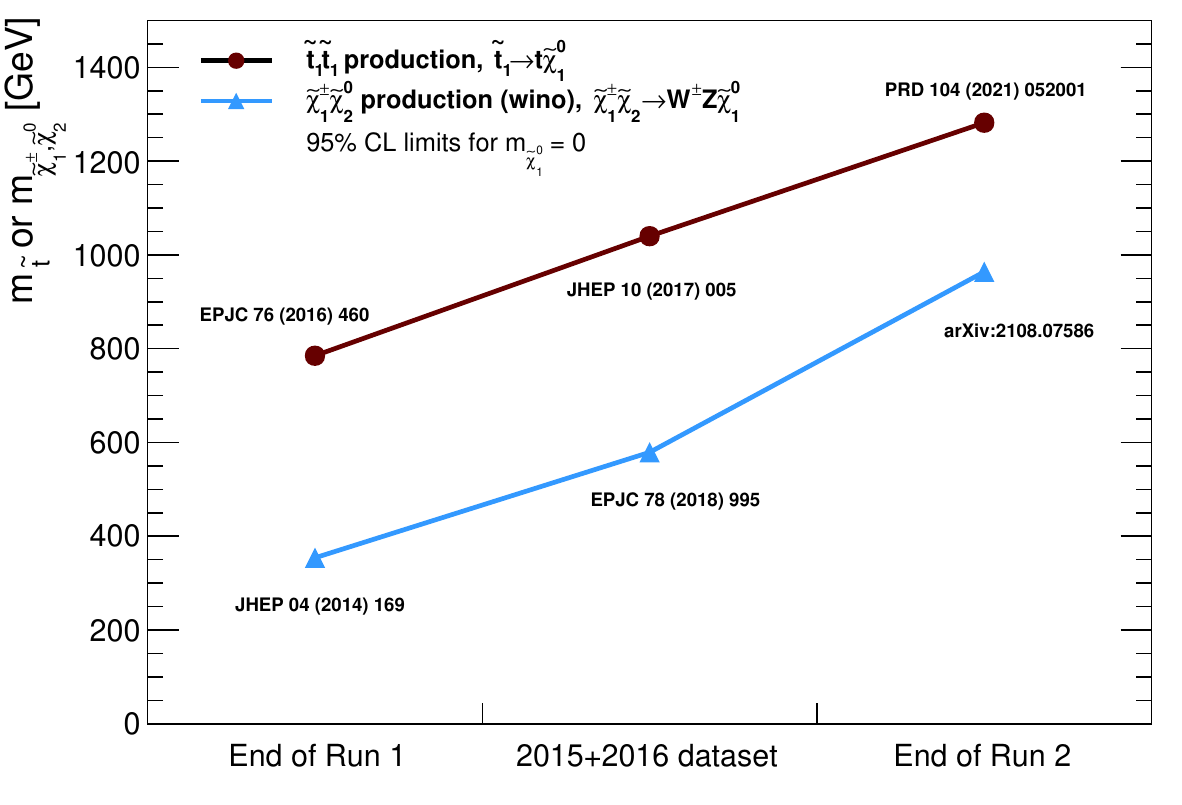}
        }
        \caption{Evolution of mass limits since the end of LHC Run~1:
          (a) pair production of gluinos~\cite{ATLAS:2014jxt,ATLAS:2016dwk,ATLAS:2017mjy,Aad:2020aze} and squarks of the first two generations~\cite{CMS:2014tzs,CMS:2016eju,CMS:2017okm,CMS:2019ybf}; (b) pair production of top squarks~\cite{CMS:2016qpc,CMS:2017mbm,CMS:2021beq} and of mass-degenerate, wino-like $\chinoonepm\ninoone$ pairs~\cite{ATLAS:2014ikz,ATLAS:2018ojr,ATLAS:2021yqv}.
          Selected results from ATLAS ($\gluino\gluino$ and $\ninotwo\chinoonepm$) and CMS ($\squark\squark$ and $\stopone\stopone$) are shown that are obtained using simplified models with direct decays to SM particles and a massless LSP.
        }
  \label{fig:limit-history}
  \end{center}
\end{figure}

The spectacular progress since LHC Run~1 in terms of constraints on masses of pair-produced supersymmetric particles is illustrated in Fig.~\ref{fig:limit-history}.
In the context of simplified models of RPC SUSY with the shortest possible decay chains and a massless LSP, the highest mass limits from searches for gluinos, squarks of the first two, or the third generation, and of partners of SU(2) gauge bosons, show improvements of several 100 GeV to almost a TeV.
The figure focuses on searches for high-momentum, prompt signatures, but improvements are even more striking for experimentally challenging signatures as they arise for small mass differences between the NLSP and the LSP, and in the presence long-lived particles.

Simplified models, introduced in Sec.~\ref{sec:models}, provide useful benchmark models for the optimization of analyses and the demonstration of their sensitivity.
At the same time, they should not be considered as a realistic incarnation of a complete, high-scale SUSY model, which is likely to include more than two or three new particles within the kinematic reach of the LHC, and multiple decay modes, including long decay chains.
Therefore, in the past, interpretations of LHC results in terms of full SUSY models were performed both by the experimental collaborations themselves, and by individual authors with strong participation of the theory community.
Interpretations outside the collaborations, including global fits to pMSSM parameters, were performed with the help of various software frameworks~\cite{Buchmueller:2011aa,Conte:2012fm,Kraml:2013mwa,Dercks:2016npn,GAMBIT:2017yxo}.
They used comparisons with limits obtained in the context of simplified models, simplified likelihoods, or reimplementations of the analyses' selection procedures, and employed different statistical approaches (see Ref.~\cite{LHCReinterpretationForum:2020xtr} for a recent summary).

Data from LHC Run~1 had been used to set limits in the context of the cMSSM where unification of scalar and fermion masses, and trilinear couplings occurs at the GUT scale.
It strongly constrains the phenomenology at the EW scale, and large parts of its parameter space were excluded by the end of LHC Run~1~\cite{ATLAS:2015gky,CMS:2015flg}.
Following the clear need for more flexibility in the description of the SUSY  spectrum, in particular the masses of partners of third generation particles, and the bino / wino / higgsino fraction of the lightest electroweakinos, the phenomenological MSSM (pMSSM) was largely adopted as a framework for further interpretations. 
Compared to the full MSSM, the number of parameters is reduced from more than 100 to 19.

At the time of writing, more results from full Run~2 dataset are still expectected, and global MSSM interpretations are not yet available.
However, the results of two specific analyses targeting production of electroweakinos have also been presented using scans of a three-parameter subspace of the MSSM.
In the ATLAS search for electroweakino production in events with boosted $W$, $Z$, or $h$ bosons~\cite{ATLAS:2021yqv}, scenarios with a set of heavy and a set of light neutralinos were investigated, where the former are wino- (higgsino-)like and the latter higgsino- (wino-)like.
The branching ratios for the heavier states depend on the MSSM parameters $M_2$, $\mu$, and $\tan{\beta}$, and results were derived by sampling this subspace while keeping all other supersymmetric particles decoupled, as reported in Sec.~\ref{sec:eweak-one-l-had}.
In the CMS search for BSM physics in events with two or three low-momentum leptons~\cite{CMS:2021edw}, the sensitivity to scenarios with a higgsino LSP is tested with a pMSSM inspired model.
Here, the parameter \tanb\ was fixed to 10, and $M_1$ and $\mu$ were scanned, with $M_2$ fixed via the relation $M_2 = 2 M_1$, inspired by unification.
Cross sections, mass spectra, and decay rates were calculated based on the pMSSM parameters.
Both analyses are described in more detail in Secs.~\ref{sec:eweak-one-l-had} and \ref{sec:eweak-two-l-os}, respectively.

A larger number of studies has explored parameter (sub-)spaces of pMSSM or equivalent models, based on results from the first substantial dataset recorded at $\sqrt{s}=13\ \TeV$ in 2016.
Examples are papers by the ATLAS and CMS collaborations~\cite{ATLAS:2017xvf,ATLAS:2017mjy,CMS:2018kag}, and by other authors~\cite{GAMBIT:2018gjo,Bagnaschi:2017tru}.
One can expect a similar set of publications in the near future, when the vast majority of individual analyses will have been published by the LHC collaborations.

%% file: tex/conclusions.tex
In this paper, we summarised status and results of direct searches for the production of new particles predicted by supersymmetric models at the end of the second "long shutdown" of the Large Hadron Collider (LHC) at CERN.
We focused on searches for the pair production of supersymmetric particles, excluding efforts to identify an extended Higgs sector as also predicted by Supersymmetry.
At this time, no compelling evidence for the existence of supersymmetric partners of the fermions and bosons of the standard model has been found, and constraints on production cross sections and masses for supersymmetric partners (sparticles) of SM particles are dominated by studies based on proton-proton collisions, recorded by the ATLAS and CMS experiments between 2015 and 2018, during LHC Run~2.
Using the higher centre-of-mass energy and integrated luminosity with respect to LHC Run~1 of 13\unit{TeV} and approximately 140\,\ifb, respectively, the two collaborations have produced a wealth of new results.

Interpretations of the currently published LHC Run~2 searches are almost exclusively based on the minimal supersymmetric extension of the SM and on the concept of simplified models as an interface between experiment and theory, where only a limited number of new states are kinematically accessible.
Limits on the product of cross section and branching ratios are shown as a function of the masses of the involved sparticles and converted into mass limits under the assumption that the predicted cross section does not depend on any other new state, and of decays proceeding exclusively via the specified channels.
The choice of the simplified models was guided by the typical mass hierarchy in different scenarios of soft SUSY breaking and by other considerations such as the naturalness of the SUSY model, and the match with the observed density of dark matter in the universe.

In \Rparity conserving scenarios with a stable LSP, and for decays to a light LSP and quark(s), mass limits for gluinos now extend beyond 2.2\unit{TeV}, and to about 1.8\unit{TeV} for eight mass-degenerate squarks of the first two generations.
Special attention has been paid to searches for partners of the top quark, as they could play an important role in the cancellation of contributions to the mass of the Higgs boson, and help to achieve the correct prediction for the density of dark matter via co-annihilation.
For high mass differences \dm with respect to the LSP, and two-body decays to $t \ninoone$, mass limits reach up about 1.3\unit{TeV}.
For lower \dm, decays would proceed via three- and four-body modes and the constraints are weaker, but still reach about 600\unit{GeV} for mass difference down to a few tens of \unit{GeV}.
These limits start to constitute a serious challenge to natural models of Supersymmetry when defined using traditional measures of naturalness. 

Searches for electroweakinos --- partners of neutral and charged higgs ("higgsinos") and electroweak gauge ("binos", "winos") bosons -- have been conducted under different assumptions on the nature of the LSP and NLSP states that determine the production cross section, decay modes, and the mass splitting between sets of nearly degenerate states.
Here, the large integrated luminosity of Run~2 resulted in a substantial increase of sensitivity with respect to Run~1.
In the most favourable scenario, members of a set of nearly degenerate wino states (\chinoonepm,\ninotwo) decay to a bino LSP (\ninoone).
The strongest mass limits for decays to an SM boson and a light LSP are now established by a search for a fully hadronic final state from the decay $\chinoonepm\ninotwo \rightarrow \Wpm h\ninoone\ninoone$ and reach beyond 1\unit{TeV}, illustrating the combined potential of a large dataset giving access to high \pt regions, and advanced analysis techniques.
Limits for decays to $WZ\ninoone\ninoone$ extend down to $\dm < \mass{Z}$, with sensitivity up to $\mass{\ninoone} \approx 250\unit{GeV}$.
Even stronger results are obtained under the assumption of decay chains involving sleptons that lead to higher lepton multiplicities in the final state.
Searches for higgsinos suffer from a smaller cross section with respect to wino pair production, and predicted mass splittings for pure higgsinos in the sub-GeV region.
Analyses using events with low-momentum leptons have achieved maximum exclusions for \mass{\ninotwo} beyond 200\unit{GeV} at $\dm(\ninotwo,\ninoone) \approx 7.5\unit{GeV}$, assuming a chargino mass halfway between \mass{\ninotwo} and \mass{\ninoone}.
These results are complemented by constraints from searches for long-lived charginos, which apply over a large range of chargino masses, with upper limits at $\dm(\chinoonepm,\ninoone) \approx 0.3\unit{GeV}$ for $\mass{\chinoonepm} = 200\unit{GeV}$.
In terms of sensitivity, a gap in terms of \dm remains between the two approaches, but they start to challenge the hypothesis of light higgsinos, which is generally considered as a robust feature of natural SUSY models.

Limits for mass-degenerate partners of charged leptons of the first two generations now reach up to $\mass{\slepton} \approx 700\unit{GeV}$. 
As for the case of the squarks, third generation sleptons attracted special attention due to their possible role in co-annihilation.
Here, masses in the range 120 to 390\unit{GeV} are excluded for degenerate partners of left- and right-handed $\tau$ leptons and a massless LSP.

A large number of results are also available for simplified models that are motivated by gauge-mediated SUSY breaking, with decay chains ending with a nearly massless, undetected gravitino.
The models used for interpretation cover pair production of gluinos, squarks, sleptons, and electroweakinos.
For strong production of gluino pairs, with decays via a lightest neutralino with bino and higgsino components, the strongest limits on \mass{\gluino} reach 2.4\unit{TeV} .
In the case when gluinos and squarks are decoupled and the lightest electroweakino states are nearly degenerate, the signal acceptance with respect to the searches described above is largely increased, and limits for degenerate higgsino states decaying to the gravitino reach up to 1\unit{TeV} under the assumption of ${\cal  B}(\tilde{H} \rightarrow h\gravino)  = 100\%$.

Many supersymmetric models predict long-lived particles because of reduced phase space in highly compressed mass spectra, weak couplings, e.g., to the gravitino, or decays via highly off-shell particles.
A variety of analysis techniques cover proper lifetimes ranging from prompt production to the multi-microsecond scale, with best sensitivities typically reached in the sub-nanosecond region where reconstruction efficiency is still high while SM backgrounds are strongly suppressed.
For long-lived gluinos, which give rise to $R$-hadrons, and $\mass{\ninoone} = 100\unit{GeV}$, mass limits now reach beyond $2.5\unit{TeV}$.
As mentioned above, searches for long-lived charginos as part of a light, almost degenerate wino multiplet, lead to a "disappearing track" signature.
For lifetimes in the nanosecond range, the best mass limits reach almost 900\unit{GeV} for a wino LSP and 750\unit{GeV} for a higgsino LSP, with lower values in the sub-nanosecond range predicted by some models.

While most of the studies mentioned above rely on an \etmiss signature expected from the production of stable, undetected LSPs, a vast programme of searches targets supersymmetric models with \Rparity violating couplings.
Here, the experimental handles are high multiplicity of jets, including those associated with $b$ quarks, full reconstruction of resonances, and displaced objects for the case of small couplings.
Within the severe, existing constraints on the size of RPV couplings, different scenarios with non-vanishing trilinear RPV couplings $\lambda'$, $\lambda''$, or $\lambda'''$ have been investigated.
Several decay modes have been considered: decays of electroweakinos via $\lambda'$, $\lambda''$, or $\lambda'''$ to three leptons, two quarks and a lepton, or three quarks, respectively; or decays of squarks or gluinos via $\lambda''$ to two or three quarks, respectively.
Depending on the assumptions on the decay modes, masses up 2.5\unit{TeV} have been probed for the pair production of gluinos, up to 1.4\unit{TeV} for top squarks, up to 1.2\unit{TeV} for sleptons, and up to 1.6\unit{TeV} for wino-like NLSPs.
Models with LSP decays through small RPV couplings have also motivated searches for long-lived particles, with a sensitivity that often exceeds the one of equivalent searches for promptly produced particles.
Finally, a few scenarios of single slepton production through a $\lambda'$ coupling have also been investigated.

Since the end of LHC Run~1, the mass reach of many searches for pair production of supersymmetric partners of gluons, quarks of the first two, or the third generation, and of SU(2) gauge bosons, has improved by hundreds of GeV.
The simplified models used for the interpretation of the various analyses are benchmarks motivated by existing constraints or considerations such as naturalness or consistency with measurements of the dark matter relic density, but one should keep in mind that the mass limits derived in this context represent optimistic scenarios in terms of branching ratios.
Nevertheless, limits on gluino and top squark production challenge the concept of naturalness for several measures of naturalness, and the new results on higgsino production start to exclude low-mass regions that are considered a robust feature of natural SUSY models.
Limits on the product of cross section and branching ratios, which are also provided, can also be used for more general interpretations.

In the next 1--2 years some more results based on the full dataset collected in LHC Run~2 can be expected from the collaborations, including combinations of multiple analyses.
At this time, all the material will be available for the interpretation in more global models such as the phenomenological MSSM, within or outside the LHC collaborations, that can also take into account the latest constraints from the Higgs sector, precision measurements, and cosmology, based on the limits obtained in the context of benchmarks, or based on a more detailed modelling of the analyses using the detailed material in the publications and their auxiliary material.
As the experimental results are signature-oriented, they can also be relevant to BSM models other than Supersymmetry.

The next data taking period at the LHC, Run~3, is about to start.
It is expected to roughly double the integrated luminosity available for high-energy proton-proton collisions at centre-of-mass energies in the range 13--14\unit{TeV}.
One can expect that the progress at the high-mass frontier will become slower, with significant improvements only at the end of Run~3.
However, during Run~2 we have seen a rapid development of new analysis techniques, such as a widespread deployment of machine learning at the object and analysis level, application of specific reconstruction and background estimation techniques for unconventional objects such as displaced or delayed photons, leptons, or jets, and the identification of the merged decay products of highly boosted, massive particles.
Certainly, the full potential of these developments has not yet been exploited and will lead to higher sensitivity even for early Run~3 searches, beyond what can be expected from the increase in luminosity or a possible, moderate change of collision energy.
Again, their impact will be the highest for signatures where one cannot rely on prompt production of high-momentum particles.
In this area, the strongest constraints for the general purpose experiments will still be at the first, hardware-based level of the trigger system.
However, progress in the capacity of the second, software-based part of this system, made possible by the deployment of heterogeneous computer farms, will allow a reconstruction and selection of events closer to the performance of the final analysis.
In addition, strategies like analyses that are entirely based on objects reconstructed in the trigger farm, or delayed processing of datasets, will allow to increase the acceptance for many critical searches.
These concepts have already been tested in LHC Run~2 and will be further developed for Run~3.

Further in the future, a decisive step will be made in the transition to the high-luminosity LHC (HL-LHC) programme (Run~4 and beyond). 
The HL-LHC will not only increase the total integrated luminosity by a further order of magnitude, compared with the situation at the end of Run~3, but the upgrade of the experiments will open new possibilities for BSM searches.
In particular, tracking information will become available at the first trigger level, with large improvements in the background suppression for isolated objects.
The new tracking capabilities, together with timing information from existing, but also new, dedicated timing detectors will enhance the possibility to select events with long-lived particles.

Despite the lack of any signal for supersymmetric particle production (or any other BSM production) at this point in the lifetime of the LHC, the effort that the collaborations have put in this research programme has produced a dramatic impact to the field. SUSY models are an inexhaustible source of key experimental signatures to be explored: they have been an inspiration for hundreds of scientists to push the performance of detectors and algorithms always further, and obtain sensitivities that before the LHC startup could only be dreamt of. The lack of a SUSY signal so far is truly reshaping the approach of the HEP community to model design, and is producing a healthy rethinking to how the SM should be extended. And while new paradigms of BSM physics are being developed, the search for SUSY continues. Run~2 has only began to scratch the interesting model parameter space for some key signatures (higgsinos and staus in particular) and certainly the collaborations continue to exceed expectations with creativity and new experimental strategies.

%% file: tex/analysisSummary_pdgFormat.tex
\section*{Summary of simplified models covered by ATLAS and CMS}

\begin{table}[htbp]
\begin{tabular}{clc|clc}
\hline
\hline
\multicolumn{4}{c}{Gluino pair production}\\
\hline
Simplified model & ATLAS Results & CMS results & Comment \\ 
(Using \href{https://pdg.lbl.gov/2021/reviews/searches-and-hypothetical-particles.html}{PDG nomenclature})  & & & \\
\hline
Tglu1A & \href{https://arxiv.org/abs/2010.14293}{2010.14293} & \href{https://arxiv.org/abs/1909.03460}{1909.03460}  & \\
	& 	& \href{https://arxiv.org/abs/1908.04722}{1908.04722} & \\
\hline
Tglu1B &  \href{https://arxiv.org/abs/2010.14293}{2010.14293}&\href{https://arxiv.org/abs/1909.03460}{1909.03460} & \\
 & \href{https://arxiv.org/abs/2101.01629}{2101.01629} & \href{https://arxiv.org/abs/2001.10086}{2001.10086} & \\
\hline 
Tglu1C & &  \href{https://arxiv.org/abs/1909.03460}{1909.03460} & \\
	& 	& \href{https://arxiv.org/abs/1908.04722}{1908.04722} & \\
\hline 
Tglu1E & \href{https://arxiv.org//abs/2008.06032}{2008.06032} & \href{https://arxiv.org/abs/2001.10086}{2001.10086} & \\
             & \href{https://arxiv.org/abs/1909.08457}{1909.08457} & & \\
\hline 
Tglu2A & & \href{https://arxiv.org/abs/1909.03460}{1909.03460} & \\
	& 	& \href{https://arxiv.org/abs/1908.04722}{1908.04722} & \\
\hline
Tglu3A & \href{https://arxiv.org//abs/2008.06032}{2008.06032} & \href{https://arxiv.org/abs/1909.03460}{1909.03460} & \\
	& & \href{https://arxiv.org/abs/1908.04722}{1908.04722} & \\
	& & \href{https://arxiv.org/abs/2001.10086}{2001.10086} & \\
	& & \href{https://arxiv.org/abs/2103.01290}{2103.01290} & \\
	& & \href{https://arxiv.org/abs/1911.07558}{1911.07558} & \\ 
\hline
Tglu3B & & \href{https://arxiv.org/abs/2001.10086}{2001.10086} & \\
	& & \href{https://arxiv.org/abs/1911.07558}{1911.07558} & \\ 
\hline
Tglu3C & & \href{https://arxiv.org/abs/2001.10086}{2001.10086} & \\
            & & \href{https://arxiv.org/abs/2103.01290}{2103.01290} & \\
\hline
Tglu4C & &  \href{https://arxiv.org/abs/2012.08600}{2012.08600} & \\
\hline
Tglu3D & & \href{https://arxiv.org/abs/2001.10086}{2001.10086} & \\
	& & \href{https://arxiv.org/abs/2103.01290}{2103.01290} & \\
\hline
Tglu1LL & & \href{https://arxiv.org/abs/1909.03460}{1909.03460} & \\
\hline
Tglu3A & \href{https://arxiv.org/abs/2106.09609}{2106.09609} & & RPV $\stopone$ decay \\
\hline
Tglu1A& \href{https://arxiv.org/abs/2106.09609}{2106.09609} & & RPV $\ninoone$ decay\\
\hline
$\gluino\gluino/\ninoone\ninoone\rightarrow \stopone\stopone$ and \stopone\stopone & & \href{https://arxiv.org/abs/2104.13474}{2104.13474} & RPV $\stopone$ decay\\
\hline
$\gluino\gluino\rightarrow gg \tilde{G}\tilde{G}$ & & \href{https://arxiv.org/abs/1906.06441}{1906.06441} & Long-lived gluinos\\
 & & \href{https://arxiv.org/abs/2012.01581}{2012.01581} & \\
 \hline
 Tglu1A & \href{https://arxiv.org/abs/2104.03050}{2104.03050} &  \href{https://arxiv.org/abs/2012.01581}{2012.01581} & Long-lived gluinos\\
\hline
Tglu3A &  & \href{https://arxiv.org/abs/2012.01581}{2012.01581} & RPV $\stopone$ decay, Long-lived $\stopone$ \\
\end{tabular}
\end{table}

\begin{table}[htbp]
\begin{tabular}{clc|clc}
\hline
\hline
\multicolumn{4}{c}{Squark pair production}\\
Simplified model & ATLAS Results & CMS results & Comment \\ 
(Using \href{https://pdg.lbl.gov/2021/reviews/searches-and-hypothetical-particles.html}{PDG nomenclature})  & & & \\
\hline
Tsqk1 & \href{https://arxiv.org/abs/2010.14293}{2010.14293} & \href{https://arxiv.org/abs/1909.03460}{1909.03460} & \\
	& \href{https://arxiv.org/abs/2102.10874}{2102.10874}	& \href{https://arxiv.org/abs/1908.04722}{1908.04722} & \\
	
Tsqk2  & \href{https://arxiv.org/abs/2101.01629}{2101.01629} & & \\
\hline
Tsqk1LL & & \href{https://arxiv.org/abs/1909.03460}{1909.03460} & \\
\hline
Tsqk3	& \href{https://arxiv.org/abs/2010.14293}{2010.14293}	&  & \\	
\hline
GGM $\tilde{q}_1\tilde{q}_1$ & & \href{https://arxiv.org/abs/1909.06166}{1909.06166} & Long-lived $\ninoone$ \\
\hline
Tstop1 & \href{https://arxiv.org/abs/2102.01444}{2102.01444} &\href{https://arxiv.org/abs/1909.03460}{1909.03460} & \\
	& \href{https://arxiv.org/abs/2012.03799}{2012.03799}	& \href{https://arxiv.org/abs/1908.04722}{1908.04722} & \\
	& \href{https://arxiv.org/abs/2004.14060}{2004.14060} & \href{https://arxiv.org/abs/2103.01290}{2103.01290} & \\
\hline
Tstop1LL &  &\href{https://arxiv.org/abs/1909.03460}{1909.03460} & \\
\hline
Tstop2 & & \href{https://arxiv.org/abs/1909.03460}{1909.03460} & \\
	& & \href{https://arxiv.org/abs/2103.01290}{2103.01290} & \\
\hline
Tstop3 & \href{https://arxiv.org/abs/2102.10874}{2102.10874} & \href{https://arxiv.org/abs/2111.06296}{2111.06296} & \\
 & \href{https://arxiv.org/abs/2102.01444}{2102.01444} &\href{https://arxiv.org/abs/2103.01290}{2103.01290} & \\
 & \href{https://arxiv.org/abs/2012.03799}{2012.03799} & & \\
 & \href{https://arxiv.org/abs/2004.14060}{2004.14060} & & \\
 \hline
 $\stopone\stopone$ & \href{https://arxiv.org/pdf/2003.11956.pdf}{2003.11956} &  \href{https://arxiv.org/abs/2012.01581}{2012.01581}  & RPV $\stopone$, Long-lived stops\\
 \hline
 $\stopone\stopone$ into $t\ninoone$, $t\ninotwo$, b$\chinoonepm$ & \href{https://arxiv.org/abs/2010.01015}{2010.01015} & & RPV $\ninoone$, $\ninotwo$, $\chinoonepm$ decays \\
  & \href{https://arxiv.org/abs/2106.09609}{2106.09609} & & \\
  \hline
  Tstop1 & & \href{https://arxiv.org/abs/2102.06976}{2102.06976} & RPV $\ninoone$ decay \\
\hline 
Tstop4 & \href{https://arxiv.org/abs/2102.10874}{2102.10874} & \href{https://arxiv.org/abs/1909.03460}{1909.03460} & \\
     & & \href{https://arxiv.org/abs/2103.01290}{2103.01290} & \\
\hline 
Tstop5 & \href{https://arxiv.org/abs/2108.07665}{2108.07665} & & \\
\hline
Tstop7 & \href{https://arxiv.org/abs/2006.05880}{2006.05880} & \href{https://arxiv.org/abs/2001.10086}{2001.10086} & \\
\hline
Tstop8 & & \href{https://arxiv.org/abs/2103.01290}{2103.01290} & \\
\hline
Tstop10 & & \href{https://arxiv.org/abs/2103.01290}{2103.01290} & \\
\hline
Tsbot1 & \href{https://arxiv.org/abs/2101.12527}{2101.12527} &  \href{https://arxiv.org/abs/1909.03460}{1909.03460} & \\
	& \href{https://arxiv.org/abs/2102.10874}{2102.10874} & \href{https://arxiv.org/abs/1908.04722}{1908.04722} & \\
\hline 
Tsbot2 & \href{https://arxiv.org/abs/1909.08457}{1909.08457} & \href{https://arxiv.org/abs/2001.10086}{2001.10086} & \\
\hline
Tsbot3 & & \href{https://arxiv.org/abs/2012.08600}{2012.08600} & \\
\hline
Tsbot4 & \href{https://arxiv.org/abs/1908.03122}{1908.03122} & & \\
\hline
Tsbot5 & \href{https://arxiv.org/abs/2103.08189}{2103.08189} & & \\
\end{tabular}
\end{table}

\begin{table}[htbp]
\begin{tabular}{clc|clc}
\hline
\hline
\multicolumn{4}{c}{Electroweakino pair production}\\
Simplified model & ATLAS Results & CMS results & Comment \\ 
(Using \href{https://pdg.lbl.gov/2021/reviews/searches-and-hypothetical-particles.html}{PDG nomenclature})  & & & \\
\hline 
Tchi1chi1C & \href{https://arxiv.org/abs/1908.08215}{1908.08215} & & \\
\hline 
TwinoLSP & & \href{http://arxiv.org/abs/arXiv:2004.05153}{2004.05153} & disappearing track signature\\  
\hline
Tchi1n2A & & \href{https://arxiv.org/abs/2106.14246}{2106.14246} & \\
\hline 
Tchi1n2E & \href{https://arxiv.org/abs/1909.09226}{1909.09226} & \href{https://arxiv.org/abs/2106.14246}{2106.14246} & \\
		& \href{https://arxiv.org/abs/2004.10894}{2004.10894} & \href{https://arxiv.org/abs/2107.12553}{2107.12553} & \\
		& \href{https://arxiv.org/abs/2106.01676}{2106.01676} & & \\
\hline 
Tchi1n2F & \href{https://arxiv.org/abs/1911.12606}{1911.12606} & \href{https://arxiv.org/abs/2111.06296}{2111.06296} & compressed states \\
		& \href{https://arxiv.org/abs/1912.08479}{1912.08479} & \href{https://arxiv.org/abs/2106.14246}{2106.14246} & \\
		& \href{https://arxiv.org/abs/2106.01676}{2106.01676} & \href{https://arxiv.org/abs/2012.08600}{2012.08600} & \\ 
\hline
Tchi1n2G & \href{https://arxiv.org/abs/1911.12606}{1911.12606} & \href{https://arxiv.org/abs/2111.06296}{2111.06296} & compressed states \\
\hline
Tchi1n2H & & \href{https://arxiv.org/abs/2106.14246}{2106.14246} & \\ 
\hline
Tchi1chi1H & \href{https://arxiv.org/abs/1908.08215}{1908.08215} & & \\
\hline
Tn1n1A & \href{https://arxiv.org/abs/2004.10894}{2004.10894} & \href{https://arxiv.org/abs/2106.14246}{2106.14246} & \\
\hline 
Tn1n1B & & \href{https://arxiv.org/abs/2106.14246}{2106.14246} & \\
	     & & \href{https://arxiv.org/abs/2012.08600}{2012.08600} & \\
\hline 
Tn1n1C & & \href{https://arxiv.org/abs/2106.14246}{2106.14246} & \\
	     & & \href{https://arxiv.org/abs/2012.08600}{2012.08600} & \\
\hline 
TwinoLSPRPV& \href{https://arxiv.org/abs/2011.10543}{2011.10543} & & Resonant search\\
\hline 
Wino NLSP & \href{https://arxiv.org/abs/2103.11684}{2103.11684} & & RPV $\ninoone$ decay\\ 
\hline 
Wino NLSP & \href{https://arxiv.org/abs/2108.07586}{2108.07586} & & All hadronic \\
\hline
Higgsino NLSP & \href{https://arxiv.org/abs/2108.07586}{2108.07586} & & All hadronic\\
\hline
Wino LSP & \href{https://arxiv.org/abs/2106.09609}{2106.09609} & & RPV $\ninoone$ decay \\
\hline
Higgsino LPS & \href{https://arxiv.org/abs/2106.09609}{2106.09609} & & RPV $\ninoone$ decay \\
\hline
Tchi1n12\_GGM & \href{https://arxiv.org/abs/2103.11684}{2103.11684} & & \\
\hline
\end{tabular}
\end{table}

\begin{table}[htbp]
\begin{tabular}{clc|clc}
\hline
\hline
\multicolumn{4}{c}{Slepton pair production}\\
Simplified model & ATLAS Results & CMS results & Comment \\ 
(Using \href{https://pdg.lbl.gov/2021/reviews/searches-and-hypothetical-particles.html}{PDG nomenclature})  & & & \\
\hline 
$\tilde{\ell}\tilde{\ell}$, $\tilde{\ell}\rightarrow \ell \ninoone$, $\ell = e,\mu$ & \href{https://arxiv.org/abs/1908.08215}{1908.08215} & \href{https://arxiv.org/abs/2012.08600}{2012.08600}  & \\
           & \href{https://arxiv.org/abs/1911.12606}{1911.12606} & & compressed states \\
           & \href{https://arxiv.org/abs/2011.07812}{2011.07812} & & Long-lived sleptons \\
\hline 
$\tilde{\tau}\tilde{\tau}$, $\tilde{\tau}\rightarrow \tau \ninoone$  & \href{https://arxiv.org/abs/1911.06660}{1911.06660} & & \\
\end{tabular}
\end{table}